\input harvmac
%
\message{S-Tables Macro v1.0, ACS, TAMU (RANHELP@VENUS.TAMU.EDU)}
%
%
\newhelp\stablestylehelp{You must choose a style between 0 and 3.}%
\newhelp\stablelinehelp{You should not use special hrules when
stretching
a table.}%
\newhelp\stablesmultiplehelp{You have tried to place an S-Table
inside another S-Table.  I would recommend not going on.}%
%
%
\newdimen\stablesthinline
\stablesthinline=0.4pt
\newdimen\stablesthickline
\stablesthickline=1pt
%
%
\newif\ifstablesborderthin
\stablesborderthinfalse
\newif\ifstablesinternalthin
\stablesinternalthintrue
\newif\ifstablesomit
\newif\ifstablemode
\newif\ifstablesright
\stablesrightfalse
%
%
\newdimen\stablesbaselineskip
\newdimen\stableslineskip
\newdimen\stableslineskiplimit
%
%
\newcount\stablesmode
\newcount\stableslines
\newcount\stablestemp
\stablestemp=3
\newcount\stablescount
\stablescount=0
\newcount\stableslinet
\stableslinet=0
%
%
%
\newcount\stablestyle
\stablestyle=0
%
%
\def\stablesleft{\quad\hfil}%
\def\stablesright{\hfil\quad}%
%
%
\catcode`\|=\active%
%
%
\newcount\stablestrutsize
\newbox\stablestrutbox
\setbox\stablestrutbox=\hbox{\vrule height10pt depth5pt width0pt}
\def\stablestrut{\relax\ifmmode%
                         \copy\stablestrutbox%
                       \else%
                         \unhcopy\stablestrutbox%
                       \fi}%
%
%
\newdimen\stablesborderwidth
\newdimen\stablesinternalwidth
\newdimen\stablesdummy
\newcount\stablesdummyc
\newif\ifstablesin
\stablesinfalse
%
%
\def\begintable{\stablestart%
  \stablemodetrue%
  \stablesadj%
  \halign%
  \stablesdef}%
\def\stablesadj{%
  \ifcase\stablestyle%
    \hbox to \hsize\bgroup\hss\vbox\bgroup%
  \or%
    \hbox to \hsize\bgroup\vbox\bgroup%
  \or%
    \hbox to \hsize\bgroup\hss\vbox\bgroup%
  \or%
    \hbox\bgroup\vbox\bgroup%
  \else%
    \errhelp=\stablestylehelp%
    \errmessage{Invalid style selected, using default}%
    \hbox to \hsize\bgroup\hss\vbox\bgroup%
  \fi}%
\def\stablesend{\egroup%
  \ifcase\stablestyle%
    \hss\egroup%
  \or%
    \hss\egroup%
  \or%
    \egroup%
  \or%
    \egroup%
  \else%
    \hss\egroup%
  \fi}%
\def\stablestart{%
  \ifstablesin%
    \errhelp=\stablesmultiplehelp%
    \errmessage{An S-Table cannot be placed within an S-Table!}%
  \fi
  \global\stablesintrue%
  \global\advance\stablescount by 1%
  \message{<S-Tables Generating Table \number\stablescount}%
  \begingroup%
  \stablestrutsize=\ht\stablestrutbox%
  \advance\stablestrutsize by \dp\stablestrutbox%
  \ifstablesborderthin%
    \stablesborderwidth=\stablesthinline%
  \else%
    \stablesborderwidth=\stablesthickline%
  \fi%
  \ifstablesinternalthin%
    \stablesinternalwidth=\stablesthinline%
  \else%
    \stablesinternalwidth=\stablesthickline%
  \fi%
  \tabskip=0pt%
  \stablesbaselineskip=\baselineskip%
  \stableslineskip=\lineskip%
  \stableslineskiplimit=\lineskiplimit%
  \offinterlineskip%
  \def\borderrule{\vrule width \stablesborderwidth}%
  \def\internalrule{\vrule width \stablesinternalwidth}%
  \def\thinline{\noalign{\hrule height \stablesthinline}}%
  \def\thickline{\noalign{\hrule height \stablesthickline}}%
  \def\trule{\omit\leaders\hrule height \stablesthinline\hfill}%
  \def\ttrule{\omit\leaders\hrule height \stablesthickline\hfill}%
  \def\tttrule##1{\omit\leaders\hrule height ##1\hfill}%
  \def\stablesel{&\omit\global\stablesmode=0%
    \global\advance\stableslines by 1\borderrule\hfil\cr}%
  \def\el{\stablesel&}%
  \def\elt{\stablesel\thinline&}%
  \def\eltt{\stablesel\thickline&}%
  \def\elttt##1{\stablesel\noalign{\hrule height ##1}&}%
  \def\elspec{&\omit\hfil\borderrule\cr\omit\borderrule&%
              \ifstablemode%
              \else%
                \errhelp=\stablelinehelp%
                \errmessage{Special ruling will not display properly}%
              \fi}%
  \def\stmultispan##1{\mscount=##1 \loop\ifnum\mscount>3
\stspan\repeat}%
  \def\stspan{\span\omit \advance\mscount by -1}%
  \def\multicolumn##1{\omit\multiply\stablestemp by ##1%
     \stmultispan{\stablestemp}%
     \advance\stablesmode by ##1%
     \advance\stablesmode by -1%
     \stablestemp=3}%
  \def\multirow##1{\stablesdummyc=##1\parindent=0pt\setbox0\hbox\bgroup%
    \aftergroup\emultirow\let\temp=}
  \def\emultirow{\setbox1\vbox to\stablesdummyc\stablestrutsize%
    {\hsize\wd0\vfil\box0\vfil}%
    \ht1=\ht\stablestrutbox%
    \dp1=\dp\stablestrutbox%
    \box1}%
%
  \def\stpar##1{\vtop\bgroup\hsize ##1%
     \baselineskip=\stablesbaselineskip%
     \lineskip=\stableslineskip%

\lineskiplimit=\stableslineskiplimit\bgroup\aftergroup\estpar\let\temp=}%
  \def\estpar{\vskip 6pt\egroup}%
  \def\stparrow##1##2{\stablesdummy=##2%
     \setbox0=\vtop to ##1\stablestrutsize\bgroup%
     \hsize\stablesdummy%
     \baselineskip=\stablesbaselineskip%
     \lineskip=\stableslineskip%
     \lineskiplimit=\stableslineskiplimit%
     \bgroup\vfil\aftergroup\estparrow%
     \let\temp=}%
  \def\estparrow{\vfil\egroup%
     \ht0=\ht\stablestrutbox%
     \dp0=\dp\stablestrutbox%
     \wd0=\stablesdummy%
     \box0}%
  \def|{\global\advance\stablesmode by 1&&&}%
  \def\|{\global\advance\stablesmode by 1&\omit\vrule width 0pt%
         \hfil&&}%
\def\vt{\global\advance\stablesmode
by 1&\omit\vrule width \stablesthinline%
          \hfil&&}%
  \def\vtt{\global\advance\stablesmode by 1&\omit\vrule width
\stablesthickline%
          \hfil&&}%
  \def\vttt##1{\global\advance\stablesmode by 1&\omit\vrule width ##1%
          \hfil&&}%
  \def\vtr{\global\advance\stablesmode by 1&\omit\hfil\vrule width%
           \stablesthinline&&}%
  \def\vttr{\global\advance\stablesmode by 1&\omit\hfil\vrule width%
            \stablesthickline&&}%
\def\vtttr##1{\global\advance\stablesmode
 by 1&\omit\hfil\vrule width ##1&&}%
  \stableslines=0%
  \stablesomitfalse}
\def\stablesdef{\bgroup\stablestrut\borderrule##\tabskip=0pt plus 1fil%
  &\stablesleft##\stablesright%
  &##\ifstablesright\hfill\fi\internalrule\ifstablesright\else\hfill\fi%
  \tabskip 0pt&&##\hfil\tabskip=0pt plus 1fil%
  &\stablesleft##\stablesright%
  &##\ifstablesright\hfill\fi\internalrule\ifstablesright\else\hfill\fi%
  \tabskip=0pt\cr%
  \ifstablesborderthin%
    \thinline%
  \else%
    \thickline%
  \fi&%
}%
\def\endtable{\advance\stableslines by 1\advance\stablesmode by 1%
   \message{- Rows: \number\stableslines, Columns:
\number\stablesmode>}%
   \stablesel%
   \ifstablesborderthin%
     \thinline%
   \else%
     \thickline%
   \fi%
   \egroup\stablesend%
\endgroup%
\global\stablesinfalse}
%

\overfullrule=0pt \abovedisplayskip=12pt plus 3pt minus 3pt
\belowdisplayskip=12pt plus 3pt minus 3pt

\noblackbox
\input epsf
\newcount\figno
\figno=0
\def\fig#1#2#3{
\par\begingroup\parindent=0pt\leftskip=1cm\rightskip=1cm\parindent=0pt
\baselineskip=11pt \global\advance\figno by 1 \midinsert
\epsfxsize=#3 \centerline{\epsfbox{#2}} \vskip 12pt
\centerline{{\bf Figure \the\figno:} #1}\par
\endinsert\endgroup\par}
\def\figlabel#1{\xdef#1{\the\figno}}

\def\np#1#2#3{Nucl. Phys. {\bf B#1} (#2) #3}

\def\IR{\relax{\rm I\kern-.18em R}}


\font\cmss=cmss10 \font\cmsss=cmss10 at 7pt
\def\rlx{\relax\leavevmode}
\def\inbar{\vrule height1.5ex width.4pt depth0pt}
\def\IC{\relax\,\hbox{$\inbar\kern-.3em{\rm C}$}}
\def\IN{\relax{\rm I\kern-.18em N}}
\def\IP{\relax{\rm I\kern-.18em P}}
\def\ZZ{\rlx\leavevmode\ifmmode\mathchoice{\hbox{\cmss Z\kern-.4em Z}}
 {\hbox{\cmss Z\kern-.4em Z}}{\lower.9pt\hbox{\cmsss Z\kern-.36em Z}}
 {\lower1.2pt\hbox{\cmsss Z\kern-.36em Z}}\else{\cmss Z\kern-.4em
 Z}\fi}
\def\IZ{\relax\ifmmode\mathchoice
{\hbox{\cmss Z\kern-.4em Z}}{\hbox{\cmss Z\kern-.4em Z}}
{\lower.9pt\hbox{\cmsss Z\kern-.4em Z}} {\lower1.2pt\hbox{\cmsss
Z\kern-.4em Z}}\else{\cmss Z\kern-.4em Z}\fi}

\def\narrowplus{\kern -.04truein + \kern -.03truein}
\def\narrowminus{- \kern -.04truein}
\def\narrowminussub{\kern -.02truein - \kern -.01truein}

\def\b{{\beta}}
\def\a{{\alpha}}

\def\e{{\epsilon}}

\def\frac#1#2{{#1\over #2}}

\def\IZ{\relax\ifmmode\mathchoice
{\hbox{\cmss Z\kern-.4em Z}}{\hbox{\cmss Z\kern-.4em Z}}
{\lower.9pt\hbox{\cmsss Z\kern-.4em Z}} {\lower1.2pt\hbox{\cmsss
Z\kern-.4em Z}}\else{\cmss Z\kern-.4em Z}\fi}
\def\IB{\relax{\rm I\kern-.18em B}}
\def\IC{{\relax\hbox{$\inbar\kern-.3em{\rm C}$}}}
\def\ID{\relax{\rm I\kern-.18em D}}
\def\IE{\relax{\rm I\kern-.18em E}}
\def\IF{\relax{\rm I\kern-.18em F}}
\def\IG{\relax\hbox{$\inbar\kern-.3em{\rm G}$}}
\def\IGa{\relax\hbox{${\rm I}\kern-.18em\Gamma$}}
\def\IH{\relax{\rm I\kern-.18em H}}
\def\II{\relax{\rm I\kern-.18em I}}
\def\IK{\relax{\rm I\kern-.18em K}}
\def\IP{\relax{\rm I\kern-.18em P}}

\font\cmss=cmss10 \font\cmsss=cmss10 at 7pt
\def\IR{\relax{\rm I\kern-.18em R}}

\def\1{{\bf 1}}
\def\3{{\bf 3}}
\def\7{{\bf 7}}
\def\2{{\bf 2}}
\def\8{{\bf 8}}

\def\hat{\widehat}
\def\quabla{{\sqcap}\!\!\!\!{\sqcup}}

\def\o{\over}
%

%
%
\def\eqnn#1{\xdef #1{(\secsym\the\meqno)}\writedef{#1\leftbracket#1}%
\global\advance\meqno by1\wrlabeL#1}
\def\eqna#1{\xdef #1##1{\hbox{$(\secsym\the\meqno##1)$}}
\writedef{#1\numbersign1\leftbracket#1{\numbersign1}}%
\global\advance\meqno by1\wrlabeL{#1$\{\}$}}
\def\eqn#1#2{\xdef #1{(\secsym\the\meqno)}\writedef{#1\leftbracket#1}%
\global\advance\meqno by1$$#2\eqno#1\eqlabeL#1$$}


\lref\DuffWD{ M.~J.~Duff, J.~T.~Liu and R.~Minasian,
``Eleven-dimensional origin of string/string duality:  a one-loop
test'', Nucl.\ Phys. {\bf B452} (1995) 261-282, hep-th/9506126. }

\lref\rBB{ K.~Becker and M.~Becker, ``${\cal M}$-Theory on
eight-manifolds,'', Nucl.\ Phys.\ {\bf B477} (1996) 155-167,
hep-th/9605053.}

\lref\DasguptaSS{ K.~Dasgupta, G.~Rajesh and S.~Sethi, ``M
theory, orientifolds and $G$-flux'', JHEP {\bf 9908} (1999) 023,
hep-th/9908088. }

\lref\beckerD{ K.~Becker and K.~Dasgupta, ``Heterotic strings
with torsion,'' JHEP {\bf 0211} (2002) 006, hep-th/0209077.}

\lref\banks{T. Banks, ``Cosmological breaking of supersymmetry?
or little Lambda goes back to the future'', hep-th/0007146.}

\lref\rBHO{E. Bergshoeff, C. Hull and T. Ortin,  ``Duality in the
type II superstring effective action'', \np{451} {1995}{547-578},
hep-th/9504081. }
 \lref\rsenorien{A. Sen,  ``F-theory and orientifolds'',
 Nucl. Phys. {\bf B475} (1996) 562-578,
hep-th/9605150.}

\lref\rstrom{A. ~Strominger, ``Superstrings with torsion'',
Nucl.\ Phys.\ {\bf B274} (1986) 253-284.}

\lref\witteno{E.~Witten, ``Some Properties of O(32)
superstrings,'' Phys.\ Lett.\ B {\bf 149} (1984) 351-356.}

\lref\sencount{A.~Sen, ``Local gauge and lorentz invariance of
the heterotic string theory,'' Phys.\ Lett.\ B {\bf 166} (1986)
300-304.}

\lref\xenwit{M.~Dine, N.~Seiberg, X.~G.~Wen and E.~Witten,
``Nonperturbative effects on the string world sheet,'' Nucl.\
Phys.\ B {\bf 278} (1986) 769-789 ; ``Nonperturbative effects on
the string world sheet. 2,'' Nucl.\ Phys.\ B {\bf 289} (1987)
319-363. }

\lref\louisL{S.~Gurrieri, J.~Louis, A.~Micu and D.~Waldram,
``Mirror symmetry in generalized Calabi-Yau compactifications,''
hep-th/0211102}

\lref\HULL{C.~M.~Hull, ``Superstring compactifications with
torsion and space-time supersymmetry,'' {\it In Turin 1985,
Proceedings, Superunification and Extra Dimensions}, 347-375;
``Sigma model beta functions and string compactifications,''
Nucl.\ Phys.\ B {\bf 267} (1986) 266-276; ``Compactifications of
the heterotic superstring,'' Phys.\ Lett.\ B {\bf 178} (1986)
357-364; ``Lectures on nonlinear sigma models and strings,''
lectures given at {\it Super Field Theories Workshop, Vancouver,
Canada}, Jul 25 - Aug 6, 1986.}

\lref\hetcit{D.~J.~Gross, J.~A.~Harvey, E.~J.~Martinec and
R.~Rohm, ``The heterotic string,'' Phys.\ Rev.\ Lett.\ {\bf 54}
(1985) 502-505; ``Heterotic string theory. 1. The free heterotic
string,'' Nucl.\ Phys.\ B {\bf 256} (1985) 253-284; ``Heterotic
string theory. 2. The interacting heterotic string,'' Nucl.\
Phys.\ B {\bf 267} (1986) 75-124.}

\lref\sensethi{P.~S.~Aspinwall, ``$K3$ surfaces and string
duality,'', hep-th/9611137; A.~Sen and S.~Sethi, ``The mirror
transform of type I vacua in six dimensions,'' Nucl.\ Phys.\ B
{\bf 499} (1997) 45-54, hep-th/9703157; J.~de Boer, R.~Dijkgraaf,
K.~Hori, A.~Keurentjes, J.~Morgan, D.~R.~Morrison and S.~Sethi,
``Triples, fluxes, and strings,'' Adv.\ Theor.\ Math.\ Phys.\
{\bf 4} (2002) 995-1186, hep-th/0103170; D.~R.~Morrison and
S.~Sethi, ``Novel type I compactifications,'' JHEP {\bf 0201}
(2002) 032, hep-th/0109197.}

\lref\robbins{K.~Dasgupta, G.~Rajesh, D.~Robbins and S.~Sethi,
``Time-dependent warping, fluxes, and NCYM,'' JHEP {\bf 0303}
(2003) 041, hep-th/0302049; K.~Dasgupta and M.~Shmakova, ``On
branes and oriented B-fields,'' hep-th/0306030.}

\lref\sewitten{N.~Seiberg, ``IR dynamics on branes and space-time
geometry,'' Phys.\ Lett.\ B {\bf 384} (1996) 81-85,
hep-th/9606017; N.~Seiberg and E.~Witten, ``Gauge dynamics and
compactification to three dimensions,'' hep-th/9607163.}

\lref\tseytlin{A.~A.~Tseytlin, ``On SO(32) heterotic - type I
superstring duality in ten dimensions,'' Phys.\ Lett.\ B {\bf
367} (1996) 84-90, hep-th/9510173; A.~A.~Tseytlin, ``Heterotic -
type I superstring duality and low-energy effective actions,''
Nucl.\ Phys.\ B {\bf 467} (1996) 383-398, hep-th/9512081.}


\lref\katzs{S.~Katz and E.~Sharpe, ``D-branes, open string vertex
operators, and Ext groups,'' Adv.\ Theor.\ Math.\ Phys.\  {\bf 6}
(2003) 979-1030, hep-th/0208104.}

\lref\rohmW{R.~Rohm and E.~Witten, ``The antisymmetric tensor
field in superstring theory,'' Annals Phys.\  {\bf 170} (1986)
454-489.}

\lref\bismut{J.-M. Bismut, ``A local index theorem for
non-K\"ahler manifolds,'' Math. \ Ann. {\bf 284} (1989) 681-699.}

\lref\hirze{F. Hirzebruch, {\it Topological Methods in Algebraic
Geometry}, Springer, Berlin, 1978.}

\lref\barsV{I.~Bars and M.~Visser, ``Number of massless fermion
families in superstring theory,'' Phys.\ Lett.\ B {\bf 163}
(1985) 118-122; ``Fermion families in superstring theory,''
USC-85/035.}

\lref\borcea{C. Borcea, ``$K3$ surfaces with involutions and
mirror pairs of Calabi-Yau manifolds,'' in {\it Mirror Manifolds
II}, ed. by B.~Greene and S.-T.~Yau.}

\lref\berkoozL{M.~Berkooz and R.~G.~Leigh, ``A D = 4 N = 1
orbifold of type I strings,'' Nucl.\ Phys.\ B {\bf 483} (1997)
187-208, hep-th/9605049.}

\lref\aspinwall{P.~S.~Aspinwall, ``Enhanced gauge symmetries and
$K3$ surfaces,'' Phys.\ Lett.\ B {\bf 357} (1995) 329-334,
hep-th/9507012.}

\lref\chs{A.~Strominger, ``Heterotic Solitons,'' Nucl.\ Phys.\ B
{\bf 343}, 167 (1990) [Erratum-ibid.\ B {\bf 353}, 565 (1991)];
C.~G.~Callan, J.~A.~Harvey and A.~Strominger, ``World Sheet
Approach To Heterotic Instantons And Solitons,'' Nucl.\ Phys.\ B
{\bf 359}, 611 (1991).}

\lref\adopt{J.~D.~Edelstein, K.~Oh and R.~Tatar,
``Orientifold, geometric transition and large N duality for SO/Sp gauge  theories,''
JHEP {\bf 0105}, 009 (2001), hep-th/0104037;
K.~Dasgupta, K.~Oh and R.~Tatar, {``Geometric
transition, large N dualities and MQCD dynamics,''} Nucl.\ Phys.\
B {\bf 610}, 331 (2001), hep-th/0105066; {``Open/closed string
dualities and Seiberg duality from geometric transitions in
M-theory,''} JHEP {\bf 0208}, 026 (2002), hep-th/0106040;
K.~Dasgupta, K.~h.~Oh, J.~Park and R.~Tatar, ``Geometric
transition versus cascading solution,'' JHEP {\bf 0201}, 031
(2002), hep-th/0110050;
K.~h.~Oh and R.~Tatar,
``Duality and confinement in N = 1 supersymmetric theories from geometric  transitions,''
Adv.\ Theor.\ Math.\ Phys.\  {\bf 6}, 141 (2003), hep-th/0112040.}

\lref\gillardetal{J.~Gillard, G.~Papadopoulos, and D.~Tsimpis,
``Anomaly, fluxes, and $(2,0)$ heterotic-string
compactifications,'' hep-th/0304126.}

\lref\spindel{Ph.~Spindel, A.~Sevrin, W.~Troost, and
A.~van~Proeyen, ``Complex structures on parallelised group
manifolds and supersymmetric sigma models,'' Phys. Lett. B {\bf
206} (1988) 71-74; ``Extended supersymmetric sigma models on
group manifolds: I,'' Nucl. Phys. B {\bf 308} (1988) 662-698;
``Extended supersymmetric sigma models on group manifolds: II,''
Nucl. Phys. B {\bf 311} (1988) 465-492.}

\lref\brin{V.~Brinzanescu and R.~Moraru, ``Holomorphic rank-2
vector bundles on non-K\"ahler elliptic surfaces,''
math.AG/0306191; ``Stable bundles on non-K\"ahler elliptic
surfaces,'' math.AG/0306192; ``Twisted Fourier-Mukai transforms
and bundles on non-K\"ahler elliptic surfaces,'' math.AG/0309031.}

\lref\ivanov{J.~Gutowski, S.~Ivanov, and G.~Papadopoulos,
``Deformations of generalized calibrations and compact
non-K\"ahler manifolds with vanishing first Chern class,''
math.DG/0205012.}

\lref\font{G.~Aldazabal, A.~Font, L.~E.~Ibanez and G.~Violero,
{``D = 4, N = 1, type IIB orientifolds,''} Nucl.\ Phys.\ B {\bf
536}, 29 (1998), hep-th/9804026.}

\lref\urangafont{G.~Aldazabal, A.~Font, L.~E.~Ibanez,
A.~M.~Uranga and G.~Violero, {``Non-perturbative heterotic D =
6,4, N = 1 orbifold vacua,''} Nucl.\ Phys.\ B {\bf 519}, 239
(1998), hep-th/9706158.}

\lref\nappiS{M.~Dine, V.~Kaplunovsky, M.~L.~Mangano, C.~Nappi and
N.~Seiberg, {``Superstring Model Building,''} Nucl.\ Phys.\ B
{\bf 259}, 549 (1985).}

\lref\glashow{H.~Georgi and S.~L.~Glashow, ``Unity Of All
Elementary Particle Forces,'' Phys.\ Rev.\ Lett.\  {\bf 32}, 438
(1974).}

\lref\slansky{R.~Slansky, ``Group Theory For Unified Model
Building,'' Phys.\ Rept.\  {\bf 79}, 1 (1981).}

\lref\curioLu{G.~Curio and A.~Krause, ``Enlarging the parameter
space of heterotic M-theory flux compactifications to
phenomenological viability,'' hep-th/0308202.}

\lref\dabhull{A.~Dabholkar and C.~Hull,
``Duality twists, orbifolds, and fluxes,''
JHEP {\bf 0309}, 054 (2003), hep-th/0210209.}

\lref\prokS{S.~Prokushkin, Private Communications.}

\lref\buchkov{E.~I.~Buchbinder and B.~A.~Ovrut,
``Vacuum stability in heterotic M-theory,'' hep-th/0310112.}

\lref\warner{M.~Gunaydin and N.~P.~Warner,
``The G2 Invariant Compactifications In Eleven-Dimensional Supergravity,''
Nucl.\ Phys.\ B {\bf 248}, 685 (1984);
P.~van Nieuwenhuizen and N.~P.~Warner,
``New Compactifications Of Ten-Dimensional And Eleven-Dimensional Supergravity On Manifolds Which Are Not Direct Products,''
Commun.\ Math.\ Phys.\  {\bf 99}, 141 (1985).}

\lref\rDJM{K. Dasgupta, D. P. Jatkar and S. Mukhi,
``Gravitational couplings and $Z_2$ orientifolds'', Nucl. Phys.
{\bf B523} (1998) 465-484, hep-th/9707224; J.~F.~Morales,
C.~A.~Scrucca and M.~Serone, ``Anomalous couplings for D-branes
and O-planes,'' Nucl.\ Phys.\ B {\bf 552} (1999) 291-315,
hep-th/9812071; B.~J.~Stefanski, ``Gravitational couplings of
D-branes and O-planes,'' Nucl.\ Phys.\ B {\bf 548} (1999)
275-290, hep-th/9812088. }

\lref\nemanja{N.~Kaloper and R.~C.~Myers,  ``The O(dd) story of
massive supergravity,'' JHEP {\bf 9905} (1999) 010,
hep-th/9901045; G.~Curio, A.~Klemm, B.~Kors and D.~Lust, ``Fluxes
in heterotic and type II string compactifications,'' Nucl.\
Phys.\ B {\bf 620} (2002) 237-258, hep-th/0106155; J.~Louis and
A.~Micu, ``Heterotic string theory with background fluxes,''
Nucl.\ Phys.\ B {\bf 626} (2002) 26-52, hep-th/0110187.}

\lref\kapulov{V.~Kaplunovsky, J.~Louis and S.~Theisen, ``Aspects
of duality in N=2 string vacua,'' Phys.\ Lett.\ B {\bf 357}
(1995) 71-75, hep-th/9506110.}

\lref\nati{N.~Seiberg, ``Observations on the moduli space of
superconformal field theories,'' Nucl.\ Phys.\ B {\bf 303} (1988)
286-304; A.~Ceresole, R.~D'Auria, S.~Ferrara and A.~Van Proeyen,
``Duality transformations in supersymmetric Yang-Mills theories
coupled to supergravity,'' Nucl.\ Phys.\ B {\bf 444} (1995)
92-124, hep-th/9502072.}

\lref\narainJ{K.~S.~Narain, ``New heterotic string theories in
uncompactified dimensions $< 10$,'' Phys.\ Lett.\ B {\bf 169}
(1986) 41-46; K.~S.~Narain, M.~H.~Sarmadi and E.~Witten, ``A note
on toroidal compactification of heterotic string theory,'' Nucl.\
Phys.\ B {\bf 279} (1987) 369-379; J.~Maharana and J.~H.~Schwarz,
``Noncompact symmetries in string theory,'' Nucl.\ Phys.\ B {\bf
390} (1993) 3-32 (1993), hep-th/9207016.}

\lref\senM{A.~Sen, ``A note on enhanced gauge symmetries in M-
and string theory,'' JHEP {\bf 9709} (1997) 001, hep-th/9707123.}

\lref\rBUSH{T. Buscher, ``Quantum corrections and extended
supersymmetry in new sigma models'', Phys. Lett. {\bf B159}
(1985) 127-130; ``A symmetry of the string background field
equations,'' Phys. Lett. B {\bf 194} (1987) 59-62; ``Path
integral derivation of quantum duality in nonlinear sigma
models'', Phys. Lett. B {\bf 201} (1988) 466-472.}

\lref\rKKL{E. Kiritsis, C. Kounnas and D. Lust, ``A large class
of new gravitational and axionic backgrounds for four-dimensional
superstrings'', Int. J. Mod. Phys. {\bf A9} (1994) 1361-1394,
hep-th/9308124. }

\lref\papado{J.~Gutowski and G.~Papadopoulos, ``AdS
calibrations,'' Phys.\ Lett.\ B {\bf 462} (1999) 81-88,
hep-th/9902034; J.~Gutowski, G.~Papadopoulos and P.~K.~Townsend,
``Supersymmetry and generalized calibrations,'' Phys.\ Rev.\ D
{\bf 60} (1999) 106006, hep-th/9905156; J.~Gutowski, S.~Ivanov
and G.~Papadopoulos, ``Deformations of generalized calibrations
and compact non-K\"ahler manifolds with vanishing first Chern
class,'' math.dg/0205012.}

\lref\gauntlett{J.~P.~Gauntlett, N.~w.~Kim, D.~Martelli and
D.~Waldram, ``Fivebranes wrapped on SLAG three-cycles and related
geometry,'' JHEP {\bf 0111} (2001) 018, hep-th/0110034.
J.~P.~Gauntlett, D.~Martelli, S.~Pakis and D.~Waldram,
``G-structures and wrapped NS5-branes,'' hep-th/0205050;
J.~P.~Gauntlett, D.~Martelli and D.~Waldram, ``Superstrings with
intrinsic torsion,'' hep-th/0302158.}

\lref\sengimon{A.~Sen, ``A non-perturbative description of the
Gimon-Polchinski orientifold,'' Nucl.\ Phys.\ B {\bf 489} (1997)
139-159, hep-th/9611186; A.~Sen, ``F-theory and the
Gimon-Polchinski orientifold,'' Nucl.\ Phys.\ B {\bf 498} (1997)
135-155, hep-th/9702061.}

\lref\gimpol{E.~G.~Gimon and J.~Polchinski, ``Consistency
conditions for orientifolds and D-manifolds,'' Phys.\ Rev.\ D
{\bf 54} (1996) 1667-1676, hep-th/9601038.}

\lref\kst{S. Giddings, S. Kachru and J. Polchinski, ``Hierarchies
from fluxes in string compactifications,'' hep-th/0105097;
S.~Kachru, M.~B.~Schulz and S.~Trivedi, ``Moduli stabilization
from fluxes in a simple IIB orientifold,'' hep-th/0201028;
A.~R.~Frey and J.~Polchinski, ``N = 3 warped compactifications,
'' Phys.\ Rev.\ {\bf D65} (2002) 126009, hep-th/0201029. }

\lref\pktspto{ S.~Gurrieri and A.~Micu, ``Type IIB theory on
half-flat manifolds,'' hep-th/0212278.}
\lref\pktsptt{P.~K.~Tripathy and S.~P.~Trivedi, Compactifications
with Flux on $K3$ and Tori,'' hep-th/0301139.}

\lref\gates{ S.~J.~Gates, ``Superspace formulation of new
nonlinear sigma models,'' Nucl.\ Phys.\ B {\bf 238} (1984)
349-366; S.~J.~Gates, C.~M.~Hull and M.~Rocek, ``Twisted
multiplets and new supersymmetric nonlinear sigma models,''
Nucl.\ Phys.\ B {\bf 248} (1984) 157-186; S.~J.~Gates, S.~Gukov
and E.~Witten, ``Two-dimensional supergravity theories from
Calabi-Yau four-folds,'' Nucl.\ Phys.\ B {\bf 584} (2000) 109-148,
hep-th/0005120.}

\lref\rkehagias{A. Kehagias, ``New type IIB vacua and their
F-theory interpretation,'' Phys. Lett. B {\bf 435} (1998) 337-342,
hep-th/9805131. }

\lref\GukovYA{ S.~Gukov, C.~Vafa and E.~Witten, ``CFT's from
Calabi-Yau four-folds,'' Nucl.\ Phys.\ B {\bf 584} (2000) 69-108,
erratum ibid {\bf 608} (2001) 477-478, hep-th/9906070. }
\lref\BeckerPM{ K.~Becker and M.~Becker, ``Supersymmetry
breaking, ${\cal M}$-theory and fluxes'', JHEP {\bf 0107} (2001)
038, hep-th/0107044. }
\lref\DineRZ{ M.~Dine, R.~Rohm, N.~Seiberg and E.~Witten,
``Gluino condensation in superstring models,'' Phys.\ Lett.\ B
{\bf 156} (1985) 55-60.}
\lref\FreyHF{ A.~R.~Frey and J.~Polchinski, ``N = 3 warped
compactifications,'' Phys.\ Rev.\ D {\bf 65} (2002) 126009,
hep-th/0201029.}
\lref\CurioAE{ G.~Curio, A.~Klemm, B.~Kors and D.~Lust, ``Fluxes
in heterotic and type II string compactifications,'' Nucl.\
Phys.\ B {\bf 620} (2002) 237-258, hep-th/0106155.}

\lref\Vafawitten{ C.~Vafa and E.~Witten, ``A one loop test of
string duality,'' Nucl.\ Phys.\ B {\bf 447} (1995) 261-270,
hep-th/9505053.}

\lref\SethiVW{ S.~Sethi, C.~Vafa and E.~Witten, ``Constraints on
low-dimensional string compactifications,'' Nucl.\ Phys.\ B {\bf
480} (1996) 213-224, hep-th/9606122.}

\lref\HananyK{ A.~Hanany and B.~Kol, ``On orientifolds, discrete
torsion, branes and M theory,'' JHEP {\bf 0006} (2000) 013,
hep-th/0003025.}

\lref\Ganor{ O.~J.~Ganor, ``Compactification of tensionless
string theories,'' hep-th/9607092.}

\lref\DMtwo{ K.~Dasgupta and S.~Mukhi,  ``A note on
low-dimensional string compactifications,'' Phys.\ Lett.\ B {\bf
398} (1997) 285-290, hep-th/9612188.}

\lref\ShiuG{ B.~R.~Greene, K.~Schalm and G.~Shiu,  ``Warped
compactifications in M and F theory,'' Nucl.\ Phys.\ B {\bf 584}
(2000) 480-508, hep-th/0004103.}

\lref\harmoni{ G.~W.~Gibbons and P.~J.~Ruback, ``The hidden
symmetries of multicenter metrics,'' Commun.\ Math.\ Phys.\ {\bf
115} (1988) 267-300; N.~S.~Manton and B.~J.~Schroers, ``Bundles
over moduli spaces and the quantization of BPS monopoles,''
Annals Phys.\  {\bf 225} (1993) 290-338; A.~Sen, ``Dyon -
monopole bound states, selfdual harmonic forms on the multi -
monopole moduli space, and SL(2,Z) invariance in string theory,''
Phys.\ Lett.\ B {\bf 329} (1994) 217-221, hep-th/9402032.}

\lref\mesO{P.~Meessen and T.~Ortin, ``An Sl(2,Z) multiplet of
nine-dimensional type II supergravity theories,'' Nucl.\ Phys.\ B
{\bf 541} (1999) 195-245, hep-th/9806120; E. Bergshoeff, C. Hull
and T. Ortin, ``Duality in the type II superstring effective
action,'' \np{451} {1995}{547-578}, hep-th/9504081; S.~F.~Hassan,
``T-duality, space-time spinors and R-R fields in curved
backgrounds,'' Nucl.\ Phys.\ B {\bf 568} (2000) 145-161,
hep-th/9907152.}

\lref\greenS{M.~B.~Green and J.~H.~Schwarz, ``Superstring
interactions,'' Nucl.\ Phys.\ B {\bf 218} (1983) 43-88.}

\lref\SmitD{ B.~de Wit, D.~J.~Smit and N.~D.~Hari Dass,
``Residual supersymmetry of compactified D = 10 supergravity,''
Nucl.\ Phys.\ B {\bf 283} (1987) 165-191; N.~D.~Hari Dass, ``A
no-go theorem for de Sitter compactifications?,'' Mod.\ Phys.\
Lett.\ A {\bf 17} (2002) 1001-1012, hep-th/0205056.}

\lref\greene{B.~R.~Greene, A.~D.~Shapere, C.~Vafa and S.~T.~Yau,
``Stringy cosmic strings and noncompact Calabi-Yau manifolds,''
Nucl.\ Phys.\ B {\bf 337} (1990) 1-36.}

\lref\polci{J.~Polchinski, ``Tensors from $K3$ orientifolds,''
Phys.\ Rev.\ D {\bf 55} (1997) 6423-6428, hep-th/9606165.}

\lref\dabholkarP{A.~Dabholkar and J.~Park, ``An orientifold of
type-IIB theory on $K3$,'' Nucl.\ Phys.\ B {\bf 472} (1996)
207-220, hep-th/9602030; ``Strings on orientifolds,'' Nucl.\
Phys.\ B {\bf 477} (1996) 701-714, hep-th/9604178; ``A note on
orientifolds and F-theory,'' Phys.\ Lett.\ B {\bf 394} (1997)
302-306, hep-th/9607041.}

\lref\nikulin{V. Nikulin, ``Discrete reflection groups in
Lobachevsky spaces and algebraic surfaces,'' pp.~654-671 in {\it
Proceedings of the International Congress of Mathematicians},
Berkeley (1986) 654.}

\lref\samref{S.~Kachru and C.~Vafa, ``Exact results for N=2
compactifications of heterotic strings,'' Nucl.\ Phys.\ B {\bf
450} (1995) 69-89, hep-th/9505105; M.~Bershadsky,
K.~A.~Intriligator, S.~Kachru, D.~R.~Morrison, V.~Sadov and
C.~Vafa, ``Geometric singularities and enhanced gauge
symmetries,'' Nucl.\ Phys.\ B {\bf 481} (1996) 215-252,
hep-th/9605200; P.~S.~Aspinwall and M.~Gross,
``Heterotic-heterotic string duality and multiple $K3$
fibrations,'' Phys.\ Lett.\ B {\bf 382} (1996) 81-88,
hep-th/9602118.}

\lref\duff{M.~J.~Duff, R.~Minasian and E.~Witten, ``Evidence for
heterotic/heterotic duality,'' Nucl.\ Phys.\ B {\bf 465} (1996)
413-438, hep-th/9601036.}

\lref\morvaf{D.~R.~Morrison and C.~Vafa, ``Compactifications of
F-Theory on Calabi--Yau threefolds -- I,'' Nucl.\ Phys.\ B {\bf
473} (1996) 74-92, hep-th/9602114; ``Compactifications of
F-Theory on Calabi--Yau threefolds -- II,'' Nucl.\ Phys.\ B {\bf
476} (1996) 437-469, hep-th/9603161.}

\lref\dasF{ K.~Dasgupta and S.~Mukhi, ``F-theory at constant
coupling,'' Phys.\ Lett.\ B {\bf 385} (1996) 125-131,
hep-th/9606044.}

\lref\PapaDI{ S.~Ivanov and G.~Papadopoulos, ``A no-go theorem
for string warped compactifications,'' Phys.\ Lett. B {\bf 497}
(2001) 309-316, hep-th/0008232.}

\lref\DineSB{M.~Dine and N.~Seiberg, ``Couplings and scales in
superstring models,'' Phys.\ Rev.\ Lett.\  {\bf 55} (1985)
366-369.}

\lref\olgi{O. DeWolfe and S. B. Giddings, ``Scales and
hierarchies in warped compactifications and brane worlds,''
hep-th/0208123.}

\lref\hellermanJ{S.~Hellerman, J.~McGreevy and B.~Williams,
``Geometric constructions of non-geometric string theories,''
hep-th/0208174.}

\lref\WIP{K. Becker, M. Becker, K. Dasgupta, work in progress.}

\lref\tatar{K.~Dasgupta, K.~h.~Oh, J.~Park and R.~Tatar,
``Geometric transition versus cascading solution,'' JHEP {\bf
0201} (2002) 031, hep-th/0110050.}

\lref\wipro{Work in progress.}

\lref\civ{F.~Cachazo, B.~Fiol, K.~A.~Intriligator, S.~Katz and
C.~Vafa, ``A geometric unification of dualities,'' Nucl.\ Phys.\
B {\bf 628} (2002) 3-78, hep-th/0110028.}

\lref\renata{K.~Dasgupta, C.~Herdeiro, S.~Hirano and R.~Kallosh,
``D3/D7 inflationary model and M-theory,'' Phys.\ Rev.\ D {\bf
65} (2002) 126002, hep-th/0203019.}

\lref\carlos{R.~Kallosh, ``N = 2 supersymmetry and de Sitter
space,'' hep-th/0109168; C.~Herdeiro, S.~Hirano and R.~Kallosh,
``String theory and hybrid inflation / acceleration,'' JHEP {\bf
0112} (2001) 027, hep-th/0110271.}

\lref\kklt{S.~Kachru, R.~Kallosh, A.~Linde and S.~P.~Trivedi,
{``De Sitter vacua in string theory,''} Phys.\ Rev.\ D {\bf 68},
046005 (2003), hep-th/0301240; C.~P.~Burgess, R.~Kallosh and
F.~Quevedo, {``de Sitter String Vacua from Supersymmetric
D-terms,''}, hep-th/0309187.}

\lref\trivsham{S.~Kachru, R.~Kallosh, A.~ Linde, J.~Maldacena,
L.~ McAllister, S.~Trivedi, work in progress.}

\lref\polchinski{ J.~Polchinski, {\it String Theory. Vol. 2:
Superstring Theory And Beyond}.}

\lref\taylor{E.~Cremmer, S.~Ferrara, L.~Girardello and A.~Van
Proeyen, ``Yang-Mills theories with local supersymmetry:
lagrangian, transformation laws and superhiggs effect,'' Nucl.\
Phys.\ B {\bf 212} (1983) 413-442; T.~R.~Taylor and C.~Vafa, ``RR
Flux on Calabi-Yau and partial supersymmetry breaking,'' Phys.\
Lett.\ B {\bf 474} (2000) 130-137, hep-th/9912152.}

\lref\guko{S.~Gukov, C.~Vafa and E.~Witten, ``CFT's from
Calabi-Yau four-folds,'' Nucl.\ Phys.\ B {\bf 584} (2000) 69-108,
[Erratum-ibid.\ B {\bf 608} (2001) 477-478], hep-th/9906070.}

\lref\hullwitten{ C.~M.~Hull and E.~Witten, ``Supersymmetric
sigma models and the heterotic string,'' Phys.\ Lett.\ B {\bf
160} (1985) 398-402; A.~Sen, ``Local gauge and lorentz invariance
of the heterotic string theory,'' Phys.\ Lett.\ B {\bf 166}
(1986) 300-304; ``The heterotic string in arbitrary background
fields,'' Phys.\ Rev.\ D {\bf 32} (1985) 2102-2112; ``Equations
of motion for the heterotic string theory from the conformal
invariance of the sigma model,'' Phys.\ Rev.\ Lett.\ {\bf 55}
(1985) 1846-1849.}

\lref\dasmukhi{ K.~Dasgupta and S.~Mukhi, ``F-theory at constant
coupling,'' Phys.\ Lett.\ B {\bf 385} (1996) 125-131,
hep-th/9606044.}

\lref\vafasen{ C.~Vafa, ``Evidence for F-theory,'' Nucl.\ Phys.\
B {\bf 469} (1996) 403-418, hep-th/9602022; A.~Sen, ``F-theory
and orientifolds,'' Nucl.\ Phys.\ B {\bf 475} (1996) 562-578,
hep-th/9605150; T.~Banks, M.~R.~Douglas and N.~Seiberg, ``Probing
F-theory with branes,'' Phys.\ Lett.\ B {\bf 387} (1996) 278-281,
hep-th/9605199.}

\lref\zwee{ M.~R.~Gaberdiel and B.~Zwiebach, ``Exceptional groups
from open strings,'' Nucl.\ Phys.\ B {\bf 518} (1998) 151-172,
hep-th/9709013.}

\lref\howi{P.~Horava and E.~Witten, ``Heterotic and type I
dynamics from eleven dimensions,'' Nucl.\ Phys.\ B {\bf 460}
(1996) 506-524, hep-th/9510209.}

\lref\howii{P.~Horava and E.~Witten, ``Eleven-dimensional
supergravity on a manifold with boundary ,'' Nucl.\ Phys.\ B {\bf
475} (1996) 94-114, hep-th/9603142.}

\lref\mps{G.~Moore, G.~Peradze and N.~Saulina, ``Instabilities in
heterotic M theory induced by open membrane instantons ,'' Nucl.\
Phys.\ B {\bf 607} (2001) 117-154, hep-th/0012104.}

\lref\cukri{G.~Curio and A.~Krause, ``G fluxes and nonperturbative
stabilization of heterotic M theory ,'' Nucl.\ Phys.\ B {\bf 643}
(2002) 131-156, hep-th/0108220.}

\lref\cukrii{G.~Curio and A.~Krause, ``Enlarging the parameter
space of heterotic M theory flux compactifications to
phenomenological viability,'' hep-th/0308202.}

\lref\wittenstrong{E.~Witten, ``Strong coupling expansion of
Calabi-Yau compactification,'' Nucl.\ Phys.\ B {\bf 471} (1996)
135-158, hep-th/9602070.}

\lref\cukriii{G.~Curio and A.~Krause, ``Four flux and warped
heterotic M theory compactifications,'' Nucl.\ Phys.\ B {\bf 602}
(2001) 172-200, hep-th/0012152.}

\lref\bckr{M.~Becker, G.~Curio and A.~Krause,``Moduli
stabilization and De Sitter vacua from heterotic M-theory'', to
appear.}

\lref\callan{ C.~G.~Callan, E.~J.~Martinec, M.~J.~Perry and
D.~Friedan, ``Strings in background fields,'' Nucl.\ Phys.\ B
{\bf 262} (1985) 593-609; A.~Sen, ``Equations of motion for the
heterotic string theory from the conformal invariance of the
sigma model,'' Phys.\ Rev.\ Lett.\  {\bf 55} (1985) 1846-1849;
``The heterotic string in arbitrary background field,'' Phys.\
Rev.\ D {\bf 32} (1985) 2102-2112.}

\lref\WittenBS{ E.~Witten, ``Toroidal compactification without
vector structure,'' JHEP {\bf 9802} (1998) 006, hep-th/9712028.}

\lref\carluest{ G.~L.~Cardoso, G.~Curio, G.~Dall'Agata, D.~Lust,
P.~Manousselis and G.~Zoupanos, ``Non-K\"ahler string backgrounds
and their five torsion classes,''
Nucl.\ Phys.\ B {\bf 652}, 5 (2003),
hep-th/0211118.}

\lref\grifhar{ P.~ Griffiths and J.~ Harris {\it ``Principles of
Algebraic Geometry,''} Wiley Classics Library.}

\lref\hanany{A.~Hanany and A.~Zaffaroni, ``On the realization of
chiral four-dimensional gauge theories using  branes,'' JHEP {\bf
9805} (1998) 001, hep-th/9801134; A.~Hanany and A.~M.~Uranga,
``Brane boxes and branes on singularities,'' JHEP {\bf 9805}
(1998) 013, hep-th/9805139.}

\lref\dasmuk{B.~Andreas, G.~Curio and D.~Lust, ``The
Neveu-Schwarz five-brane and its dual geometries,'' JHEP {\bf
9810} (1998) 022, hep-th/9807008; K.~Dasgupta and S.~Mukhi,
``Brane constructions, conifolds and M-theory,'' Nucl.\ Phys.\ B
{\bf 551} (1999) 204-228, hep-th/9811139.}

\lref\zwigab{A.~Johansen, ``A comment on BPS states in F-theory
in 8 dimensions,'' Phys.\ Lett.\ B {\bf 395} (1997) 36-41,
hep-th/9608186; M.~R.~Gaberdiel and B.~Zwiebach, ``Exceptional
groups from open strings,'' Nucl.\ Phys.\ B {\bf 518} (1998)
151-172, hep-th/9709013; M.~R.~Gaberdiel, T.~Hauer and
B.~Zwiebach, ``Open string-string junction transitions,'' Nucl.\
Phys.\ B {\bf 525} (1998) 117-145, hep-th/9801205.}

\lref\gsw{M.~B.~Green, J.~H.~Schwarz and E.~Witten, {\it
Superstring Theory},  Vol. 1, 2; J.~Polchinski, {\it String
Theory}, Vol. 1, 2.}

\lref\papers{A.~Bergman, K.~Dasgupta, O.~J.~Ganor,
J.~L.~Karczmarek and G.~Rajesh, ``Nonlocal field theories and
their gravity duals,'' Phys.\ Rev.\ D {\bf 65} (2002) 066005,
hep-th/0103090; K.~Dasgupta and M.~M.~Sheikh-Jabbari,
``Noncommutative dipole field theories,'' JHEP {\bf 0202} (2002)
002, hep-th/0112064.}

\lref\GP{ E.~Goldstein and S.~Prokushkin, ``Geometric model for
complex non-K\"ahler manifolds with SU(3) structure,''
hep-th/0212307.}

\lref\toappear{ K.~Becker, M.~Becker, K.~Dasgupta, P.~S.~Green,
..., work in progress.}

\lref\MeessenQM{ P.~Meessen and T.~Ortin, ``An Sl(2,Z) multiplet
of nine-dimensional type II supergravity theories,'' Nucl.\
Phys.\ B {\bf 541} (1999) 195-245, hep-th/9806120.}

\lref\bem{P.~Bouwnegt, J.~Evslin and V.~Mathai, ``T-duality:
topology change and H flux'', hep-th/0306062.}

\lref\HUT{ C.~M.~Hull and P.~K.~Townsend, ``The two loop beta
function for sigma models with torsion,'' Phys.\ Lett.\ B {\bf
191} (1987)  115-121; ``World sheet supersymmetry and anomaly
cancellation in the heterotic string,'' Phys.\ Lett.\ B {\bf 178}
(1986) 187.}

\lref\SenJS{ A.~Sen, ``Dynamics of multiple Kaluza-Klein
monopoles in M and string theory,'' Adv.\ Theor.\ Math.\ Phys.\
{\bf 1} (1998) 115-126, hep-th/9707042. }

\lref\sav{ K.~Dasgupta, G.~Rajesh and S.~Sethi, ``M theory,
orientifolds and G-flux,'' JHEP {\bf 9908} (1999) 023,
hep-th/9908088.}

\lref\BeckerNN{ K.~Becker, M.~Becker, M.~Haack and J.~Louis,
``Supersymmetry breaking and $\alpha'$-corrections to flux
induced  potentials,'' JHEP {\bf 0206} (2002) 060,
hep-th/0204254.}

\lref\maeda{K.~i.~Maeda, ``Attractor in a superstring model: the
Einstein theory, the Friedmann universe and inflation,'' Phys.\
Rev.\ D {\bf 35} (1987) 471-479.}

\lref\rohmwit{R.~Rohm and E.~Witten, ``The antisymmetric tensor
field in superstring theory,'' Annals Phys.\  {\bf 170} (1986)
454-489.}

\lref\kg{S.~Gukov, S.~Kachru, X.~Liu, L.~McAllister, to appear.}

\lref\eva{E.~Silverstein, ``(A)dS backgrounds from asymmetric
orientifolds,'' hep-th/0106209; S.~Hellerman, J.~McGreevy and
B.~Williams, ``Geometric constructions of nongeometric string
theories,'' hep-th/0208174; A.~Dabholkar and C.~Hull, ``Duality
twists, orbifolds, and fluxes,'' hep-th/0210209.}

\lref\bbsb{K.~Becker and M.~Becker, ``Supersymmetry Breaking,
M-theory and fluxes'' JHEP {\bf 0107} (2001) 038, hep-th/0107044.}

\lref\bbdg{K.~Becker, M.~Becker, K.~Dasgupta and P.~S.~Green,
``Compactifications of heterotic theory on non-K\"ahler complex
manifolds: I,'' hep-th/0301161.}

\lref\ibars{I.~Bars, D.~Nemeschansky and S.~Yankielowicz,
``Compactified superstrings and torsion,'' Nucl.\ Phys.\ B {\bf
278} (1986) 632-656; I.~Bars, D.~Nemeschansky and S.~Yankielowicz,
``Torsion in superstrings,'' SLAC-PUB-3775, presented at the {\it
Workshop on Unified String Theories}, Santa Barbara, Calif., Jul
29-Aug 16, 1985; I.~Bars, ``Compactification of superstrings and
torsion,'' Phys.\ Rev.\ D {\bf 33} (1986) 383-388.}

\lref\candle{P.~Candelas, G.~T.~Horowitz, A.~Strominger and
E.~Witten, ``Vacuum configurations for superstrings,'' Nucl.\
Phys.\ B {\bf 258} (1985) 46-74; A.~Strominger and E.~Witten,
``New manifolds for superstring compactification,'' Commun.\
Math.\ Phys.\  {\bf 101} (1985) 341-361.}

\lref\lenny {L.~Susskind ``The anthropic landscape of string
theory,'' hep-th/0302219.}

\lref\mikei{M.~R.~Douglas, ``The statistics of string/M theory
vacua'' JHEP {\bf 0305} (2003) 046, hep-th/0303194.}

\lref\mikeii{S.~Ashok and M.~R.~Douglas,`` Counting flux vacua'',
hep-th/0307049.}

\lref\bdine{T.~Banks and M.~Dine, ``Is there a string theory
landscape'', hep-th/0309170.}

\lref\weba{J.~Bagger and J.~Wess, {\it ``Supersymmetry and
Supergravity,''} Princeton University Press.}

\lref\becons{ M.~Becker and D.~Constantin, ``A note on flux
induced superpotentials in string theory,'' hep-th/0210131.}

\lref\witsuper{E.~Witten, ``New issues in manifolds of $SU(3)$
holonomy,'' Nucl.\ Phys.\ B {\bf 268} (1986) 79-112.}

\lref\douglas{M.~R.~Douglas, ``The statistics of string/M theory
vacua,'' hep-th/0303194.}

\lref\senF{A.~Sen, ``Orientifold limit of F-theory vacua,''
Phys.\ Rev.\ D {\bf 55} (1997) 7345-7349, hep-th/9702165;
``Orientifold limit of F-theory vacua,'' Nucl.\ Phys.\ Proc.\
Suppl.\  {\bf 68} (1998) 92-98 [Nucl.\ Phys.\ Proc.\ Suppl.\
{\bf 67} (1998) 81-87], hep-th/9709159.}

\lref\gopmuk{R.~Gopakumar and S.~Mukhi, ``Orbifold and
orientifold compactifications of F-theory and M-theory  to six
and four dimensions,'' Nucl.\ Phys.\ B {\bf 479} (1996) 260-284,
hep-th/9607057.}

\lref\origa{S.~Chakravarty, K.~Dasgupta, O.~J.~Ganor and
G.~Rajesh, ``Pinned branes and new non Lorentz invariant
theories,'' Nucl.\ Phys.\ B {\bf 587} (2000) 228-248,
hep-th/0002175; K.~Dasgupta, G.~Rajesh, D.~Robbins and S.~Sethi,
``Time-dependent warping, fluxes, and NCYM,'' JHEP {\bf 0303}
(2003) 041, hep-th/0302049.}

\lref\toappear{ K.~Becker, M.~Becker, K.~Dasgupta, E.~Goldstein,
P.~S.~Green and S.~Prokushkin, work in progress.}

\lref\anna{A.~Fino and G.~Grantcharov, ``On some properties of
the manifolds with skew-symmetric torsion and holonomy SU(n) and
Sp(n),'' math.dg/0302358.}

\lref\bg{K.~Behrndt and S.~Gukov, ``Domain walls and
superpotentials from M theory on Calabi-Yau three-folds,'' Nucl.\
Phys.\ B {\bf 580} (2000) 225-242, hep-th/0001082.}

\lref\luest{G.~L.~Cardoso, G.~Curio, G.~Dall'Agata and D.~Lust,
``BPS action and superpotential for heterotic string
compactifications with fluxes,'' hep-th/0306088.}

\lref\bbdp{K.~Becker, M.~Becker, K.~Dasgupta and S.~Prokushkin,
``Properties of heterotic vacua from superpotentials,''
Nucl.\ Phys.\ B {\bf 666}, 144 (2003),
hep-th/0304001.}

\lref\bbdg{K.~Becker, M.~Becker, K.~Dasgupta and P.~S.~Green,
``Compactifications of heterotic theory on non-Kaehler complex
manifolds.  I,'' JHEP {\bf 0304} (2003) 007, hep-th/0301161.}

\lref\disgreone{J.~Distler and B.~Greene, ``Aspects of (2,0)
string compactifications,'' Nucl. Phys. {\bf B304} (1988) 1-62.}

\lref\ct{C.~Callan and L.~Thorlacius, ``Sigma models and string
theory,'' in {\it Particles, Strings, and Supernovae}, the
proceedings of TASI 1988.}

\lref\gswtwo{M.~Green, J.~Schwarz, and E.~Witten, {\it Superstring
theory}, volume 2.}

\lref\mathaipriv{V. Mathai, private communication.}

\lref\liyau{J.~Li and S.-T.~Yau, ``Hermitian Yang-Mills
connections on non-K\"ahler manifolds,'' in {\it Mathematical
Aspects of String Theory}, World Scientific, 1987.}

\lref\tomaone{M.~Toma, ``Stable bundles on non-algebraic surfaces
giving rise to compact moduli spaces,'' C. R. Acad. Sci. Paris
S\'er. I Math. {\bf 323} (1996) 501-505.}

\lref\tomatwo{M.~Toma, ``Compact moduli spaces of stable sheaves
over non-algebraic surfaces,'' Doc. Math. {\bf 6} (2001) 11-29.}

\lref\kcsub{E.~Sharpe, ``K\"ahler cone substructure,'' Adv. Theor.
Math. Phys. {\bf 2} (1999) 1441-1462, hep-th/9810064.}

\lref\shamitjohn{S.~Kachru and J.~McGreevy, ``Supersymmetric
three-cycles and supersymmetry breaking,'' Phys. Rev. {\bf D61}
(2000) 026001, hep-th/9908135.}

\lref\ce{E.~Calabi and B.~Eckmann, ``A class of compact, complex
manifolds which are not algebraic,'' Ann. of Math. (2) {\bf 58}
(1953) 494-500.}

\lref\lutian{P.~Lu and G.~Tian, ``The complex structure on a
connected sum of $S^3 \times S^3$ with trivial canonical bundle,''
Math. Ann. {\bf 298} (1994) 761-764.}

\lref\reid{M.~Reid, ``The moduli space of 3-folds with $K=0$ may
nevertheless be irreducible,'' Math. Ann. {\bf 278} (1987)
329-334.}

\lref\wall{C.~T.~C.~Wall, ``Classification problems in
differential topology, V:  on certain 6-manifolds,'' Inv. Math.
{\bf 1} (1966) 355-374.}

\lref\blumZ{J.~D.~Blum and A.~Zaffaroni, {``An orientifold from F
theory,''} Phys.\ Lett.\ B {\bf 387}, 71 (1996), hep-th/9607019.}

\lref\fmw{R.~Friedman, J.~Morgan, and E.~Witten, ``Vector bundles
and F-theory,'' Comm. Math. Phys. {\bf 187} (1997) 679-743,
hep-th/9701162.}

\lref\dtthree{E.~Sharpe, ``Discrete torsion,'' hep-th/0008154.}

\lref\dtrev{E.~Sharpe, ``Recent developments in discrete
torsion,'' Phys. Lett. B {\bf 498} (2001) 104-110,
hep-th/0008191.}

\lref\dtshift{E.~Sharpe, ``Discrete torsion and shift orbifolds,''
Nucl. Phys. B {\bf 664} (2003) 21-44, hep-th/0302152.}

\lref\evaed{E.~Silverstein and E.~Witten, ``Criteria for conformal
invariance of $(0,2)$ models,'' Nucl. Phys. B {\bf 444} (1995)
161-190, hep-th/9503212.}

\lref\candelasetal{P.~Berglund, P.~Candelas, X.~de~la~Ossa,
E.~Derrick, J.~Distler, and T.~Hubsch, ``On the instanton
contributions to the masses and couplings of $E_6$ singlets,''
Nucl. Phys. B {\bf 454} (1995) 127-163, hep-th/9505164.}

\lref\chrised{C.~Beasley and E.~Witten, ``Residues and worldsheet
instantons,'' hep-th/0304115.}

\lref\bbh{K.~Becker, M.~Becker, M.~Haack, and J.~Louis,
``Supersymmetry breaking and $\alpha'$-corrections to flux
induced potentials,'' hep-th/0204254.}

\lref\ersrev{E.~Sharpe, ``Lectures on D-branes and sheaves,''
hep-th/0307245.}

\lref\tian{G.~Tian and S.~T.~Yau, ``Three-Dimensional Algebraic
Manifolds With $c_1 = 0$ And $\chi = -6$.''}

\lref\candelas{P.~Candelas, P.~S.~Green and T.~Hubsch, ``Finite
Distances Between Distinct Calabi-Yau Vacua: (Other Worlds Are
Just Around The Corner),'' Phys.\ Rev.\ Lett.\  {\bf 62}, 1956
(1989); P.~Candelas, P.~S.~Green and T.~Hubsch, ``Rolling Among
Calabi-Yau Vacua,'' Nucl.\ Phys.\ B {\bf 330}, 49 (1990).}

\lref\andy{A.~Strominger, ``Massless black holes and conifolds in
string theory,'' Nucl.\ Phys.\ B {\bf 451}, 96 (1995),
hep-th/9504090.}

\lref\gms{B.~R.~Greene, D.~R.~Morrison and A.~Strominger,
``Black hole condensation and the unification of string vacua,''
Nucl.\ Phys.\ B {\bf 451}, 109 (1995), hep-th/9504145.}

\lref\mirror{B.~R.~Greene and M.~R.~Plesser, ``Duality In
Calabi-Yau Moduli Space,'' Nucl.\ Phys.\ B {\bf 338}, 15 (1990);
P.~Candelas, X.~C.~De La Ossa, P.~S.~Green and L.~Parkes, ``A
Pair Of Calabi-Yau Manifolds As An Exactly Soluble Superconformal
Theory,'' Nucl.\ Phys.\ B {\bf 359}, 21 (1991); P.~Candelas,
X.~C.~De la Ossa, P.~S.~Green and L.~Parkes, ``An Exactly Soluble
Superconformal Theory From A Mirror Pair Of Calabi-Yau
Manifolds,'' Phys.\ Lett.\ B {\bf 258}, 118 (1991).}

\lref\kstt{S.~Kachru, M.~B.~Schulz, P.~K.~Tripathy and
S.~P.~Trivedi, ``New supersymmetric string compactifications,''
JHEP {\bf 0303}, 061 (2003), hep-th/0211182.}

\lref\gaugL{G.~L.~Cardoso, G.~Curio, G.~Dall'Agata and D.~Lust,
``Heterotic String Theory on non-Kaehler Manifolds with H-Flux and
Gaugino Condensate,'' hep-th/0310021.}

\lref\Kgukov{S.~Gukov, S.~Kachru, X.~Liu and L.~McAllister,
``Heterotic Moduli Stabilization with Fractional Chern-Simons
Invariants,'' hep-th/0310159.}

\Title{\vbox{\hbox{hep-th/0310058} \hbox{UMD-PP-03-69}
\hbox{SU-ITP-03/24} \hbox{ILL-(TH)-03-07}}} {\vbox{ \vskip-2.5in
\hbox{\centerline{Compactifications of Heterotic Strings}}
\hbox{\centerline{on Non-K\"ahler Complex Manifolds: II}}}}

\vskip-.4in


\centerline{\bf Katrin Becker} \centerline{\it Department of
Physics, University of Utah, Salt Lake City, UT 84112-0830}
\vskip.01in \centerline{\bf Melanie Becker, Paul S. Green}
\centerline{\it Department of Phys \&  Maths, University of
Maryland, College Park, MD 20742-4111} \vskip.01in
\centerline{\bf Keshav Dasgupta} \centerline{\it Department of
Physics, Varian Lab., Stanford University, Stanford CA 94305-4060}
\vskip.01in \centerline{\bf Eric Sharpe} \centerline{\it
Department of Mathematics, University of Illinois, 1409 W. Green
St., MC-382}

\vskip.03in

\centerline{\bf Abstract}

We continue our study of heterotic compactifications on
non-K\"ahler complex manifolds with torsion. We give further
evidence of the consistency of the six-dimensional manifold
presented earlier and discuss the anomaly cancellation and
possible supergravity description for a generic non-K\"ahler
complex manifold using the newly proposed superpotential. The
manifolds studied in our earlier papers had zero Euler
characteristics. We construct new examples of non-K\"ahler
complex manifolds with torsion in lower dimensions, that have
non-zero Euler characteristics. Some of these examples are
constructed from consistent backgrounds in F-theory and therefore
are solutions to the string equations of motion. We discuss
consistency conditions for compactifications of the heterotic
string on smooth non-K\"ahler manifolds and illustrate how some
results well known for Calabi-Yau compactifications, including
counting the number of generations, apply to the non-K\"ahler
case. We briefly address various issues regarding possible
phenomenological applications.

\vskip.05in

\Date{Oct 2003}

\centerline{\bf Contents}\nobreak\medskip{\baselineskip=12pt
 \parskip=0pt\catcode`\@=11  

\noindent {1.} {Introduction} \leaderfill{2} \par 
\noindent \quad{1.1.} {A Brief History} \leaderfill{2} \par 
\noindent \quad{1.2.} {Recent Advances} \leaderfill{3} \par 
\noindent \quad{1.3.} {Phenomenological Aspects} \leaderfill{6} \par 
\noindent \quad{1.4.} {Organization of the paper} \leaderfill{8} \par 
\noindent {2.} {Generalities} \leaderfill{11} \par 
\noindent \quad{2.1.} {Review of Consistency Conditions for Heterotic Compactifications} \leaderfill{11} \par 
\noindent \quad{2.2.} {Gauduchon Metrics and $\alpha'$ Corrections} \leaderfill{12} \par 
\noindent \quad{2.3.} {Stability and the DUY Equation} \leaderfill{13} \par 
\noindent \quad{2.4.} {Number of Massless Fermion Families} \leaderfill{15} \par 
\noindent \qquad{2.4.1.} {Review of Calabi-Yau Results} \leaderfill{15} \par 
\noindent \qquad{2.4.2.} {Non-K\"ahler Backgrounds} \leaderfill{16} \par 
\noindent \quad{2.5.} {A Note on Orbifolds in Flux Backgrounds} \leaderfill{17} \par
\noindent \quad{2.6.} {Superpotential and the Torsional Equations} \leaderfill{18} \par 
\noindent \qquad{2.6.1.} {Solution to the Anomaly Equation to all orders in $\alpha'$} \leaderfill{20} \par 
\noindent \qquad{2.6.2.} {Supergravity Description} \leaderfill{25} \par 
\noindent {3.} {New Six-Dimensional Compactifications} \leaderfill{30} \par 
\noindent \quad{3.1.} {Dimensional Reduction to Eight Dimensions} \leaderfill{31} \par 
\noindent \quad{3.2.} {Compactifications of the Heterotic String on a Background Conformal to $K3$} \leaderfill{34} \par 
\noindent {4.} {Fluxes versus Branes: A Different Perspective} \leaderfill{45} \par 
\noindent {5.} {$T^2$ Bundles on $K3$: An Alternative Description} \leaderfill{49} \par 
\noindent \quad{5.1.} {Consistency Check via Duality Chasing} \leaderfill{50} \par 
\noindent \qquad{5.1.1.} {Summary of Known Results} \leaderfill{50} \par 
\noindent \qquad{5.1.2.} {Steps for Generating the Background} \leaderfill{51} \par 
\noindent \qquad{5.1.3.} {Precise Supergravity Analysis} \leaderfill{52} \par 
\noindent \quad{5.2.} {An Example with a $U(1)$ Bundle and a Nonzero Number of Generations} \leaderfill{60} \par 
\noindent {6.} {Smooth Examples with Nonzero Euler Characteristics} \leaderfill{66} \par 
\noindent \quad{6.1.} {Connected Sums of $S^3 \times S^3$} \leaderfill{66} \par 
\noindent \quad{6.2.} {Flops of Calabi-Yau's} \leaderfill{67} \par 
\noindent \qquad{6.2.1.} {A few necessary conditions to pull bundles through flops} \leaderfill{68} \par 
\noindent \qquad{6.2.2.} {Flops on elliptically-fibered Calabi-Yau manifolds} \leaderfill{72} \par 
\noindent {7.} {Some Orbifold Examples} \leaderfill{74} \par 
\noindent \quad{7.1.} {The Duality Chains} \leaderfill{74} \par 
\noindent \quad{7.2.} {Analysis of Type IIB Background} \leaderfill{78} \par 
\noindent \quad{7.3.} {Analysis of the Heterotic Background} \leaderfill{82} \par 
\noindent {8.} {Discussion and Future Directions} \leaderfill{84} \par
\noindent \quad{8.1.} {New Issues on Non-K\"ahler Manifolds and Open Questions} \leaderfill{85} \par 
\noindent \quad{8.2.} {De Sitter spacetimes} \leaderfill{87} \par  
\noindent Appendix {A.} {Massless Spectra in Non-K\"ahler Compactifications} \leaderfill{91} \par 
\noindent \quad{\hbox {A.}1.} {Heterotic Sigma Model with (0,1) Supersymmetry and Torsion} \leaderfill{92} \par 
\noindent \quad{\hbox {A.}2.} {Heterotic Sigma Model with (0,2) Supersymmetry and Torsion} \leaderfill{92} \par 
\noindent \quad{\hbox {A.}3.} {Interpretation} \leaderfill{94} \par 
\noindent \quad{\hbox {A.}2.} {Bismut's Index Theorem} \leaderfill{96} \par 
\catcode`\@=12 \bigbreak\bigskip}



\newsec{Introduction}

\subsec{A Brief History}

Compactifications of heterotic strings on Calabi-Yau (CY)
manifolds have been an active area of research ever since they
were proposed in \candle. Many of the basic structures of
superstring compactifications and its relation to the topological
properties of the Calabi-Yau manifolds were formulated in the
classic work of \candle. Soon after this, Tian and Yau \tian\
gave the first CY manifold that would give a realistic three
generation model in four dimensions with minimal supersymmetry.
Other examples (using complete intersections of CY manifolds)
followed immediately. It was also realized simultaneously that
many phenomenological properties could be easily studied using
simple techniques of algebraic geometry. The phenomenological
aspects of the Calabi-Yau construction led to some significant work in
the early eighties continuing to the present.  Despite the
enthusiasm, it was realized that Calabi-Yau compactifications
suffer from various weaknesses. The first one of them was, of
course, the degeneracy problem. There are thousands of Calabi-Yau
manifolds that could be potential solutions to string theory,
giving rise to new vacua that are not realized in nature. A
second rather important problem appears, if one takes into
account that deformations of a given Calabi-Yau manifold, give
rise to many uncontrolled moduli. These moduli are basically the
K\"ahler and complex structure moduli governed by the Hodge
numbers $h_{11}$ and $h_{21}$ respectively, that would remain
unfixed at tree level, so that no predictions for the coupling
constants of the standard model could be made. Although there
were many indirect arguments, that non-perturbative effects would
eventually fix many of these moduli, it was never shown
explicitly how many of them would get fixed by such effects
because non-perturbative effects are many times hard to evaluate
explicitly.
One particularly important K\"ahler modulus,
that was difficult to stabilize
was the radial modulus of the CY. It was conjectured by Dine and
Seiberg \DineSB, that because of this uncontrolled modulus, all
Calabi-Yau manifolds will eventually attain an infinite radius
making naive {\it compactification} unrealistic. Many arguments
to stabilize the radial modulus by generating a superpotential
failed, because it could be shown easily, that the no-scale
structure (at least at the supergravity level) of the scalar
potential remains unbroken. Progress in this direction has only
been made recently, as we will discuss a little later.

Returning to our discussion about the degeneracy problem, one
interesting step towards solving this issue was proposed by
Candelas and his collaborators \candelas\ some time ago. In this
paper it was argued, that many of the CY moduli spaces are {\it
connected} via conifold singularities. One can go from one CY to
the other by shrinking a two-cycle and blowing up a three-cycle,
in other words, via a so called conifold transition. This provided
a way to connect Calabi-Yau manifolds with different topological
data. Many interesting developments in this direction
followed once non-perturbative
effects in string theory were understood in the mid nineties.
Examples of these are:  the resolution of conifold singularities
via black holes from the discovery that the previous
conifold transitions are smooth transitions, if one takes black
hole condensation into account \andy,\gms\ and mirror symmetry
\mirror, to name a few. However, despite these
major advances, one basic problem still remained: the moduli space
problem.

\subsec{Recent Advances}

Some early advances in understanding the moduli space problem came
from the works of \HULL,\ibars,\rstrom\ and \SmitD, where
heterotic compactifications in the presence of a torsion, that is
not closed were introduced and analyzed for the first time. The
torsion in these theories is generated by a background three-form
${\cal H}$ flux \foot{In these compactifications there also
appeared the so called ``warp factor'' for the first time in the
physics literature (to our knowledge), that has become very  popular
in recent times in particle phenomenology since the work of
Randall and Sundrum. See also \warner\ for an early discussion of warp factor 
in $AdS$ compactifications.}. 
It was argued therein, that the ${\cal H}$
torsion relaxes the K\"ahler condition by making $dJ \ne 0$,
where $J$ is the fundamental two-form. The manifold could still
remain complex if the torsion three-form is related to the
fundamental two-form $J$ by the torsional equation ${\cal H} = i
(\bar\del - \del)J$. An explicit form of these manifolds was not
presented in the works of \HULL,\ibars,\rstrom\ and \SmitD, as it
was difficult to realize the non-K\"ahlerity directly from the
string equation of motion\foot{One explicit four-dimensional
example, that is conformal to K3 was constructed in \rstrom\
though.}. Furthermore, it was required, that these manifolds
should have an $SU(3)$ holonomy with vanishing first Chern class to
preserve minimal supersymmetry in four dimensions. In these early
works, it was already anticipated that, if these manifolds were
explicitly constructed, they would overcome the Dine-Seiberg
runaway problem because the radial modulus would get stabilized.
But because of the technical difficulties mentioned above,
progress in this direction was not made for many years, until the
discovery of string dualities in the mid nineties.

An explicit solution to this problem came, rather unexpectedly,
from a slightly different direction. This involved the study of
M-theory compactifications on a four-fold in the presence of
fluxes \rBB. In \rBB\ it was shown rather clearly, for the
first time, how a four-fold background changes in the presence of
the $G$-fluxes of M-theory. The background is determined by the
warp factor and constraint equations, which relate the background
to the fluxes. Soon after this it was argued in \guko, that these
constraint equations can be derived from a superpotential that
fixes all the complex structure moduli and some K\"ahler
structure moduli. The radial modulus was however not fixed. The
explicit form of the scalar potential for the moduli fields of the
dual type IIB theory was presented in the first reference of \kst.
At this point it became evident, that string
compactifications with fluxes lead us one step closer to finding
the solution to one of the most important open problems in string
theory.

However, in order to describe the four-dimensional real world in
which we live, we are interested in considering phenomenologically
interesting models, in other words, we are forced to understand
four-dimensional compactifications of the heterotic string with
non-vanishing fluxes and derive the corresponding potential for
the moduli fields along the lines described above. In this way, we
would get one step closer to determining the coupling constants
of the standard model. We can do this using string dualities.

This led to the observation that a specific
background of \rBB, which is a $K3 \times K3$  four-fold,
can have an orientifold description in the
type IIB theory. Duality chasing this background, it was shown in
\sav\ and  \beckerD, that there exists a heterotic dual, which is
generically non-K\"ahler. The torsion originates from the M-theory
$G$ fluxes. This gave an explicit non-K\"ahler manifold in the
heterotic theory. By construction the manifold was also compact
and since all the analysis involved U-dualities, it was a
specific solution to string theory. This showed that
compactifications of the heterotic string on non-K\"ahler
manifolds are indeed consistent, something that was not known in
the earlier literature mentioned at the beginning of this
section.

Topologically, the manifolds considered in \sav, \beckerD\ 
 are non trivial $T^2$ bundles over
a $K3$ base. A detailed mathematical analysis of these manifolds
was presented recently in \GP\ and \bbdg. What remained now was
to verify whether all the conditions proposed in
\HULL,\ibars,\rstrom,\SmitD\ worked for the explicit background.

In \beckerD, a slightly simplified version of the non-K\"ahler
manifold, proposed in \sav, was taken. In this model the orbifold
limit of the base $K3$ was considered. It was shown that the torsional
equations are satisfied and the manifold was indeed non-K\"ahler.
The radial modulus also gets stabilized along with all the complex
structure moduli and some of the K\"ahler structure moduli by
balancing the fluxes with the non-K\"ahler nature of the manifold
\bbdp. A somewhat similar fixing of the complex structure and the K\"ahler
structure moduli was argued to happen for the type IIB case in the
first two references of \kst.

The non-K\"ahler background discussed above can be found by
minimizing the superpotential computed in \bbdg, \bbdp,\luest, as the
torsional constraint can be derived from this superpotential. All
these solutions were shown to have zero Euler characteristics
\GP,\bbdg. The superpotential fixes all the complex structure
moduli, some K\"ahler structure moduli (the precise number is not
yet known) and the radial modulus. However the dilaton modulus
remains unfixed (at tree level). This situation can be compared
to the situation in type IIB theory. In type IIB, as discussed in
\kst, the dilaton is fixed
but the radius is not. Whereas on the heterotic side, the radius
is fixed but the dilaton is not. A more detailed discussion of the
type IIB side for the non-K\"ahler heterotic example discussed in
detail in our work is given in \pktsptt, while compactifications
on the $T^6/Z_2$ manifold can be found in \kstt\ (See also \dabhull\ for
an alternative argument on moduli stabilisation).

In this paper we continue to study the heterotic theory
compactified on non-K\"ahler manifolds. We give an alternative
derivation of the non-K\"ahler manifolds that were studied
earlier and construct new examples of manifolds that have non-zero
Euler characteristics. As we mentioned earlier, in four
dimensions, all the manifolds constructed so far in
\sav,\beckerD,\GP,\bbdg,\bbdp\ have zero Euler characteristics.
This is not a problem by itself, as zero Euler characteristic does
not imply zero number of generations\foot{This will be elaborated
more in later sections.}. However it is interesting to see,
whether it is possible to construct non-K\"ahler manifolds in the
heterotic theory that could have non-zero Euler characteristics.
Part of the fascination lies in the fact, that this will provide
new manifolds for string compactifications, which would be
phenomenologically more attractive and also a new fascinating
area for mathematics.

To finish this sub-section, let us comment briefly some more on
the mathematical aspect on non-K\"ahler complex manifolds with
torsion. Many of the techniques applied to Calabi-Yau's also
apply, to some extent, to non-K\"ahler compactifications. For
example, one can formally count massless spectra in non-K\"ahler
compactifications, as we discuss in appendix A, but as we cannot
go to an $\alpha'=0$ limit, the results of such a computation
should be interpreted with caution.  A somewhat more solid
result should be the number of generations, which can also be
computed and turns out to have the same form as in Calabi-Yau
compactifications. Also, in Calabi-Yau compactifications, the
Donaldson-Uhlenbeck-Yau equation can be translated into the
mathematical notion of stability. For non-Kahler
compactifications, an analogous translation can also be
performed, albeit under certain restrictions on the metric. In
addition to discussing how standard methods from the analysis of
heterotic strings on Calabi-Yau's can be applied to non-K\"ahler
compactifications, we shall discuss some smooth examples of
non-K\"ahler compactifications in later sections.

\subsec{Phenomenological Aspects}

For phenomenological applications we can use some of the results
(with some modifications) that are used for ordinary CY compactifications.
Here we will follow the discussions given
in \witsuper, though with one key difference: we no longer
 {\it can } embed the spin-connection into the gauge connection.
 What we require is an
 $SU(5)$ or $SO(10)$ gauge group from a unified point of view, which
 can eventually yield us $SU(3)_C \otimes SU(2)_L \otimes U(1)_Y$ gauge
 group
 in
 four dimensions. This can be achieved by giving vacuum expectation values
to some of the
  scalars in four dimensions (recall that the flux quantization keeps some
of the K\"ahler moduli unfixed).
 We also require that there should be Yukawa couplings that are
 phenomenologically useful.
 As pointed out by \nappiS\ the string induced phenomenological models
  should
 be able to
 (a) produce {\it massive} quarks and leptons, (b) produce almost
massless
 neutrinos and (c) show a very slow proton decay.
The unbroken $E_6$ gauge group can have particles that would
mediate proton decay. Therefore to have
 very slow decays we would require superheavy particles.

This brings us to the
 the first simple case of $E_8$ breaking to $E_6 \times SU(3)$. This is
 the
 case considered so far in the Calabi-Yau literature. We can
 easily rule out the situation where $A = \omega$
 as we do not have the standard embedding of spin connection into the gauge
 connection. But we can have a different embedding of the $SU(3)$
 gauge bundle that does not satisfy the standard equation. One would have to
check
 the
 precise way we can embed the $SU(3)$ bundle and still satisfy the DUY
 equation.
 The next interesting possibility would be to have an $SU(4)$ bundle on the
 non-K\"ahler manifold. This
 would mean that we have an $SO(10)$ gauge group in space-time. The
 $SO(10)$ gauge group is, of course, very useful for a unified model.
 However, the choice of $SO(10)$ gives rise to many moduli from the gauge
 bundle that are difficult to stabilize. We have to make sure that many
 of the particles that would in general allow proton decay pick up
 a
 very large mass without ruining the consistency of the whole picture.
 This is in principle possible but the analysis looks a little contrived.
 We
 can do better than this by having an $SU(5)$ bundle on the
 non-K\"ahler space, implying that the $E_8$ group breaks to
 $SU(5)
 \times SU(5)$. This situation is nice in the sense that the $SU(5)$
 group is one of the groups considered in grand unified theories \glashow.
We now need to see how the
 irreducible representations of $E_8$ decompose under $SU(5)$. This is easy
to work out
 from the $E_8$ Dynkin diagram

 \vskip.2in

 \centerline{\epsfbox{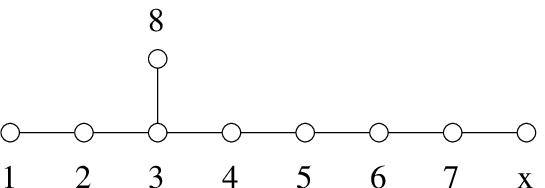}}

 \vskip.2in

 \noindent where we have used x to denote the extended root of $E_8$. The
 ${\bf 248}$ of $E_8$ can be easily seen to decompose under the $A_4$
 Dynkin diagrams as
 \eqn\twfoei{{\bf 248} = ({\bf 24}, {\bf 1}) + ({\bf 1}, {\bf 24}) +
 ({\bf
 10}, {\bf 5}) + (\bar{\bf 10}, \bar{\bf 5}) +
 ({\bf 5}, \bar{\bf 10}) +
 (\bar{\bf 5}, {\bf 10})}
 where we have the ${\bf 24}$, ${\bf 10}$, ${\bf 5}$ of $SU(5)$ and their
 conjugates.  The various
 unrequired moduli will then become massive. Whether the choice of
 background
 fluxes can do this job, remains to be checked. We leave this for
 future work.

 Consider next the possibility of an $SU(6)$ gauge bundle. This would imply
 that the $E_8$ breaks to $SU(2) \times SU(3) \times SU(6)$. This is
 almost (but not quite) the gauge group that we need, because of the
 absence of the $U(1)$ factor. If we ignore the $U(1)$, the ${\bf 248}$
 of
 $E_8$ can be shown to decompose as
 \eqn\susixde{\eqalign{{\bf 248} =& ~({\bf 3}, {\bf 1}, {\bf 1}) +({\bf
 1},
 {\bf 8}, {\bf 1}) +({\bf 1}, {\bf 1}, {\bf 35})
 +({\bf 1}, {\bf 3}, \bar{\bf 15}) + ({\bf 1}, \bar{\bf 3}, {\bf 15})
 \cr
  & ~~~~~~~~~~~~~ + ({\bf 2}, {\bf 3}, {\bf 6}) + ({\bf 2}, \bar{\bf 3},
 \bar{\bf 6}) + ({\bf 2}, {\bf 1}, {\bf 20})}}
 It is easy to recognize the standard model structure here (without the
 $U(1)$). The first two are the ${\bf 11}$ vector bosons and the next
 fields are the Higgs in ${\bf 35}$ representation.
 The rest of the fields are the singlet quarks
 and
 the weak doublet quarks.

 To finish this discussion, one can consider an $SU(2)$ bundle on the
non-K\"ahler
 space. This breaks $E_8$ to $E_7 \times SU(2)$. Phenomenology with
 $E_7$ is interesting. It has a maximal subgroup of $SU(3) \times SU(6)$.
 Now
 $SU(6)$ has many subgroups, one of them being $SU(3) \times SU(3)
 \times U(1)$. This would imply, that we have three $SU(3)$ subgroups and
one
 $U(1)$ subgroup. Identifying one $SU(3)$ as the color group of
 the standard model, the other $SU(3) \times SU(3)$ breaking down to a
 diagonal $SU(2)$ might give us a reasonable phenomenological model.
 Again,
 how this breaking appears explicitly in our set up with fluxes needs to
 be
 worked out (see \curioLu\ for some recent discussion of the phenomenological aspects of
flux compactifications).
All these issues are beyond the scope of the present work
 and
 will be addressed elsewhere.

\subsec{Organization of the paper}

This paper is organized as follows. In section 2 we discuss
properties of generic non-K\"ahler manifolds. We include the study
of Gauduchon metrics and the corresponding $\alpha'$ corrections.
We discuss the existence of vector bundles on these manifolds,
their stability and the solutions to the Donaldson-Uhlenbeck-Yau
(DUY) equation, and also redo computations of the number of
generations in this non-K\"ahler context.
We also discuss
some more properties of the superpotential computed in \bbdg,
\bbdp and \luest. We show, that iterative techniques are powerful
enough to reproduce the background values of the three-form
fields to any arbitrary order in $\alpha'$ (when the underlying space
is an orbifold). The knowledge of the
{\it complex} three-form field and the superpotential to all
orders in $\alpha'$ allows us to make predictions for the ten
dimensional low energy supergravity lagrangian of the heterotic
theory order by order in $\alpha'$. We compare the terms of the
four dimensional lagrangian, that can be obtained from a
dimensional reduction with the recent computation of  \luest\ and
find agreement. However, our approach is more powerful, as it
allows us to make {\it predictions} for the form of the ten
dimensional effective action of the heterotic string to all
orders in $\alpha'$.

Next we discuss more specific examples. In section 3, we study
non-K\"ahler compactifications to six space-time dimensions. We
elaborate the earlier studied example of torsional $K3$ manifold
appearing in the work of Strominger \rstrom. In section 3.1 we
show that the conformal $K3$ example cannot be thought of as the
dimensional reduction (to eight dimensions) of the previously
studied model \sav,\beckerD, \GP, \bbdg\ due to F-theory
monodromies and radial modulus stabilization in the heterotic
theory. In section 3.2 we elaborate the conformal $K3$ example by
duality chasing it to a IIB Gimon-Polchinski type model. Due to
subtleties of fluxes in six-dimensions, we consider only non
compact examples (though we show that our analysis easily
overcomes the no-go theorems in six dimensions). The mapping to
the type IIB models (and then to F-theory model) is subtle due to
the two-fold ambiguities in the orientifolding action. We discuss
this issue and consider the (3,243) model in F-theory, that is
directly related to our case. We also evaluate the precise
F-theory three fold, that corresponds to the torsional $K3$ with fluxes.

The analysis done in section 3 can be given an {\it alternative}
description purely in terms of branes instead of fluxes. In
section 4 we elaborate on this and argue, that the background
metric (given in terms of the warp factors for the manifolds with
fluxes) can be written in terms of the metric of branes sitting at
a point on the manifold. This can be done by relating the harmonic
functions appearing in this context to the warp factors
previously used. Using this we give an alternative picture of the
six dimensional compactifications by replacing the torsion with
$NS5$ branes. We show that this can also be done for the four
dimensional compactification by replacing the three form fluxes
(in the type IIB picture) with $D3$ branes, or the torsion on the
non-K\"ahler manifold by {\it wrapped} $NS5$ branes.

We then discuss non-K\"ahler compactifications to four spacetime
dimensions in section 5 and study some aspects of the non-K\"ahler
manifolds of \sav,\beckerD,\GP,\bbdg\ using different techniques.
The advantages of this alternative perspective are two-fold: one, we
can show the consistency of the manifolds studied earlier in a
different way and two, this gives us a way to study non-K\"ahler
compactifications with non-constant three form fluxes. Recall that
all the earlier examples were developed using constant three form
background fluxes. Section 5.1 is devoted to this aspect. We show
that in the type IIB theory there exists a {\it brane-box}-like
configuration, with the sides of the box made with $NS5$ branes,
embedded in a non-trivial Taub-NUT background that can give rise
to the non-K\"ahler manifold that we studied earlier. We show how
the two $NS5$ branes are responsible to twist the fiber torus on
the dual side. We also show that the gauge bundles (that we
expect on the heterotic side) appear naturally if we incorporate
the F-theory monodromies carefully.

In section 5.2 we consider the number of generations for these
type of compactifications. We use the formula derived earlier in
section 2 for the toy example of $U(1)$ bundles. We explicitly
study the {\it local} case of DUY equations for this bundle and
compute the solution to this equation. The usual DUY equations get
modified by the addition of a constraint on ${\rm tr} ~F \wedge
F$ from the torsion. Solving these set of equations give us a
possible embedding. We use this to evaluate the number of
generations for our case taking into account the possibilities of
(a) non trivial warp factors, (b) torsion and (c) non- standard
embedding.

The non-K\"ahler manifolds studied earlier and in section 5
all have zero Euler characteristic.  In sections 6 and 7 we give
examples of non-K\"ahler compactifications down to four spacetime
dimensions, that have non-zero Euler characteristics. In section 6
we provide some smooth examples. We study the examples of
connected sums of $S^3 \times S^3$ manifolds in section 6.1. In
section 6.2 we study examples of non-K\"ahler manifolds that
could be generated from flops of Calabi-Yau manifolds. We discuss
in detail how bundles on these manifolds follow through flops and
construct examples of elliptically fibered non-K\"ahler manifolds
that appear from flopping an elliptically fibered Calabi-Yau.

In section 7 we give some more examples of compactifications with non-zero
Euler characteristics. These examples are of orbifold kind and can
be duality chased from a consistent orientifold background of
F-theory. In section 7.1 we give some details on the duality
chasing from consistent four-folds in F-theory. These orientifold
examples are subtle again because of ambiguities in the
orientifolding action. In section 7.2 we analyze the corresponding
type IIB background keeping the axion-dilaton and the Euler
characteristic of the four-fold arbitrary (to avoid the
subtleties). We give a detailed discussion of the moduli fixing
in the type IIB picture by solving all the background equations
generated by minimizing the superpotential with respect to the
axion-dilaton modulus and the complex structure of the manifold
$\tau$. We specify a possible background that gives rise to a
complex manifold in type IIB theory. In section 7.3 we go to the
heterotic theory by making a set of U dualities. The manifold on
the heterotic side is generically non-K\"ahler and has non-zero
Euler characteristics. We determine the fibration structure of
the manifold and specify the torsion explictly.

Section 8 is dedicated to discussions. We discuss various
unresolved issues, that appeared in earlier sections and study
some aspects of phenomenology and de-Sitter spacetimes.

In the appendix A we use a sigma model description of the
heterotic string to calculate the massless spectrum appearing in
our theory. Even though, some of the assumptions entering the
calculation do not necessarily have to strictly hold for our
case, we still hope to obtain some qualitative information from
this calculation.

\vskip 0.5cm

{\bf Note Added I}: As the draft was being written there appeared
a paper discussing the gaugino condensate that would fix the
dilaton modulus in the heterotic theory \gaugL.
Similar arguments have been reached independently by S. Prokushkin \prokS.
We have also been
informed that some related results have been worked out
for the CY case in \Kgukov.

\vskip 0.5cm

{\bf Note Added II}: In recent times there have appeared several
articles on the net \lenny,\mikei\ and \mikeii, discussing the
question on how many vacua there exist in string theory
compactifications with fluxes. Since fluxes satisfy the Dirac
quantization condition one would think, in principle, that there
is still an infinite number of string vacua, labeled by a discrete
integer, so that the predictive power of string theory gets lost.
However, this is not strictly correct and different solutions to
this problem have been proposed. One of them involves the
anthropic principle, as discussed in \lenny. A different approach
was taken by \mikei\ and \mikeii. In \mikei\ it is discussed how
many vacua string theory actually has, and what this means for
the predictivity, raising precisely questions such as whether the
number is finite or infinite, and explaining how getting good
estimates could help make predictions (or, depending on the
results, could suggest that certain predictions cannot be made).
In \mikeii\ the number of type II supersymmetric flux vacua is
counted and it is shown, that under a simple physically motivated
constraint this number can actually be finite.

However, we have an even more optimistic point of view, that a
selection principle will hopefully be found that picks out a
small number of vacua (and maybe just one) out of this large
number of vacua. A nice discussion on these issues has recently
appeared in {\bdine}.

\newsec{Generalities}

We start with some general properties of non-K\"ahler
compactifications. Our aim is to build a framework that can be
used to study specific examples later on. We will mainly be
interested in compactifications of the heterotic string on these
type of manifolds.

\subsec{Review of Consistency Conditions for Heterotic
Compactifications}

Let us briefly review some necessary conditions for
consistent non-K\"ahler
compactifications. Recall that for
compactifications without flux, unbroken supersymmetry implies
the existence of a complex structure and a K\"ahler structure,
together with a nowhere-zero holomorphic top form. These are the
conditions required for a space to be Calabi-Yau and they are
formulated in terms of a covariantly constant spinor on the
internal manifold. Conversely, given a Calabi-Yau manifold it can
be shown that one is guaranteed to have a corresponding
covariantly constant spinor.

A similar situation appears for compactifications on manifolds
with torsion that were considered by Strominger in \rstrom. The
compactification manifold turns out to be complex and have a
nowhere-zero holomorphic top form, but the manifold is not
K\"ahler. Rather, derivatives of the fundamental (1,1) form
derived from the metric are related to the torsion background.

So far we have only discussed constraints on the underlying
background manifold with torsion. Additional constraints exist on
the non-abelian gauge field of the heterotic theory. For both,
Calabi-Yau compactifications and the non-K\"ahler
compactifications discussed here, the gauge field must be
holomorphic, meaning
\eqn\holodu{
F_{i j}  =  F_{ \overline{\imath} \overline{\jmath} }  =  0,
}
i.e the (0,2) and the (2,0) part vanishes,
and it must also satisfy the partial differential equation
\eqn\hoduye{
g^{i \overline{\jmath}} F_{i \overline{\jmath}}  =  0,
}
where $F_{i \overline{\jmath}}$ is the curvature of the gauge
field. This partial differential equation is known as the
Donaldson-Uhlenbeck-Yau equation and it will be discussed it in
greater detail in a later section.

Moreover, in both cases the anomaly cancellation condition
\eqn\dhequa{
d{\cal H}  =  {\rm tr}~R \wedge R - {1\over 30} {\rm Tr}~F \wedge F,
}
must be satisfied. However, in contrast to Calabi-Yau
compactifications of the heterotic string, in a the non-K\"ahler
case  it is not consistent to embed the spin connection into the
gauge connection, see \PapaDI, \bbdg.

Finding explicitly non-K\"ahler backgrounds for the heterotic
string is, as should by now be clear, somewhat more complicated
than finding explicit Calabi-Yau compactifications. A list of
conditions that non-K\"ahler complex manifolds have to satisfy to
lead to supersymmetric theories was given in \rstrom. To check
the consistency of the non-K\"ahler compactifications we are
interested in (which are all supersymmetric), we will use string
duality arguments.

\subsec{Gauduchon Metrics and $\alpha'$ Corrections}

There are two analogues of the K\"ahler condition that will play
an important role in the mathematical analysis relevant to many
parts of this paper. First, one says that a complex hermitian
manifold is strong K\"ahler torsion if the metric has the
property that the $(1,1)$ form \eqn\gaumet{ J~ = ~i g_{i
\overline{\jmath}}~ dz^i \wedge d \overline{z}^{
\overline{\jmath}} } obeys $\overline{\partial} \partial J = 0$.
Furthermore, the complex hermitian manifold is said to be {\it
Gauduchon} if $J$ has the property $\overline{\partial} \partial
J^{n-1} = 0$, where $n$ is the complex dimension of the manifold.
These conditions are ubiquitous in relevant mathematics, and
appear at numerous points in this paper. Physically, the strong
K\"ahler torsion condition is equivalent to the condition that $d
{\cal H} = 0$.

Strictly speaking, the physical metric in a non-K\"ahler
compactification will never be strong K\"ahler torsion; see for
example \PapaDI. Nevertheless, we can still consistently use
results about strong K\"ahler torsion and Gauduchon metrics so
long as $\alpha'/r^2$ is small ($r$ being the
radius of the six manifold), the same regime in which
target-space supergravity is a good approximation. In particular,
the warp factors mentioned above are $1 + {\cal O}(\alpha'/r^2)$,
and $d {\cal H} = {\cal O}(\alpha'/r^2)$. If $\alpha'/r^2$ is
small, then the warp factors are approximately 1, and $d {\cal H}$
is approximately zero. Thus, our program is to expand
perturbatively about strong K\"ahler torsion metrics. Since the
size of the internal manifold is determined by the fluxes (which
in principle can be pretty large) the quantity $\a'/r^2$ can be
sufficiently small so that quantum corrections can be consistently
incorporated.

\subsec{Stability and the DUY Equation}

In both K\"ahler and non-K\"ahler compactifications the heterotic
gauge field must satisfy the Donaldson-Uhlenbeck-Yau (DUY)
equation:
$$
g^{i \overline{\jmath}} F_{i \overline{\jmath}}  =  0.
$$
as we mentioned above.
This constraint arises as a D-term in heterotic theories
\witsuper, as can be seen by dimensional reduction of the
ten-dimensional gaugino supersymmetry variation
\eqn\fmunu{
\delta \chi  =  F_{\mu \nu} \Gamma^{\mu \nu} \epsilon
}
to a D-term-type piece in the supersymmetry variation of the
low-energy gaugino:
\eqn\delch{
\delta \chi  =  F_{i \overline{\jmath}} g^{ i \overline{\jmath}}
\epsilon.}
The fact that the DUY equation is realized as a D-term can also
be seen via realizations of K\"ahler cone substructure \kcsub, as
will be discussed below.

Now, solving partial differential equations can be very hard, but
we can convert this into a problem in algebraic geometry. On a
K\"ahler manifold, (holomorphic) gauge fields that satisfy the
DUY equation are equivalent to ``Mumford-Takemoto stable''
holomorphic vector bundles. In the following we will denote by
${\cal E}$ the holomorphic vector bundle defining the gauge
bundle. Moreover, we denote the K\"ahler form by $J$, and for a
holomorphic vector bundle on an $n$-dimensional K\"ahler manifold
$X$, define the quantity $\mu({\cal E})$ (known as the ``slope''
of ${\cal E}$) by
\eqn\mumford{ \mu({\cal E}) =  { \int_X J^{n-1} \wedge c_1( {\cal
E} ) \o {\rm rk}~ {\cal E} }.}
Then, ${\cal E}$ is said to be ``Mumford-Takemoto stable'' (and
the associated gauge field satisfies the DUY partial differential
equation) if for all `suitable' subsheaves $ {\cal F}$,
\eqn\takemo{
\mu( {\cal F} )   <  \mu( {\cal E} ).}
The formulation of stability as an inequality can be very useful
from the algebraic geometry point of view, unlike the formulation
as a partial differential equation. For example, the construction
of bundles on elliptically-fibered Calabi-Yau manifolds described
in \fmw\ are designed to give stable bundles, at least when the
fibers are much smaller than the base.

On non-K\"ahler manifolds, the subject of this paper, there is an
analogous statement. Assume that the metric is Gauduchon, and
define $J$ to be the (no longer closed) $(1,1)$ form associated
to the metric in the usual way, then the gauge field associated to
a holomorphic vector bundle ${\cal E}$ will satisfy the DUY
partial differential equation if and only if ${\cal E}$ is stable
in the sense that $\mu({\cal F}) < \mu({\cal E})$, for `proper'
${\cal F}$ and $\mu$ defined just as above. See \liyau, \tomaone,
\tomatwo\ for more information on relevant mathematics.

Stability clearly depends upon the metric. For heterotic
compactifications on K\"ahler manifolds, this metric dependence
is well-understood \kcsub. Briefly, the K\"ahler cone breaks up
into subcones, with a different moduli space of bundles in each
subcone (related by birational transformations).  As one
approaches a chamber wall, some stable bundles become semistable
and then unstable. If one has compactified on a stable bundle
that is becoming semistable, then in the low-energy effective
theory, one gets a perturbatively enhanced $U(1)$ gauge symmetry,
under which some of the formerly neutral moduli become charged.
D-terms realize the change in moduli space. (Thus, again we see
that stability is realized physically via D-terms in the
low-energy effective theory.) (For bundles on D-branes, this
phenomenon is mirror to a phenomenon involving the behavior of
special Lagrangian submanifolds under changes of complex
structure; see \shamitjohn.) We expect that closely related
phenomena should happen in non-K\"ahler compactifications, though
unfortunately there do not seem to be any relevant mathematical
studies of such phenomena for non-K\"ahler cases.

For most of this paper we shall not consider stability. A more detailed
discussion will be relegated to future publication.

 \subsec{Number of Massless Fermion
Families} Let us turn to a slightly different subject and discuss
some generic phenomenological issues. In the following we will
discuss the counting of the number of massless fermion families in
K\"ahler and non-K\"ahler compactifications.

\vskip.1in

\centerline{$\underline{\rm {\bf Review ~of ~Calabi-Yau
~results}}$}

\vskip.1in

\noindent We begin by reviewing the counting of the number of
generations in a standard large-radius compactification of the
heterotic string on a Calabi-Yau manifold. A detailed discussion
of this counting in the perturbative regime can be found in
\disgreone. To be brief, the massless modes are counted by sheaf
cohomology groups, which are realized as $\bar\del$-cohomology on
bundle-valued differential forms, which are in one-to-one
correspondence with the BRST-invariant massless right-moving
Ramond-sector states.  (More generally, vertex operators always
look like bundle-valued differential forms, except in open
strings \katzs, \ersrev\ where nontrivial boundary conditions can
realize spectral sequences involving bundle-valued differential
forms.) For example, if the gauge bundle ${\cal E}$ has $SU(3)$
structure group, breaking an $E_8$ to an $E_6$, then the number
of ${\bf 27}$'s is counted by $H^1(X, {\cal E} )$ and the number
of ${\bf \overline{27} }$'s is counted by $H^1(X, {\cal E}^{\vee}
)$ where the $\vee$ superscript denotes the dual (an asterisk is
sometimes used instead). So, the net number of generations is
counted by \eqn\numgen{ {\rm dim} H^1(X, {\cal E}^{\vee} ) ~ - ~
{\rm dim }H^1(X, {\cal E} ).} Similar remarks hold for gauge
bundles of other rank; We will omit a detailed discussion.

For the cases that we are interested in, the condition $c_1({\cal
E})=0$ is satisfied. We will assume this condition henceforth
since it is also required for examples with torsion. For stable
bundles ${\cal E}$ on Calabi-Yau threefolds $X$ with $c_1({\cal
E})=0$, it is typically the case that
\eqn\cabycase{ H^0(X, {\cal E} ) ~ = ~ 0 ~ = ~ H^0(X, {\cal
E}^{\vee} ).}
Furthermore, Serre duality on a Calabi-Yau threefold $X$ implies
that
\eqn\serdual{\eqalign{ & H^2(X, {\cal E} )~ \cong ~ H^1(X, {\cal
E}^{\vee} )^\ast \cr & H^3(X, {\cal E} ) ~ \cong ~  H^0(X, {\cal
E}^{\vee} )^\ast,}}
where the asterisk $\ast$ indicates the dual group. This has the
same dimension as the group without asterisk. Since we are only
interested in dimensions, we will ignore the $\ast$'s from now on.
Hence, the number of generations can be counted by \eqn\numgeond{
N_{\rm gen} = \sum_i~ (-)^i {\rm dim }~ H^i(X, {\cal E} ).} Now,
we apply the Hirzebruch-Riemann-Roch theorem \hirze, which says
\eqn\hirsay{ \sum_i~ (-)^i {\rm dim }~H^i(X, {\cal E} )~ = ~
\int_X {\rm Td}(TX) \wedge {\rm ch}({\cal E}).} Simplifying the
right-hand side for a holomorphic vector bundle ${\cal E}$ with
$c_1({\cal E})=0$ on a Calabi-Yau, we compute
\eqn\resofnog{ N_{\rm gen} = \sum_i (-)^i {\rm dim }H^i(X, {\cal
E} ) ~ = ~ {1\o 2} \int_X c_3( {\cal E} ).}
Despite the factor of half in the above result, the number of
generations does turns out to be an integer. Thus, we see that
the number of generations is proportional to the third Chern
class of the gauge bundle.

\vskip.1in

\centerline{$\underline{\rm {\bf Non-Kahler ~Backgrounds }}$}

\vskip.1in

\noindent In a large-radius non-K\"ahler compactification, the
analysis above can be repeated, with a few minor modifications.

One approach is to calculate massless spectra on the world-sheet,
following both \disgreone\ and \rohmW, as we review in
appendix~A. The massless spectrum is counted by a type of
$H$-twisted sheaf cohomology, or more precisely, $(\bar\del +
H)$-twisted cohomology, closely analogous to the pattern laid
down in both \disgreone\ and \rohmW. Applying an index theorem by
Bismut we find that the number of generations is again given by
half the third Chern class, just as in standard large-radius
Calabi-Yau compactifications (and also under the same conditions
as in large-radius Calabi-Yau compactifications). Alternatively,
one can work in target-space supergravity, following \barsV, and
again one recovers the same result for the number of generations.

In order to help illustrate these ideas, we will discuss in
section 5.2 the number of generations in a simple example. This
example consists of a heterotic compactification involving a
$U(1)$ bundle over a non-K\"ahler manifold given by a
$T^2$-bundle on $K3$.

\subsec{A Note on Orbifolds in Flux Backgrounds}

For much of this paper, we will be considering fluxes in orbifold
backgrounds. In the following we would like to review a subtlety
which appears when trying to make sense of degrees of freedom in
such orbifolds.

Degrees of freedom appearing in orbifolds such as orbifold Wilson
lines, discrete torsion, shift phases, ``localized fluxes,'' {\it
etc.}, are all ultimately due to an ambiguity in defining group
actions on fields with gauge invariances. Given any one group
action, one can combine the group action with a set of gauge
transformations to get another group action. This is precisely
the historical description of orbifold Wilson lines, and applies
in other cases as well.

To define an orbifold, one must always make {\it some} choice of
group action. In a background with no fluxes, or `trivial' fluxes,
there are always group actions, and more to the point, there is a
canonical group action, and so it makes sense to speak of
``turning on'' and ``turning off'' various degrees of freedom --
when the degrees of freedom are ``turned off,'' one is using the
canonical trivial group action on the fields.

\noindent However, when one has backgrounds with nontrivial fluxes, there
are two subtleties that can come into play:

\item{1.} Group actions might not exist.

\item{2.} When they do exist, none can be considered a canonical trivial
choice, and so there is no way to ``turn off'' the degrees of
freedom.

\noindent See \dtthree, \dtrev, and \dtshift\ for discussions of these
subtleties in the case of discrete torsion, described as a group
action on the $B$ field.

Let us consider two simple examples in which group actions do not
exist.

First consider an $S^1$ action on the principal $U(1)$ bundle over
$S^2$ corresponding to the Hopf fibration. As everyone who has
taken a quantum mechanics class knows, one must rotate the $S^2$
twice for the total space of the bundle, namely $S^3$, to rotate
once. If we consider an $S^1$ action on the $S^2$ base that
rotates about a pole, then in order to be able to lift the group
action to a group action on the bundle ({\it i.e.} for the lift
of the group to be the same group), since a rotation by 360
degrees on the base is equivalent to the identity, we must
require that a rotation of the total space by 360 degrees also be
the identity. Since that cannot happen here, we see that the
$S^1$ action on the base does not lift to the bundle.

Second, consider a nontrivial $B$ field on $T^3$ classified by $1
\in H^3(T^3,{\bf Z})$.  Consider the group $S^1$ again, acting on
the $T^3$ by rotations of one of the circles. We can find out
whether any group actions exist by looking at the fiber in
equivariant cohomology of the map $H^3_G(T^3,{\bf Z}) \rightarrow
H^3(T^3,{\bf Z})$. In the present case, the group acts freely, so
$H^*_G(T^3,{\bf Z}) = H^*(T^3/G, {\bf Z})$. But $T^3/G = T^2$,
and $T^2$ has no degree~3 cohomology, hence this group action on
the base space does not lift to the $B$ field at all.

Neither this nonexistence issue, nor the non-canonical-property
issue, will arise in any of the orbifolds in this particular
paper. However, in general when considering orbifolds in flux
backgrounds, it is extremely important to take these subtleties
into account.

\subsec{Superpotential and the Torsional Equation}

We now discuss some issues regarding the superpotential and its
relation to the torsional equation. We shall discuss the
iterative procedure developed in \bbdg\ and \bbdp\ to determine
the complex three-form $G$, which enters into our closed (i.e. to
{\it all} orders in $\alpha'$) expression for the superpotential.
We shall also see, that our superpotential allows us to make
predictions for higher order interactions in the heterotic
string, that have not been known in the literature so far, when we
discuss the supergravity description.

Recall that the torsion of the heterotic background is partially
generated by switching on three-form fluxes $d{\cal B}$. In our
earlier studies, we switched on three-form backgrounds that were
constants. This choice of constant fluxes fixes the complex
structure of the manifold to $i$ and also fixes some of the
K\"ahler structure moduli by generating a superpotential and a
D-term. One important K\"ahler structure moduli, that is fixed is
the radial modulus (see \bbdp\ and \bbdg).

We now briefly review some generic results, which apply for
compactifications on more general manifolds that may or may not
be complex, 
than the particular
example discussed earlier in detail in \sav,\beckerD. The
superpotential and D-term for the heterotic theory compactified
on a non-K\"ahler complex manifold with torsion and vanishing
Euler characteristic were first constructed in \bbdg\ and later,
in a more elaborate way in \bbdp\ (see also \luest\ for an
alternative derivation of the superpotential). Contrary to the
usual expectation (and from earlier papers), the superpotential
is complex and can be written explicitly in terms of the real
three-form ${\cal H}$ and the fundamental two-form
$J$ as \eqn\superinhet{ W = \int_{{\cal M}_6} ({\cal H} + i dJ)
\wedge \Omega,} where $J$ may not be 
integrable for generic choice of fluxes, and $\Omega$ is the holomorphic $(3,0)$ form on
the non-K\"ahler manifold ${\cal M}_6$ defined wrt the almost complex structure J. 

The above form of the superpotential is of the generic form $\int
G \wedge \Omega$, which further implies, that the three-form
appearing in the superpotential of the heterotic theory should
actually be a complex one. This three-form captures most of the
essential properties of the heterotic compactification on
non-K\"ahler complex manifolds. Of course this does not imply
that the three-form, that would appear in the torsional equation
should be complex; rather the complex three-form $G$, that is
anomaly free and gauge invariant is more useful to study the
dynamics on ${\cal M}_6$. Observe, that for K\"ahler manifolds
$dJ = 0$ and therefore the superpotential \superinhet\ takes the
usual form given in the literature (see e.g. \bg\ and \becons).

The torsional equation derived first by Hull and Strominger in
\HULL\ and \rstrom\ respectively follows easily from the
superpotential \superinhet\ by imposing the imaginary
self-duality (ISD) condition on the complex three-form $G$ as
$\ast_6 G = i G$, where $\ast_6$ is the Hodge duality operation
on the internal six manifold. The above relation gives us
\bbdp\foot{Our equation differs from \rstrom\ by a sign. In
\rstrom\ the torsion is taken as $-{\cal H}$.} \eqn\tore{ {\cal
H} = - \ast dJ = i(\bar\del - \del)J,} as the torsional equation
for the background. The fact, that the background given in
\sav,\beckerD,\bbdg\ and \bbdp\ satisfies the torsional equation
has been shown explicitly in \beckerD\ without the gauge fields
and in \bbdg\ with the gauge fields. One important point, that
needs to be mentioned now is, that even though the background has
been derived following duality chasing of a consistent type IIB
background (which is one of the constant coupling scenarios of
F-theory compactifications) the naive T-duality calculation
applied to the type IIB superpotential {\it does not} reproduce
the superpotential, that we get in \superinhet. In fact, the
naive T-duality will give us the imaginary part to be proportional
to the spin-connection. As we saw in \bbdp, the complex
three-form, that actually solves the anomaly equation is given by
\eqn\geff{ G = {\cal H} + i \omega_{eff},} where $\omega_{eff}$
is an effective spin-connection. In \bbdp\ it was shown
explicitly (at least for a conformally CY background), that
$\omega_{eff} = dJ$. This fact cannot be derived using simple
duality chasing arguments, because these duality rules hold only
to the lowest order in $\alpha'$.

\vskip5cm

\noindent \centerline{$\underline{\rm {\bf Solution ~to ~the ~
Anomaly ~Equation ~to ~all ~Orders ~in ~ \alpha'}}$}

\vskip.1in

\noindent The goal of this section is to use the iterative
procedure developed in \bbdp\ to determine the value of the
complex three-form $G$. We shall see, that the simple ansatz made
in formula (3.5) of \bbdp\ is not needed in order to find the
solution for $G$. Recall, that the basic idea presented in
\bbdp, is that in the presence of torsion the complex field $G$,
captures more properties of the non-K\"ahler compactifications
than the real three-form ${\cal H}$. However, the real three-form
is still important because it appears in the supersymmetry
relations and the torsional constraints. In fact, as mentioned
below, the complex three-form can be written in terms of the real
one. Therefore, one should interpret the complex three-form as
being constructed out of the real three-form plus other
contributions. However, any arbitrary combination of forms will
not be anomaly free. Therefore, we have to look for a
combination, that is anomaly free and gauge invariant. Some
aspects of this have been discussed in \bbdg\ and \bbdp, where
the anomaly equation was solved to cubic order in ${\alpha}'$.
Here we will show, that the iterative techniques can be extended
to all orders in ${\alpha}'$ without making any approximations.

Incorporating all orders of $\alpha'$ it is easy to show, that
the generic form of the three-form is given by the expression
\eqn\genG{G = (a~{\cal H} + \star_6 A) + i~(dJ + B),} where
${\cal H}$ is the real root of the anomaly equation, $J$ is the
fundamental two-form on the non-K\"ahler space, as defined
earlier and $a$ is a constant. Also, $A$ and $B$ are in general
functions of $\omega$, the torsion-free spin-connection and $f$
is the flux density defined in our previous papers \bbdg\ and
\bbdp. In this form we expect $G$ to be anomaly free and gauge
invariant. Therefore, up to possible gauge invariant terms the
above ansatz for $G$  should solve the anomaly equation.
Considering only the first orders in $\alpha'$ of this equation,
we had previously found a cubic equation, where it was shown
explicitly \bbdp, that $A = B = 0$ is a solution. Therefore $G$
is given by the expression \eqn\gcubic{G = a~{\cal H} + i~dJ.}
Here we are ignoring the warp factors. Now in the presence of $A$
and $B$, which may appear to higher orders in ${\alpha}'$, the
imaginary self dual (ISD) condition on our background will imply
(ignoring possible constants) \eqn\isd{{\cal H} = - \star_6~dJ -
\star_6~(A + B).} There could be two possibilities now: (a) The
torsional constraint derived earlier and mentioned above is {\it
invariant} to all orders in $\alpha'$. In that case we expect $A
= - B$; or (b) The torsional constraint receives corrections. In
that case the corrections would be proportional to $\star_6~(A +
B)$.

In the following discussions we will try to see how much can be
said regarding the above two possibilities using iterative
techniques. We begin with the anomaly equation \eqn\anomaeq{ G +
{\alpha'\o 2} {\rm Tr} \left( \omega \wedge d{\tilde G} - {1 \o
3} \omega \wedge {\tilde G} \wedge {\tilde G} + {\tilde G} \wedge
{\cal R}_{\omega} - {1 \o 2} {\tilde G} \wedge d{\tilde G} + {1
\o 6} {\tilde G} \wedge {\tilde G} \wedge {\tilde G} \right) = f,}
where ${\cal R}_{\omega}$ is defined in \bbdp\ and ${\tilde G}$ is
a one-form created out of three-form $G$ by using the vielbeins.
We will not take any ${\cal O}(\alpha')$ corrections to the above
equation into account and assume that the supersymmetry
transformations receive corrections to higher orders in
$\alpha'$. We now demand, that the equation \anomaeq\ should be
solved by a complex $G$. Let us therefore use the following
ansatz for the three-form \eqn\comthree{G = (H_0 + \alpha' H_1 +
\alpha'^2 H_2 + ...) + {i\o \alpha'^m} (h_0 + \alpha' h_1 +
\alpha'^2 h_2 + ...),} where $H_i$ and $h_i$ are independent of
$\alpha'$ and we have kept an overall factor of $\alpha'^m$. Here
$m$ is a constant to be determined soon which appears as a
prefactor in front of the imaginary part. This will be important
for finding the functions $H$ and $h$. Furthermore, the vielbeins
will also have an order by order expansion as \eqn\vielbein{e =
e_0 + \alpha' e_1 + \alpha'^2 e_2 + \alpha'^3 e_3 + ...,}
implying, that any expression constructed out of these should
also have an order by order expansion in $\alpha'$. This means,
in particular, the spin-connection and $dJ$ should be expressed
in such a series \eqn\spianddj{\eqalign{&\omega = \omega_o +
\alpha' \omega_1 + \alpha'^2 \omega_2 + \alpha'^3 \omega_3 +
...,\cr & J = J_0 + \alpha' J_1 + \alpha'^2 J_2 + \alpha'^3 J_3 +
...}} In the analysis below, for simplicity, we will only
consider terms that are zeroth order in $e$ and $\omega$ i.e
$e_0$ and $\omega_o$. It is  not very difficult to take all
orders into account, but we will not do so here.

\vskip5cm

\noindent (a) $\underline{\rm Finding~h_0}$

\vskip.1in \noindent The equation satisfied by $h_0$ can be easily
determined by equating the imaginary parts of \comthree\ after
substituting into \anomaeq. This gives us the following equation
\eqn\hzero{\eqalign{{i h_0 \o \alpha'^m}  =  {\alpha'\o 2} {\rm
Tr} {\Big (} & - {i \o \alpha'^m}~ \omega_o \wedge d {\tilde h}_0
+ {2 i \o 3 \alpha'^m}~ \omega_0 \wedge {\tilde H}_0 \wedge
{\tilde h}_0 - {i \o \alpha'^m}~{\tilde h}_0 \wedge {\cal
R}_{\omega_o} + {i \o 2\alpha'^m}~ {\tilde H}_0 \wedge d {\tilde
h}_0 + \cr & + {i \o 2\alpha'^m}~ {\tilde h}_0 \wedge d{\tilde
H}_0 + {i \o 6 \alpha'^{3m}}~ {\tilde h}_0 \wedge {\tilde h}_0
\wedge {\tilde h}_0 +{i \o 6 \alpha'^{3m - 3}}~ {\tilde h}_1
\wedge {\tilde h}_1 \wedge {\tilde h}_1 + ... {\Big )}.}} To
determine $m$ we need to identify the powers of $\alpha'$ on both
sides of the expression. It is easy to discard the powers of
$\alpha'$ that go as $m-1$ just by comparing the LHS and RHS of
the above equation. However, we see, that there could be, in
principle, an infinite series of terms suppressed by powers of
$\alpha'$, that grow as $3m-n$, where $n$ is an integer. But this
ambiguity is easily resolved, if we consider the fact that, to
the lowest order in $\alpha'$, we have to reproduce a {\it cubic}
equation, as discussed in \bbdg\ and \bbdp. This fixes the value
of $m$ to $m = {1\o 2}$, as we obtain $m = 3m -1$ from the
previous expression. Once $m$ is determined, we see that the
equation satisfied by $h_0$ is \eqn\solforh{ h_0 - {1\o 12}
{\tilde h}_0 \wedge {\tilde h}_0 \wedge {\tilde h}_0 = 0.} For
the simplest case, that we considered in our previous papers we
took only one component of $h_0$ into account and absorbed the
traces of the holonomy matrices and the constants into the
definition of $t$, the size of the six manifold. Doing this will
tell us, that
$$h_0 = \sqrt {t^3},$$ exactly as predicted in \bbdg\ and \bbdp.
But in general one needs to solve \solforh\ to determine all the
components of $h_0$. Observe, that \solforh\ is an algebraic
equation and therefore the solutions are straightforward to
obtain. In fact, we will show, that the iterative technique, that
we are applying, gives solutions in terms of algebraic equations
only.

\vskip.1in

\noindent (b) $\underline{\rm Finding~H_0}$

\vskip.1in

\noindent To determine $H_0$ we need to equate $\alpha'$
independent terms in the above equation. One has to be careful
though with the imaginary parts, since they come with different
powers of $\alpha'$. Taking this into account, the equation
satisfied by $H_0$ is given by \eqn\HZero{H_0 + {\alpha' \o 2}
{\rm Tr}~ \left( {1\o 3\alpha'}~ \omega_0 \wedge {\tilde h}_0
\wedge {\tilde h}_0 + {1\o 2\alpha'}~ {\tilde h}_0 \wedge
d{\tilde h}_0 - {1\o 2\alpha'}~ {\tilde H}_0 \wedge {\tilde h}_0
\wedge {\tilde h}_0 \right) = f,} where a factor of 3 in the
fourth term appears because the three copies of ${\tilde H}_0
\wedge {\tilde h}_0 \wedge {\tilde h}_0$ contribute equally. The
above expression is basically the polynomial equation, that one we
should solve to get the functional form of $H_0$. Observe, that
all the derivatives act on $h_0$, whose value we have already
determined. We can now use the simplification, that was made in
\bbdg\ and \bbdp\ (see the simple ansatz written in (3.5) of
\bbdp). This will give \eqn\Hsimpl{ H_0 = -{f \o 2} + {1\o 4} {\rm
Tr}~({\tilde h}_0\wedge d{\tilde h}_0 ) + {1\o 6} {\rm Tr}~
(\omega_0 \wedge {\tilde h}_0 \wedge {\tilde h}_0).} The first
term is exactly the expression derived in our earlier papers. The
other terms should be regarded as corrections.

\vskip.1in

\noindent (c) $\underline{\rm Finding~h_1}$

\vskip.1in

\noindent To determine the functional form of $h_1$, we need to
equate terms, that are proportional to $i \sqrt{\alpha'}$. Again,
we need to be careful, because the real and the imaginary parts of
$G$ will mix non-trivially. The equation satisfied by $h_1$ is a
little more involved (as expected) but again is given in terms of
a polynomial equation as \eqn\hone{\eqalign{h_1 & - {1\o 4}
{\tilde h}_0 \wedge {\tilde h}_0 \wedge {\tilde h}_1 = -{1\o 2}
{\rm Tr}~ (\omega_0 \wedge d{\tilde h}_0)
 + {1\o 3}{\rm Tr}~ (\omega_0 \wedge {\tilde H}_0 \wedge {\tilde h}_0)
\cr & - {1\o 2} {\rm Tr}~({\tilde h}_0 \wedge {\cal
R}_{\omega_o}) + {1\o 4} {\rm Tr}~( {\tilde H}_0 \wedge d{\tilde
h}_0 + {\tilde h}_0 \wedge d{\tilde H}_0) - {1\o 4} {\rm Tr}~
{\tilde H}_0 \wedge {\tilde H}_0 \wedge {\tilde h}_0.}} The RHS
of this equation contains expressions, that we have already
determined. The LHS is written in terms of $h_0$ and $h_1$, as an
algebraic equation, whose solution will determine $h_1$
completely. It is again interesting to use the approximation,
that we have been performing above. This gives us
\eqn\honeappr{h_1 = -{3\o 8} {f^2 \o \sqrt{t^3}} + ...,} as we
might have expected. The explicit value appearing above comes
from the cubic term, ${\tilde H}_0 \wedge {\tilde H}_0 \wedge
{\tilde h}_0$, when we use the lowest approximation for $H_0$
derived in \Hsimpl. The dotted terms in \honeappr\ can be easily
estimated from the known values of $H_0$ and $h_0$. We can
similarly proceed to determine all the other terms in the
definition of $G$ \comthree. Although straightforward, the higher
order terms increasingly become more and more involved.

\vskip5cm

\noindent (d) $\underline{\rm Finding~the~real~root}$

\vskip.1in

\noindent Now instead of computing higher order terms of $G$ in
\comthree, let us consider the real root of \comthree. Our aim is
now to see, whether we can write the complex root $G$ as \genG.
For this purpose we take the following ansatz for the real root
${\cal H}$ \eqn\realroot{{\cal H} = {\cal H}_0 + \alpha' {\cal
H}_1 + \alpha'^2 {\cal H}_2 + \alpha'^3 {\cal H}_3 + \dots .} This
real root will be the one, that appears in the torsional equation
and therefore the components ${\cal H}_i$ are to be determined
exactly. We can indeed achieve this from our previous analysis.
Plugging this ansatz for ${\cal H}$ into the anomaly equation
\anomaeq\ we get \eqn\realH{\eqalign{& {\cal H}_0 = f, \cr & {\cal
H}_1 = {1\o 2} {\rm Tr}~ \left( \omega_0 \wedge {\tilde f} \wedge
{\tilde f} +  {\tilde f} \wedge {\cal R}_{\omega_0} + {1\o 2}
{\tilde f} \wedge d{\tilde f} - {1\o 6} {\tilde f} \wedge {\tilde
f} \wedge {\tilde f} \right),}} and further terms of higher
order. In fact, using the simplified approximation we can
immediately see, that the familiar values of ${\cal H}_0$ and
${\cal H}_1$ are reproduced. As usual, the other terms should be
regarded as corrections. Observe also, that using the real root
${\cal H}$ we can easily see, that the complex root does follow
from the ansatz \genG\ as \eqn\ansatG{ G = -{{\cal H} \o 2} +
\star_6 A + i~ {\cal F},} where $A$ can be determined from the
above analysis. We also require the imaginary part of \ansatG\
i.e. ${\cal F}$, to resemble the imaginary part of \genG. This
will determine $B$. From here we see that $A \ne - B$ to this
order in $\alpha'$ and therefore the torsional equation should
receive corrections in $\alpha'$ to this order.

This is an important result, which to this order differs from the
expectation of \rstrom, where only the leading order in $\alpha'$
of the supersymmetry constraints was evaluated and no higher order
corrections in $\alpha'$ were expected to appear in the solution
to these constraints. As opposed to \rstrom\ we are working with
a complex three-form $G$ \eqn\ai{G=d{\cal B}+{\alpha'}
\left[{\Omega}_3({\omega}-{1\over 2}{\tilde G})-{\Omega}_3({\cal A})
\right].} As should by now be clear, we observe that the three-form
itself appears in the Chern-Simons term, something that was not
taken into account in \rstrom\ (because this term is higher order
in $\alpha'$). Thus, as opposed to \rstrom, we have to solve
iteratively for $G$ to obtain the solution for $G$, as we have
done above. This procedure can be performed to an arbitrary order
in $\alpha'$, but gets more cumbersome to higher orders. It would
be interesting to find a closed expression, that solves the
determining equation for G \ai\ to all orders in $\alpha'$. This
closed expression in $\alpha'$ would tell us for sure, whether
the solution from the supersymmetry constraints receives
corrections in $\alpha'$. It would then be interesting to verify if
$\epsilon_{1,2}$ in
\eqn\starega{\ast_6 A + \epsilon_1 ~=~ iA + \epsilon_2 ~=~ -iB,}
are non-zero. Here $\epsilon_{1,2}$ measure the deviation from ISD.
Vanishing $\epsilon_{1,2}$
would imply that {\it both} the potential and the torsional
equation remain uncorrected to all orders in $\alpha'$. Though it
is possible that $\epsilon_2 = 0$ once we have the closed form of $G$, but
$\epsilon_1 = 0$ looks implausible. In fact the analysis done by \BeckerNN\
would imply, after dualising to the heterotic theory, that $\epsilon_1$ is
non-zero and the potential receives correction. The analysis performed here is
of course consistent with the expectation that the potential should be corrected,
but the precise mapping of our result presented here and the $\alpha'$
corrections derived in  \BeckerNN\ needs to be worked out.
This issue is beyond the scope of this
paper and will be addressed in future work. However, before moving ahead, we
would like to discuss one other issue that is related to the potential. The existence
of $\alpha'$ corrections to the potential (from the calculations done above) tells us that
the no-scale structure will be broken. These corrections are precisely from the $t$ dependences
in ${\cal H}$ such that $D_t W$ will be non-zero. Both ${\cal H}$ and $J$ will have $\alpha'$
dependences (as we saw earlier), and therefore the potential computed from $\vert G \vert^2$ will
incorporate these.

\vskip.1in

\noindent \centerline{$\underline{\rm {\bf Supergravity ~
Description}}$}

\vskip.1in

\noindent The goal of this section is to show, that our
superpotential \superinhet, and the complex $G$ \genG,
 allows us to predict higher order
interactions of the effective action of the heterotic string,
that have not been known in the literature so far. Thus, on a
slightly different note, let us try to analyze the supergravity
description in terms of the complex three-form $G$, that we
derived above. As we mentioned earlier, considering only the
cubic form of the anomaly relation gave us an effective
three-form given by \geff, which is written in terms of an {\it
effective} spin-connection $\omega_{eff}$. Now that we have an
expansion in orders of $\alpha'$ for the vielbein, we can use
this to find the form of $\omega_{eff}$. This has already been
done in \bbdp, where it was shown, that $\omega_{eff} = dJ$, at
least to the cubic order in the anomaly equation. The analysis
above gives the value of $G$, that can be extended to all orders
in $\alpha'$ \genG. From ${\cal N}=1$ supersymmetry in four
dimensions we know that the scalar potential (without taking
D-terms into account) is of the form \eqn\fodpot{ V = -
\int_{{\cal M}_6} G^+ \wedge \ast {\overline {G^+}} \equiv
e^\kappa \left[g^{i\bar j} D_iW~{\overline {D_{\bar j}W}} - 3
\vert W \vert^2 \right],} where we have used $G^+$ to write the
ISD part of $G$ and $\kappa$ denotes the K\"ahler potential for
the metric $g_{i{\bar j}}$ on the moduli space. Notice, that only
the superpotential $W$ is known at this point and it is given by
our formula \superinhet, but the exact form for the
K\"ahler potential needs to be worked out. We could in principle use the above
formula (to the lowest order in $\alpha'$) 
to determine the K\"ahler potential by comparing both
sides of the above equation using our explicit form of the
superpotential. To all orders in $\alpha'$ the issue is involved 
since the RHS of \fodpot\ would change because of higher order corrections \BeckerNN.
This will be left for work in the future.

The above form of the scalar potential was also found recently in
\luest. The complex three-form $G$ derived in the previous section
can be decomposed into ISD and imaginary anti self dual (IASD)
parts as \eqn\gdeco{ G = G^+ + G^-.} However, only the ISD part
contributes to the scalar potential. In fact, in \bbdp\ we used
the ISD part of the three-form $G$ to extract the scalar
potential from the kinetic term of the heterotic lagrangian. More
precisely, we derived the scalar potential for the radial modulus
and ignored all the other moduli. Up to order $\alpha'^2$ the
result can be written as \eqn\potforrad{ V(t) = {t^3 \over
\alpha'} + {C \alpha' f^4 \over t^3} + {D \alpha'^2 f^6 \over
t^6} + {\cal O}(\alpha'^3),} where $t$ is the radial modulus and
$C,D$ are numerical constants. For the example studied in \bbdp\
these constants are $C = -2$ and $D = 7$. We have however kept
$C$ and $D$ arbitrary because of the possible $\alpha'$
corrections, that could appear in the anomaly equation, when we
solve iteratively for $G$, as was done at the beginning of this
section. The radial modulus can be shown to be generically fixed
at a value given by \eqn\radfix{ t_o = n \left[\alpha'\vert f
\vert^2 \right]^{1/3},} where $n$ is a constant. As was also
shown in \bbdp, the potential \potforrad\ has a minimum and the
minimum shifts to larger values, as we take large flux densities.
To the order the analysis was done in \bbdp, the potential at the
minimum was positive $V(t_o) > 0$. This however does not imply
that supersymmetry is broken. The analysis to all orders in
$\alpha'$ should yield $V = 0$, because in our analysis we are
searching for supersymmetric field configurations. This could be
explicitly checked, if a closed expression for the K\"ahler
potential appearing in \fodpot\ would be known. Once this is
known, it should be possible to see, if the rather interesting
possibility of finding solutions to the equations of motion which
break supersymmetry and lead to a vanishing cosmological constant
along the lines of \bbsb\ and \renata\ appears in this context. At
the same time it will certainly be interesting to find solutions
to the equations of motion, which break supersymmetry and lead to
a positive cosmological constant. Work in this direction in the
context of the heterotic and type IIB theory have recently appeared in
\gaugL\ and \kklt\ respectively. Again, we will leave these issues for the time being and
shall address them in future work.

There are two important things, that need to be clarified
regarding \radfix. First, from the torsional equation \tore, we
see that the real three-from ${\cal H}$ should be proportional to
the scaling of the two-form $J$. One the other hand, from the
Bianchi identity we observe that, $d{\cal H}$ should scale as a
constant by scaling the metric \luest. This would imply, that the
radius is fixed to a different value than the one given above in
\radfix. Why are we getting two different answers? The reason is
actually easy to understand. First of all, from the Bianchi
identity, the real three-form scales as a constant only to the
lowest order in $\alpha'$. But to higher orders, which are not
discussed in \luest, the real three-form is given by equation
\realH, i.e \eqn\rehgiveby{{\cal H} = f - {\alpha' f^3 \over t^3}
+ {\cal O}(\alpha'^2),} where we are ignoring some constant
factors. In deriving this we have taken a very simple scaling
behavior of the metric, namely $g_{ij} \to t ~g_{ij}$. This
ansatz is too simple and it does not take into account the fact,
that the metric has a fibration structure, that depends on the
fluxes \sav,\beckerD. In other words, the cross term in the
metric is not taken into account. The simple ansatz however, is
good enough to study the radial stabilization. On the other hand,
from the torsional equation \tore, we see that the RHS depends on
the two-form $J$, which comes from the cross terms in the metric.
Therefore, the scaling behavior of $J$, to the order that we
considered, is $J = 0 + {\cal O}(t)$. This gives us the following
relation \eqn\folrelg{ f - {\alpha' f^3 \over t^3} + {\cal
O}(\alpha'^2) = 0 + {\cal O}(t).} Solving this equation will
reproduce \radfix\ up to some constant factors. In this way we
can reconcile all the approaches done to get the value of the
radial modulus for the non-K\"ahler manifolds considered herein.
To the order that we did the analysis, the radius is proportional
to the flux density. The second issue is related. Since we took
the flux density to be large enough, the radius $t_o$ of the
manifold should be very big and therefore this allows a
supergravity description. This unfortunately fails for the
following reasons:

\item{1.} The finite (but large) radius of the six manifold $t_o$
actually gets fixed at sizes $t_{\rm fiber} = \sqrt {\alpha'}$
and $t_{\rm base} = \sqrt{\vert f \vert}$. Therefore, the base
could be large enough but the fiber is always fixed at a radius of
order $\alpha'$ \bbdp. This further implies, that the curvature
scalar is of order 1 \luest.

\item{2.} In the presence of fluxes the Betti numbers have been
calculated in \GP,\bbdg. These topological numbers differ from
the ones in the absence of fluxes, implying that the supergravity
description may not capture the full physics here.

\noindent However, the absence of a low energy supergravity
description does not prohibit us to study many properties of the
background. In fact, in \bbdg\ and \bbdp\ many important
properties were derived without actually using a supergravity
description. The lagrangian, that we expect in four dimensions
can be derived from the explicit form of $G$, that has to be used
in the expression for the superpotential. Notice the very
important fact, that our superpotential {\it cannot} be derived
from a dimensional reduction of the ten dimensional supergravity
action of the heterotic string, as an infinite number of $\alpha'$
corrections would be needed.

{}From the above discussion one may arrive at the following
interesting observation. It is rather plausible that the ten
dimensional heterotic lagrangian can be rewritten in terms of $G$
and the gauge field $F$, to all orders in $\alpha'$ as
\eqn\hetlag{\eqalign{S & =\int d^4x~\sqrt{g_4} \int_{{\cal M}_6}
d^6x~ e^{-2\phi} \left[ 2 \vert G \vert^2 + {\rm Tr}~ F^2 +
\sum_{m,n,p} a_{mnp} G^m~F^n~R^p \right] + \cr
&~~~~~~~~~~~~~~~~~~+ \int d^4 x ~\sqrt{g_4} e^{-2\phi} \vert \del
\phi \vert^2,}} where $F = d{\cal A} + {\rm Tr}~{\cal A} \wedge
{\cal A}$ and the tensors are contracted in a way to get scalars
(with suitable traces of course). Observe, that the lagrangian is
written in terms of squares of $G$ and $F$ and therefore
expanding, for example, the $G$ term should reproduce a four
dimensional lagrangian, that will have terms, that are BPS-like
squares. For the terms, that arrive from dimensional reduction
this has recently been verified in \luest\ by an explicit
calculation. Let us see how this works in some detail.

The explicit form of $G$ is given by \genG. Using this, the {\it
kinetic term} will imply the following combination of the real
three-form and other terms \eqn\handother{S_1=\int
d^4x~\sqrt{g_4}\int_{{\cal M}_6} ({\cal H} + \ast_6 A) \wedge
\ast_6 ({\cal H} + \ast_6 A),} where we haven't specified the
precise form of $A$. This can be derived from the order by order
expansion in $\alpha'$ discussed earlier. The above expression
does not take the warp factor into account. In the presence of
the warp factor we assume, that some powers of the dilaton should
appear in \handother. For some of the terms it is indeed possible
to predict the factors of the dilaton, that could appear in the
supergravity lagrangian. One precise contribution is given by
\eqn\jterm{S_2= \int d^4x~\sqrt{g_4}\int_{{\cal M}_6} e^{-4\phi}
d(e^{2\phi}J) \wedge \ast d(e^{2\phi}J),} where the above powers
of the dilaton are fixed from the arguments presented in \bbdp\
(see equation (5.23) of that paper). The above two terms (or at
least some parts of them) have recently appeared in \luest\ (see
section 2 of this paper), where they are shown to appear from the
supergravity lagrangian.

{\it However here we have shown something more}. Our analysis does
not rely on the existence of any specific ten dimensional
heterotic effective lagrangian. Our claim is: the complex $G$
flux captures most of the dynamics of the heterotic
compactification with fluxes. This flux is anomaly free and gauge
invariant and allows us to make {\it predictions} for the form of
the ten dimensional heterotic lagrangian to all orders in
$\alpha'$. Therefore the kinetic term appearing in \hetlag\
should give rise to all the terms, that would appear by reducing
the heterotic lagrangian over the compact six manifold ${\cal
M}_6$ to all orders in $\alpha'$. Similarly the gauge kinetic term
should reproduce
\eqn\gauter{S_3 = \int d^4x~\sqrt{g_4}
\int_{{\cal M}_6} \left[{\rm Tr}~F_{ab}^2  +
{\rm Tr}~F_{\bar a \bar b}^2 + {\rm Tr}~
(g^{a \bar b}F_{a \bar b})^2 \right],}
upto possible relative numerical factors. Vanishing of this term of course
implies the DUY type of equation for the background. In fact we have
already shown in \bbdp\ that the D-term $\int F \wedge J \wedge J$
should reproduce this. This is no
surprise as this is similar to the K\"ahler CY case also. The interaction
terms in \hetlag\ can give rise to many possible terms. A sample would be
\eqn\interter{ S_4 = \int d^4x~\sqrt{g_4}\int_{{\cal M}_6}
\left[ J \wedge dH + {\rm Tr}~R_{ab}^2 +
{\rm Tr}~R_{\bar a \bar b}^2 + {\rm Tr}~(g^{a \bar b}R_{a \bar b})^2 +
...\right]}
Some of
the above terms have also appeared in \luest. The curvature $R$ is measured with
respect to the modified connection.  Observed that the way we derived our
result, could in principle predict higher order terms in the supergravity
lagrangian. It will be interesting to verify the existence of these directly
using the reduction to lower dimensions of the ten dimensional theory
{\it a-la} \luest. This looks difficult in practice because
we would need many higher order corrections to the
low energy lagrangian to perform a reduction similar to \luest.

To summarize, our ansatz for the lagrangian \hetlag\ reproduces
the terms in the four dimensional action that can be derived from
a dimensional reduction of the ten dimensional heterotic
lagrangian as done in \luest. The gravitational and the three form
kinetic terms conspire together to give us the $\vert G \vert^2$
term in \hetlag. However, our formula provides much more
information. Namely we can predict additional contributions, for
example the $\int B \wedge \ast B$ term from \genG, and other
possible cross-terms, that cannot be obtained from the dimensional
reduction of the ten dimensional heterotic action, that is known
at this point.

\newsec{New Six-Dimensional Compactifications}
So far we have studied generic properties of complex non-K\"ahler
manifolds. In the following we would like to illustrate these
properties in examples. Some specific examples have already been
constructed in \sav,\beckerD,\bbdg,\bbdp. These are
compactifications to four dimensions and, as discussed in \bbdg,
they have several properties that are of vital phenomenological
interest. However, all the manifolds considered so far have
vanishing Euler characteristics. In the following we would like
to present new examples of backgrounds in which the internal
manifolds have non-vanishing Euler characteristics. In this
section we will present new six-dimensional compactifications
while the four-dimensional case will be left for a later section.
Some aspects of six-dimensional compactifications have already
appeared in \rstrom.

Our goal here is is to exploit the F-theory/heterotic duality to
find the possible three-folds on the F-theory side, that will
give rise to such kind of six-dimensional compactifications on
the heterotic side. We will briefly discuss the role of discrete
fluxes as has appeared before in \sensethi. The compactifications
to six dimensions studied in \rstrom\ are rather useful to verify
some of the properties of the non-K\"ahler manifolds developed
above. In fact, the conformally $K3$ example studied in \rstrom\
is a consistent model for a compactification to eight dimensions
i.e, when we Kaluza-Klein reduce our heterotic model over the
fiber torus. Let us elaborate this issue first. This will help us
to switch gears smoothly towards new examples.

\subsec{Dimensional Reduction to Eight Dimensions}

To study the dimensional reduction of the heterotic string to
eight dimensions, we will use the fibration structure of the
non-K\"ahler manifold worked out earlier in \sav,\beckerD\ with a
choice of constant fluxes and perform a Kaluza-Klein reduction
over the fiber. The fiber about which we will reduce is given in
terms of the base coordinates $z^1, z^2$ as \eqn\fiber{ dz + 2
i~{\bar z}^2 dz^1 - (4 + 2i) ~ {\bar z}^1 dz^2,} where $z$ is the
complex coordinate of the fiber. The existence of cross terms
above guarantees the existence of some $U(1)$ gauge fields on the
base $T^4/{\cal I}_4$. These $U(1)$ fields appear in addition to
the non-abelian gauge fields already existing in the theory. After
reduction we obtain an eight dimensional theory, that is compactified
on $K3$. In fact, from the choice of metric, that we had in
\sav,\beckerD\ it is clear that the four-dimensional manifold is
conformal to $K3$, with the conformal factor given by the warp
factor $\Delta$. Before dimensional reduction the two-form $J$ of
the six-dimensional manifold is given by \eqn\twenone{ J =
e^{2\phi}\left[\sum_{i=1}^4 \alpha_i {\cal T}_i + \sum_{j =
1}^{16} \beta_j {\cal B}_j \right] + (dz + \gamma) \wedge (d{\bar
z} + \bar\delta),} where $\gamma$ and $\delta$ are some generic
functions of the coordinates of the base $K3$. The coefficient
$e^{2\phi}$ is related to the warp factor $\Delta$. The four
$(1,1)$ forms ${\cal T}_i$ are the forms, that survive globally on
the orbifold $T^4/{\cal I}_4$. The rest of the sixteen $(1,1)$
forms ${\cal B}_j$ are the ones at the sixteen fixed-points of
the orbifold. Therefore, in the orbifold limit the two form $J$
becomes \eqn\jnowis{ J = \sum_{i,j =1}^2 e^{2\phi}
\alpha_{ij}~dz^i \wedge d{\bar z}^j +(dz + a ~{\bar z}^2 dz^1 - b
~ {\bar z}^1 dz^2) \wedge  (d{\bar z} + {\bar a} ~{z}^2 d{\bar
z}^1 - {\bar b} ~ {z}^1 d{\bar z}^2),} where $\alpha_{ij}$ are
constants and $a,b$ can be determined from \fiber.
 Once the two-form $J$ is determined by \twenone, we can obtain
 the three-form flux
 by using the torsional equation ${\cal H} = i (\bar\del - \del) J$
 and it is given by
\eqn\bachfiel{{\cal H} = (\bar\del - \del)\gamma \wedge d{\bar z}
- (\bar\del - \del){\bar\delta} \wedge dz + (\bar\del - \del)
\gamma \wedge
 \bar\delta +
(\bar\del - \del) \phi \wedge {\cal N}_{1,1}.} This is a generic
combination of (2,1) $\oplus$ (1,2) forms and ${\cal N} _{1,1}$
are the constant (1,1) forms on $T^4$. There is another
combination given by $(\bar\del + \del)J$, which determines the
anomaly condition for the given heterotic background. This generic
background on $K3$ should satisfy yet another consistency
condition coming from the fact, that the holomorphic $(2,0)$ forms
on $K3$ are constants. In terms of the notations used, this
 condition is
\eqn\cononk{ {\bar\del}^\dagger J -i~\bar\del ~\phi = -
 ({\del}^\dagger J + i ~\del ~\phi).}
This is an important condition, the derivation of which was given
in \rstrom, in the sense that it will tell us the background, that
we derived for the case with constant fluxes can indeed be
extended to {\it any} arbitrary choice of fluxes. The fact, that
only the base is conformally Calabi-Yau is inherent in the
relation \cononk. To see this assume the complex dimension of the
space to be $n$ and the metric to be conformal to the CY metric
with the conformal factor given by $\Delta^2 = e^{2\phi}$. In
terms of torsion classes, ${\cal W}_i$, this would mean \gauntlett\  (see also
sec. 4.1 of \rstrom) \eqn\implywhat{ (2n-2){\cal W}_5 - (-1)^n
2^{n-2} n {\cal W}_4 = (-1)^n 2^{n-2} (n-2) {\cal W}_4,} whose
only solution is $n = 2$, when ${\cal W}_5 = {\cal W}_4 = 2
d\phi$. This implies, that the base has to be conformally
Calabi-Yau i.e $K3$, as was pointed out in \rstrom. This tells
us, that the metric that we derived earlier in \sav,\beckerD\ can
be thought of as an exact solution (if we also replace the simple
fibration structure \fiber\ with a more generic one \twenone).
This is consistent with the fact, that these solutions are derived
directly from F-theory at constant coupling.

At this point one might wonder, whether the conformal $K3$
solution can be obtained directly as a torsional $K3$
compactification of the ten-dimensional theory, i.e without going
through the dimensional reduction over the fiber. One method to
achieve this would be, if we take the fiber to be non-compact.
Imagine this is a plausible scenario (we will soon show, that for
the kind of configuration we have, this is not possible). A
necessary condition for this interpretation to work is, that the
warp factor depends on the coordinates of the $K3$ base only.
This works perfectly for our case,
 because of the inherent T-dualities we were
constrained to keep the warp factor independent of the fiber
directions. The above encouraging steps might lead us to think
erroneously, that the duality chasing arguments can be helpful to
obtain a compactification on the torsional $K3$ directly.
However, this is misleading as it is {\it not} possible in this
scenario to decompactify the fiber direction. A decompactified
fiber on the heterotic side would imply, that the $T^2/Z_2$
manifold on the type IIB side should be of zero size. This is, in
principle, no problem because even if we turn on fluxes, the
no-scale structure of the type IIB supergravity theory does not
guarantee any radial stabilization. However, the reason why this
wouldn't work is because the existence of F-theory monodromies.
These monodromies give rise to extra charges on the type IIB side
and therefore lead to a finite (and non-zero) torus. This further
implies, that the fiber on the heterotic side should have a
finite size.

We might be able to overcome this by taking a non-compact torus
directly in the type IIB theory. One crucial step would be to see,
whether this is possible in M-theory. The manifold, that we
require in M-theory will be $K3 \times N$, where $N$ is a
non-compact manifold, that is a $T^2$ fibration over a
non-compact base. We also require the torus fibration to
degenerate at only one point of the base. This construction looks
plausible enough, as it allows a non-zero $G$-flux. Furthermore,
the $\int X_8$ term does not have to vanish, as it is no longer
identified with any topological number. But the above
construction fails due to two subtle reasons.

\item{1.} The manifold in F-theory will
also be $K3 \times N$ and now we will have non-trivial
monodromies around the point where the F-theory torus
degenerates. Since the fiber torus degenerates at only one point
of the base, the monodromy around it will tell us, that in the
type IIB theory there will be a self-dual four form source whose
charge is given precisely by \rDJM \eqn\fourform{ {1\over 24}~p_1
\wedge \delta(z).} Here $p_1$ is the first Pontryagin class and
$z$ is the complex coordinate of the non-compact base. Going to
the heterotic side will imply, that the equation satisfied by the
heterotic three-form is \eqn\hetano{ {\alpha' \over 48}{\rm
tr}~\left[\Omega_3\left(\omega - {1\over 2}{\cal H}\right)\right]
- {\cal H} = 0,} where $\omega$ is the torsion free spin
connection and $\Omega_3$ is the torsional Chern-Simons form.
This equation is anomalous and therefore needs additional
contributions. The additional contributions are precisely the
ones which gave us the {\it six}-dimensional non-K\"ahler
manifold. Therefore, from the duality chasing it is not possible
to get only conformal $K3$ as our solution.

\item{2.} Even though the fluxes seem to fix the {\it total} volume, the
individual sizes of the base and the fiber also do get fixed in
the process due to the F-theory monodromies \bbdp. This would
imply that the only possible size of the fiber would be
proportional to $\alpha'$. Therefore it is not possible to take
the zero size or the decompactifying limits of the non-K\"ahler
manifolds derived in \sav,\beckerD.

\noindent A slightly different setup would be to add {\it
discrete} fluxes to a four-dimensional (or six-dimensional)
compactification \sensethi. But before we  address the issue of
discrete fluxes, let us discuss the heterotic compactification on
$K3$ with generic fluxes.

\subsec{Compactifications of the Heterotic String on a Background
Conformal to $K3$}

In this section we will study the $SO(32)$ (or the $E_8 \times
E_8$) heterotic compactifications with fluxes and describe the
dual F-theory geometries associated with it. In the absence of
fluxes the heterotic (or type I) theory has an F-theory dual given
by a non-trivial torus fibration over an $F_n$ base \morvaf. Here
$n$ could be $0,1$ or $4$. These three choices of $n$ are related
to the three different choices of vector bundles on the type I
side. In order to achieve our goal of finding the F-theory dual
for heterotic compactifications with fluxes, we need to go to the
point, where the F-theory manifold can have an {\it orientifold}
description. This is crucial for various reasons. The orientifold
description provides a good description of the theory even for
the regions, where the F-theory description is not sufficient. As
long as we keep the coupling constant small we can trust the
orientifold description \sensethi. Furthermore, the orientifold
description provides us a way to go smoothly from one theory to
another using U-dualities, as we saw earlier in \sav,\beckerD.
Using T-dualities we can go from the orientifold description in
the type IIB theory (valid at weak coupling) to the type I
description (which can also be made at weak coupling). Whenever
we have an orientifold description, there will be involved a
generic operation of the form: $(-1)^{F_L} \cdot \Omega \cdot
\sigma,$ where $\sigma$ is the inversion operation (it will soon
be clear for our case, $\sigma$ is basically the Nikulin
involution \nikulin).

Let us begin with the F-theory dual for the compactification of
the heterotic on $K3$. We will take the simplest example, when
the F-theory dual is given by $F_0$. The Weierstrass equation
governing the background has the generic form \eqn\weiequa{y^2 =
x^3 + f(u,v)~x + g(u,v),} where $u$ and $v$ are the coordinates
of the two $P^1$'s. The manifold \weiequa\ is an elliptically
fibered Calabi-Yau manifold, if $f(u,v)$ is a polynomial of
degree (8,8) and $g(u,v)$ is a polynomial of degree (12,12).
Observe, that the above equation, taking into account the right
powers of $f$ and $g$, has 243 complex deformations \morvaf.
There is also an obvious exchange symmetry of the two $P^1$'s.
This exchange symmetry is related to the heterotic/heterotic
duality in six dimensions \duff. We will soon elaborate some more
on this, when we discuss the conformal $K3$ example. Let us first
define the quantities $f$ and $g$, which will take us to the
orientifold point. In fact, this has already been done in some
detail in \morvaf\ and \senF. We define
\eqn\fandg{\eqalign{f(u,v) =~ & {\cal A}_1 ~\prod_{i=1}^8
(u-u_i)(v-v_i) + {\cal A}_2 ~\prod_{j=1}^4 (u-{\hat
u}_j)^2(v-{\hat v}_j)^2\cr g(u,v) = ~ & {\cal A}_3~
\prod_{{}^{i=1}_{j=1}}^{{}^{i=8}_{j=4}} (u-u_i)(u-{\hat
u}_j)(v-v_i)(v-{\hat v}_j)  + {\cal A}_4~\prod_{i=1}^4 (u-{\hat
u}_i)^3 (v-{\hat v}_i)^3 +  \cr
 + ~ & {\cal A}_5~\prod_{k=1}^{12} (u- {\tilde u}_k)(v - {\tilde
v}_k).}} The above choice of $f$ and $g$ is the  most generic
one, if we keep all the constants ${\cal A}_i$'s to be arbitrary.
The next step is to evaluate the discriminant of the Weirstrass
equation. The vanishing of this discriminant spans the following
curve \eqn\discurve{ \sum_{m,n,p} {\cal C}_{mnp} \prod_{i,j,k}
(u-{\hat u}_i)^m (u - u_j)^n (u - {\tilde u}_k)^p (v-{\hat
v}_i)^m (v - v_j)^n (v - {\tilde v}_k)^p = 0,} where ${\cal
C}_{mnp}$ are constants. This is the most generic curve for the
system and is valid at all points in the F-theory moduli space.
The analysis of the above equation is rather complicated, because
at any generic point the solutions of the above equation do not,
in general, relate to perturbative D-branes. The non-trivial
monodromies about any of the solutions of the above equation will
tell us, what kind of objects we are dealing with. In the above
equation we have also summed over powers of $m,n$ and $p$. The
choices of $m,n,p$ are not arbitrary as they have to obey the
following equation \eqn\mnpeq{ m + 2n + 3p = 6,} the integer
solutions of which we should be summing over in \discurve. The
constants ${\cal C}_{mnp}$ are determined in terms of the ${\cal
A}_i$ as expected and their values are correlated to the powers
of $m,n,p$ in an appropriate way. In fact, we will use some
specific values of ${\cal C}_{mnp}$ to go to the orientifold
limit. Following the work of \senF\ one can easily show, that the
above equation \discurve\ gives the orientifold description only,
when \eqn\cvalue{ {\cal C}_{410}~ = ~{\cal C}_{111} ~= ~{\cal
C}_{002}~ = ~{\cal C}_{600}~ =~ {\cal C}_{030} ~ = ~ 0,} while
${\cal C}_{220}$ and ${\cal C}_{301}$ are the only non-vanishing
constants. It is, however, not necessary, that both of the ${\cal
C}$ should be non-zero. As discussed in \senF, we can have a
consistent orientifold background, if the only non-zero ${\cal
C}$ is ${\cal C}_{220}$. The key point, which takes the
background \discurve\ to an orientifold background appears, when
we allow a maximum of two non-zero ${\cal C}$'s, the generic
F-theory equation breaks into {\it two} different hypersurfaces.
The monodromies around each of these surfaces will tell us,
whether we have orientifold planes or $D7$ branes. In fact, it is
enough to extract the hypersurface corresponding to the $D7$
branes only, as the other curve would factorize out of \discurve.
{}From \discurve\ the hypersurface corresponding to the $D7$ branes
is of the generic form \eqn\hypergen{ {\cal C}_{220}~{\cal F} +
{\cal C}_{301} ~{\cal G} ~ = ~ 0,} where ${\cal F},{\cal G}$ are
polynomials of order 16 in $u,v$, the precise form of which can
be easily extracted from \discurve. For the generic hypersurface
\discurve\ the story is well known \vafasen. The singular points
that do not have any interpretation in terms of $D7$ branes or
$O7$ planes are in fact ($p,q$) seven branes. Their combined
effect will be like a {\it non-dynamical} orientifold plane.
The hypersurface
\discurve\ at its orientifold limit \hypergen\ will therefore
imply that our manifold in type IIB will be $K3/ (-1)^{F_L}\cdot
\Omega \cdot \sigma$ where $\sigma$ describes a Nikulin
involution \nikulin\ of the form $(r,a,\delta) = (2,2,0)$ (see
\nikulin\ for the notations). Making two T-dualities (or
equivalently, a mirror transformation) will take us to type I
theory on mirror $K3$ which is S-dual to heterotic on the mirror
$K3$. Thus following this chain of dualities we eventually reach
the heterotic compactifications for the corresponding F-theory
model.

The above discussions are valid for the case of vanishing fluxes.
Before including fluxes we still need to verify that the
supergravity description is a valid approximation on the type IIB
and the heterotic sides. In $d$ dimensions the supergravity
approximation will be valid if the $d$-dimensional string coupling
is weak and the internal volumes are large.

The volume of the base $P^1 \times P^1$ in F-theory is $v_1 v_2$,
where $v_i$ are the volumes of each of the $P^1$'s. On the type I
side the coupling constants are 
\eqn\typeonecase{ g_I = {\alpha' g_B \over v_1}, \qquad  v_1^I =
{\alpha'^2 \over v_1}, \qquad  v_2^I = v_2\qquad {\rm and } \qquad
g_I^{(6)} = {g_B \over \sqrt{v_1 v_2}},} 
where $g_B$ denotes the ten-dimensional type IIB string coupling
constant. To get to the type I side we have made two T-dualites
along the two directions of one $P^1$ (which is of the form
$T^2/Z_2$ and $Z_2$ involves orientifold operation). On the
heterotic side we have
\eqn\hetcasesi{g_{het} = {v_1 \over \alpha' g_B}, \qquad v_1^{het}
= {\alpha' \over g_B}, \qquad  v_2^{het} = {v_1 v_2 \over \alpha'
g_B}\qquad {\rm and} \qquad  g_{het}^{(6)} =
 {1\over
\alpha'} \sqrt{v_1 \over v_2}.}
{}From here it becomes clear that the six-dimensional heterotic
coupling constant is given by the ratio of the volumes of the two
$P^1$'s. This is of course consistent with the literature
\morvaf. Furthermore, by interchanging $v_1$ and $v_2$ one
obtains another theory which was already predicted in \duff. In
order to analyze if the supergravity approximation is valid we
assume
\eqn\assume{ v_1 \to \epsilon^{-\alpha}, \qquad v_2 \to
\epsilon^{-\beta}\qquad{\rm with } \qquad  \epsilon \to 0, \qquad
\a,\b \in \IN}
The above equation implies large internal volumes. Next we will
insert the scaling behaviors \assume\ into \typeonecase\ and
\hetcasesi\ to determine how the different coupling constants
scale with $\e$. In order to do this we keep $g_B$ constant. The
various scalings are
\eqn\scali{v_I \to \epsilon^{\alpha - \beta},\qquad  g_I^{(6)} \to
\epsilon^{(\alpha + \beta)/2 }, \qquad  v_{het} \to
\epsilon^{-\alpha - \beta} \qquad{\rm and} \qquad g_{het}^{(6)}
\to  \epsilon^{(\beta - \alpha)/2 },}
where $v_I$ and $v_{het}$ denote the total six-dimensional
volumes. If we choose $\beta > \alpha$ we can have large
six-dimensional volumes and weak couplings on the type I and
heterotic sides. From this point of view it seems as if the
supergravity description can be applied on the two sides of the
duality. In order to support this claim we will check the
validity of the supergravity approximation explicitly later.

Our next question is, what happens in the presence of fluxes in
the heterotic side? An ansatz for this has already been given in
\rstrom. This is the conformal $K3$ example. From the above
analysis, done in the absence of fluxes, we can now try to infer
the corresponding F-theory duals when we have fluxes switched on
in the heterotic side. The ten dimensional metric ansatz for our
case will be \eqn\metans{ds^2 = ds^2_{012345} +
\Delta^m~ds^2_{K3}, ~~~~~~~~e^\phi = \Delta,}where $\phi$ is the
heterotic dilaton and $\Delta$ is the warp factor. We have kept
the conformal factor as an integer power $m$ of the warp factor.
We will argue in the next section that $m = 2$ is realised in this
set-up. The above ansatze slightly generalises the one given in
\rstrom. In this section we will {\it not} argue the reason for
this ansatze. There could be a possibility of a different
solution in the heterotic theory, but as far as the metric
\metans\ goes, this solves the background torsional equation as
shown by \rstrom\ (see sec. 4.1 of \rstrom). We will start from
this situation. We will also take the orbifold limit of $K3$ and
therefore replace the metric by an almost flat metric (with
sixteen fixed points).

At the orbifold point we would expect the $J$-form to be given
only by the ${\cal T}_i$ part of \twenone. Therefore comparing
our result to
 \bachfiel\ this will immediately tell us that the background three-form
flux on $K3$ is given by: \eqn\torflisg{{\cal H} = {\cal N}_{1,1}
\wedge \del \Delta + {\cal M}_{1,1}\wedge  \bar\del \Delta,}
where ${\cal N}_{1,1}$ and ${\cal M}_{1,1}$ are proportional to
$(1,1)$ forms on $T^4$. They are  functions of the warp factor
and are related by the torsional constraints. This is the
background responsible for the torsion in our space. Observe that
the three-form can exist on $K3$ even though there are no three
cycles. This kind of form is similar to the four-form in \rBB\
which is proportional to the warp factor there. Combining
\torflisg\ with \metans\ will more or less give us the complete
background for the torsional $K3$ case. What remains are the
solution of the warp factor $\Delta$ and the choice of vector
bundle for this background. These two are intimately related,
because the equation for the warp factor is determined in terms
of the vector bundle \rstrom. We will discuss this issue soon and
for the time being we assume that both the vector bundle and the
warp factor are determinable for this system. In terms of
components, \torflisg\ can be written as \eqn\torcom{{\cal
H}_{\bar a a b} = \del_b g_{a \bar a}, ~~~~~~ {\cal H}_{a \bar a
\bar b} = \del_{\bar b} g_{a \bar a},} where we have chosen our
complex structure as $i$ and $a,b$ are complex coordinates. For
the choice of flat metric (in the orbifold limit of $K3$) we have
components ${\cal H}_{1 \bar 1 2}$ and ${\cal H}_{2 \bar 2 1}$
and their complex conjugates. From the relation \torcom, these
two components are proportional to $\del_2 \Delta^m$ and $\del_1
\Delta^m$ respectively. However in the kind of models that we are
studying, we have to make sure that the warp factor is
independent of one of the set of complex coordinates\foot{Because
these will be the two duality directions.} $-$ say for example
$\del_1 \Delta^m = \del_{\bar 1} \Delta^m = 0$ $-$ and so we will
only consider the background three forms that are of the form
$\del_2 \Delta^m, \del_{\bar 2} \Delta^m$. In other words, we need
the three-forms as ${\cal H}_{1 \bar 1 2}$ and ${\cal H}_{1 \bar
1 \bar 2}$. Under duality chasing this implies we have two RR
one-forms in type IIB. On the other hand, choosing the other two
three-forms, namely ${\cal H}_{2 \bar 2 1}$ and ${\cal H}_{2 \bar
2 \bar 1}$, we will get two three-forms in type IIB by duality
chasing that would survive the orientifold projections. This
would seem problematic because the surviving components are not
the ones that we should be interested in. This issue is a little
subtle, and so let us elaborate it in some details. To start
with, we didn't take the orientifolding action carefully in the
above analysis. In fact the type IIB picture is actually
compactified on $T^4$ with the orientifold action given by $(1,g)
\times (1,h)$ where $g,h$ are \eqn\ganahare{g = (-1)^{F_L}\cdot
\Omega \cdot \sigma_1, ~~~~~~h = (-1)^{F_L}\cdot \Omega \cdot
\sigma_2, ~~~~~~ gh = \sigma_1 \sigma_2~ (-1)^{F_L + F_R},} where
$\sigma_i$ are the orbifold action of the two $T^2$ respectively.
This is the Gimon-Polchinski orientifold action \gimpol, whose
F-theory limit is discussed in \sengimon. The action \ganahare,
actually involve orientifold as well as pure orbifold action and
therefore the bulk states of the type IIB supergravity should
survive all the actions. Various fixed points of the orientifold
actions $g$ and $h$ are shown in the figure below:

\vskip.2in

\centerline{\epsfbox{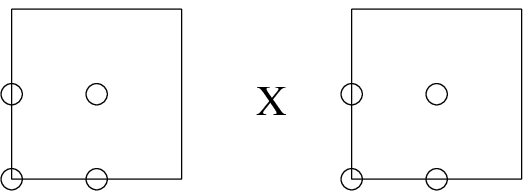}}\nobreak \centerline{{\bf Fig.
1}: {\it The location of orientifold planes on $P^1 \times P^1$
at the points $0,{1\o 2},{\tau_i\o 2},{\tau_i+1\o 2}$.}}

\vskip.2in

\noindent It is easy to work out the multiplets in six dimensions.
There are in fact various contributions coming from the bulk and
the brane states. At the orientifold point there would be set of
$O7$ planes and $D7$ branes that are placed orthogonal to the
another set of equivalent branes and planes. The total multiplets
therefore will come from the untwisted sectors, twisted sectors,
DN open strings and the brane-orientifold states. The multiplets
are  arranged in terms of the susy spectra for $D = 6, {\cal N} =
1$ multiplets. It is given by: \eqn\mulsix{ (g_{\mu\nu},
B^+_{\mu\nu}, \psi_\mu) ~\oplus~ (B^-_{\mu\nu}, \chi^R, \phi)
~\oplus~ (A^a_\mu, \psi^{aL}) ~\oplus~ N~(4\phi, \psi^R)} where
$L,R$ denote the various chiralities of six dimensional fermions
and $\pm$ denote the self-dual and the anti self-dual parts of
the two form fields. In \mulsix, the first one is the gravity
multiplet, the second one is the tensor multiplet\foot{In this
model it is only possible to get one tensor multiplet and
therefore the number of hypermultiplets $H$ is $H = V + 244$
where $V$ is the number of vector multiplets. To get more than
one tensor multiplets (say for example nine tensor multiplets) we
need to compactify type IIB theory on $T^4$ modded out by the
group $G =$ ($1, gh, \Omega \cdot S, \Omega\cdot gh \cdot S$)
where $S$ involve a half-shift of the $T^4$ coordinates $z^{1,2}$
\dabholkarP.}, the third one is the vector multiplet in the
adjoint representation of the enhanced gauge group $G$ at various
points of the moduli space and the last ones are $N$ copies of
hypermultiplets in various representations. To evaluate $N$ for
our case we need to know what is the enhanced gauge symmetry
allowed in our setup. As discussed in the literature, the
Gimon-Polchinski model has many branches of enhanced symmetry.
These branches are basically of the form $G = U(m)^n \times
U(m')^{n'}$, with $m, m', n, n'$ integers. In terms of $m, m', n$
and $n'$, $N$ in \mulsix\ is given by: \eqn\Nisgive{N = 4({\bf 1,
1}) \oplus n n'({\bf m,m'}) \oplus 2n \left({\bf {{m(m-1)\over
2},1}} \right) \oplus 2n' \left({\bf {1, {m'(m'-1)\over 2}}}
\right) \oplus 16 ({\bf 1,1}),} where the first is from the
untwisted sector, the second is from the DN open strings between
the branes, the third and the fourth are from the antisymmetric
representations of the gauge group $G$ and the last ones are from
the 16 fixed points of $gh$ in \ganahare. Taking only the bulk
states, the number of scalars that we get is seventeen. Eight of
which come from the $B$-fields and the rest come from the metric,
the axion-dilaton and the four form fields. Thus the $B_{RR}$
contribute four scalars. These scalars (along with the axion) in
type IIB theory should be the ones that determine the torsional
background in the heterotic side. As an example, in the heterotic
theory we can have a ${\cal B}$-field of the form ${\cal B}_{12}$
with the corresponding three-form $${\cal H}_{ \bar 2 1 2} =
\del_{\bar 2} {\cal B}_{12} = g_{\bar 2 [1,2]} \sim \del_1
\Delta^m.$$ \noindent  This would be a background that survives
all the orientifold and orbifold projections. However we now have
to assume that the warp factor depends on all the coordinates.
This would naively mean that the usual Buscher duality wouldn't
work here, as the rules require us to make all the fields
independent of the duality directions. We are making two
T-dualities along one of the $P^1$ directions that convert
\eqn\conever{ (1,g) \times (1,h)
~~~{}^{T_{P^1}}_{\longrightarrow} ~~~ (1,\Omega) \times (1,gh),}
so naively we have to keep all fields independent of the $P^1$
coordinates. However we can assume that the dependence is very
mild and therefore three-form background is very small, implying
that the corrections to the Buscher rules would be small.  This is
possible because the three-form is {\it not} in cohomology and
therefore is not quantised. At every point of the internal space
we would now require the magnitude of the warp factor to be
small. If we put $m =2$ in \metans, then the equation satisfied
by the warp factor will be \rstrom \eqn\eqsatwa{{\quabla}
\Delta^2 = {\rm tr}~R_{ab}R^{ab} - {1\over 30} {\rm
Tr}~F_{ab}F^{ab}} where ${\quabla}$ and the curvature are
measured wrt the unwarped metric. The above equation has
solutions if the usual integrability condition is satisfied. We
have to make sure that the warp factor have no $\pm r^{-n}$ type
of singularities (where $r$ is the radial coordinate in this
space) and $\vert\vert \Delta^2 \vert\vert$ to be small
everywhere. This in particular implies that the orbifold limit
that we are considering should be completely {\it non-singular}.

Until now the discussion was motivated by parallel examples of
non-K\"ahler complex manifolds constructed in \sav,\beckerD,\bbdg.
These examples were obtained from M-theory backgrounds in the
presence of primitive (2,2)-forms, after performing some string
dualities. The existence of fluxes was guaranteed by topological
considerations and the structure of the heterotic three-form was
determined by the primitive forms. Here we would like to start
with an heterotic background and reverse the string dualities to
get to M-theory. In the following, we will consider the most
general ${\cal H}$-fluxes allowed on the heterotic side. First,
we will determine the conditions necessary for the existence of
fluxes in a $K3$ compactification of the heterotic string. In
order to do this we need some terms in the low-energy effective
action of the heterotic string (we will use the notations of
\tseytlin)
\eqn\treelaghet{S = -{1\over 8} \int d^{10}x \sqrt{g} e^{-2\phi}
\Bigg[ R + 4 (\del \phi)^2 -{1\o 12}{\cal H}^2 + {1\o 8}( {\rm
tr} F^2 - {\rm tr} R^2 ) +e^{2\phi}J_1+\dots \Bigg].}
Here $J_1$ is a polynomial of fourth order in the Riemann tensor
and gauge fields. The dots contain other higher order terms whose
explicit form can be found in \tseytlin. This ten-dimensional
lagrangian can be reduced to four dimensions to obtain the
previously discussed lagrangian \hetlag. What we would like to
obtain is the six-dimensional low-energy effective action in the
presence of ${\cal H}$-fluxes. The existence of these fluxes is
constrained by the integrability condition
\eqn\integcond{ \int_{K3} e^{-2\phi} {\cal H}_{\mu\nu\rho} {\cal
H}^{\mu\nu\rho} + {3 \alpha'\o 2} \int_{K3} e^{-2\phi} \left(
{\rm tr} ~F_{\mu\nu}F^{\mu\nu} - R^{ab}_{\mu\nu} R^{ba\mu\nu}
\right) = 0,}
where we have ignored higher order corrections. {}From the
integrability condition we see that, in general, the ${\cal
H}$-flux on $K3$ will be non-vanishing because of the presence of
curvature related terms and higher order corrections. This avoids
the no-go theorem for compact manifolds\foot{For non-compact
manifolds this is no longer the issue. From the discussion above
in \integcond, putting fluxes on compact manifolds is somewhat
subtle. To avoid further complications, we will consider
specifically {\it non-compact} manifolds unless mentioned
otherwise. These manifolds (and their duals) are in the same
universality class as the non-compact Gimon-Polchinski
orientifolds. We will therefore continue to call them as
Gimon-Polchinski ($GP$) models.}. Equation \integcond\ can also
be interpreted as the equation of motion of the dilaton. The
presence of ${\cal H}$-flux and other fields will imply a
non-constant dilaton. This is, of course, consistent since the
dilaton is proportional to the warp factor.

The considerations discussed above gave us a way to see (in the
heterotic theory) the existence of non-zero three-forms on $K3$
(with components ${\cal H}_{\mu\nu\rho}$, where $\mu,\nu, \rho$,
in general, span all the four coordinates of $K3$) and non-zero
dilaton $\phi$. In the type I theory, this three-form will be the
RR three-form of type I compactified on $K3$ with complex
coordinates ${\cal H}_{1\bar 1 2}, {\cal H}_{1 \bar 1 \bar 2},
{\cal H}_{1 2 \bar 2}, {\cal H}_{\bar 1 2 \bar2}$. For simplicity
we will consider real coordinates and assume that the coordinates
of $K3$ are $x^{6,7,8,9}$ with $z^1 = x^6 + i x^7$ and $z^2 = x^8
+ i x^9$. The next question would be to ask for the corresponding
fields in the type IIB side by compactifyting two generic
directions x and y and making T-dualities along these directions.
If our type I theory has fields ${\cal H}_{{\rm x} \mu\nu}$ and
${\cal H}_{{\rm x y}\nu}$, then after two T-dualities we will
have a RR three-form and an axion background that are related to
the type I three-forms in the following way:
\eqn\tduiib{H^{RR}_{{\rm y}\mu\nu} = {\cal H}_{{\rm x}\mu\nu},
~~~~~~ F^{RR}_\mu = {\cal H}_{{\rm xy} \mu},} where $H^{RR}$ and
$F^{RR}$ are the RR three and one $-$ forms of type IIB theory.
As discussed earlier, type IIB background is actually a
Gimon-Polchinski orientifold whose F-theory limit is well
defined, in other words, the axion-dilaton background is well
defined. In any generic point the RR charge is not neutralised
locally and therefore in the type IIB side we should expect a
non-trivial three-form and an axio-dilaton background. An
important thing to observe here is that by choosing ${\cal H}_{1
2 \bar 2} = {\cal H}_{\bar 1 2 \bar 2} = 0$ directly in the
heterotic theory, we can have a consistent background in type IIB
theory with {\it only} a non-trivial axio-dilaton. This choice of
background overcomes all the earlier problems that we had,
namely, the warp factor can be made independent of $z^1, {\bar
z}^1$ directions by having ${\cal H}_{1 \bar 1 2} \sim \del_2
\Delta^m$.

Let us now see how this works for the explicit case of \metans.
Making an S-duality of the background given in \metans\ will give
us the corresponding type I background. The torsion three-form
will be replaced by the RR three-form of the type I theory. One
can easily show that the corresponding type I background follows
from the following background in type IIB theory (we put $m = 2$
in \metans): \eqn\metiintwob{ds^2 = \Delta^{-1} ds^2_{012345} +
\Delta~ds^2_{P^1} + \Delta^{-1} ds^2_{P^1},} where we have
explicitly shown the metric of the two non-compact $P^1 \equiv
{R^2\o {\cal I}_2}$. Observe that the type IIB manifold has
retained its basic form but the two $P^1$'s are now scaled
differently. The type IIB coupling constant, $g_B$, is no longer
a constant number as in \sav,\beckerD, rather now it has
dependence on the warp factor. The dependence is easy to infer,
and is given by $g_B = \Delta^{-2}$, where we are ignoring
constant factors. This therefore implies that the
Gimon-Polchinski model is at an orientifold limit with coupling
constant $\tau$ given by \eqn\gptau{ \tau \equiv \tilde\phi +
i~g^{-1}_B = B_{\rm xy} + i~\Delta^2,} where $\tilde\phi$ is the
RR scalar (axion). Notice that ${\rm Im}~\tau > 0$ and therefore
this presumably incorporates possible non-perturbative
corrections to the $D7$ branes and $O7$ planes background. This
is because  near an orientifold seven plane the behaviour of
$\tau$ naively would be \eqn\taubeor{\tau = -{2\o i\pi} {\rm ln}
\left(u - {\tilde u}_i \right)} where ${\tilde u}_i$ is the
position of one of the orientifold plane on the $P^1$ with
coordinates $u$ (similar behaviour is expected for the other
$P^1$). But this background of axio-dilaton should receive
correction because $\tau \to - i \infty$ as we approach the
orientifold plane. The correct behaviour therefore should be
\gptau. Switching on blow-up models at the orientifold
singularities fuses two intersecting orientifold planes to a
complex hyperbola \sengimon. In the figure below we have
represented a pair of $O7$ planes that become fused together.

 \vskip.2in

\centerline{\epsfbox{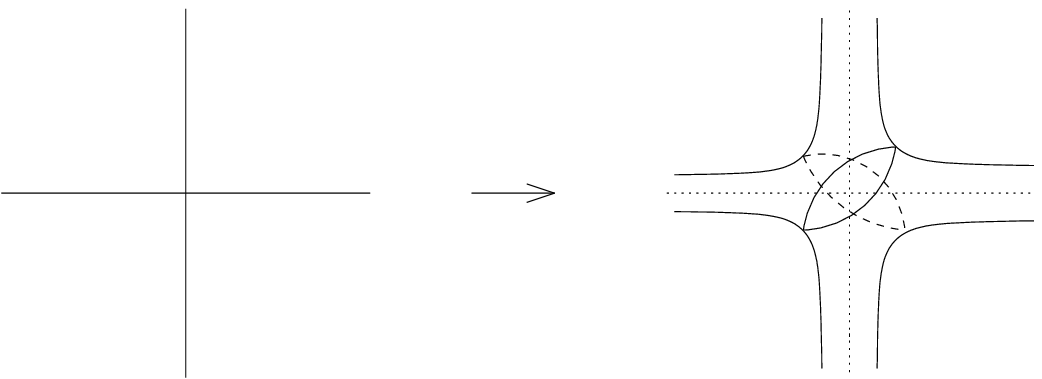}}

\vskip.2in \noindent The above backround in type IIB theory can be
easily lifted to F-theory. In the absence of fluxes we know that
the F-theory background is a torus fibation over a $P^1 \times
P^1$ base. What happens now?  Again the scalings in our present
case will give us the following background in F-theory:
\eqn\backinF{ds^2 = ds^2_{\rm spacetime} + \Delta^2 ds^2_{P^1} +
ds^2_{P^1} + \Delta^2 dx^2 + \Delta^{-2} dy^2,} where $x,y$ are
the coordinates of the fiber. Since both F-theory and M-theory
span the same moduli space, this metric will also correspond to
the metric in M-theory. The gauge fluxes will correspond to
localised $G$-fluxes in M-theory.

The issue of the vector bundle is now important. From the
F-theory side there are many enhanced gauge symmetry point. The
simplest one is the $SU(2)^8 \times SU(2)^8$ point.  This is
because at no point in the moduli space of the theory a single
brane can move freely. The minimum number of branes that can move
together in this theory is two, giving us the above mentioned
gauge group. In the heterotic side the vector bundle has to
satisfy \eqn\vecbund{F_{ab} = F_{\bar a \bar b} = g^{a \bar
b}F_{a \bar b} = 0, ~~~~ {\rm tr}~F \wedge F = {\rm tr}~R \wedge
R - i \bar\del \del J,} Assuming $\bar\del \del J = 0$, then from
\eqsatwa\ the warp factor will satisfy $\quabla \Delta^2 = 0$.
Since we also require the warp factor to be small everywhere, we
can take it to be a linear function of the coordinates $z^i$,
where $z^i$ are defined on patches. This would imply that
globally the three-form ${\cal H}$, satisfying \torcom, will be a
constant form. The background therefore is given by:
\eqn\consisol{\eqalign{&\Delta^2 = c_o + A~z^1 + B~z^2 + {\rm
c.c}, ~~~~~ e^\phi = \Delta \cr & {\cal H} =  A ~ dz^1 \wedge
dz^2 \wedge d{\bar z}^2 + B~dz^1 \wedge d{\bar z}^1 \wedge dz^2 +
{\rm c.c},}} where $c_o$ is a constant that would determine the
size of the conformal $K3$ space. Putting $A = 0$ will reproduce
the $GP$ model with non-trivial $\tau$. Now since $\quabla$ is
measured wrt the unwarped metric, $\Delta^2$ will satisfy the
warp factor equation. Therefore the constant three-form would
follow from there. We also have to make sure that $\Delta^2$ is
small everywhere. This is possible because our space is compact
and therefore $z^1, z^2$ are measured only in a small patch.
Assuming $\vert z^1 \vert^2 + \vert z^2 \vert^2 \le \epsilon$, we
can have $\Delta^2$ small everywhere. The equation \consisol\
will therefore be the torsional background for the conformal $K3$
in the heterotic theory.

Before we go any further we should clarify the role of the
discrete fluxes here. This has already been done in \sensethi\ so
we will be brief. The idea is that the various embedding of
instantons on $K3$ in the type I side is related to the existence
of discrete $B_{NS}$ fluxes in the type IIB side. More
explicitly, the duality chain that we followed is:
\eqn\dualchain{{\rm Type~I~on} ~~~{T^4\over G_1}~~~
{}^{~\sim}_{\longrightarrow} ~~~ {\rm Type~IIB~on}~~~{T^4 \over
G_2}~~~ {}^{T_{P^1}}_{\longrightarrow}~~~ {\rm
Type~IIB~on}~~~{T^4\over G_3},} where the first and the second
theories are equivalent; whereas the second and third are related
by two T-dualities. The type I theory that we have is S-dual to
heterotic theory compactified on the same manifold. The various
groups $G_i$ are given by \eqn\dualgroup{G_1 = (1, gh), ~~~ G_2 =
(1,\Omega) \times (1,gh), ~~~ G_3 = (1,g) \times (1,gh) = (1,g)
\times (1,h),} where we have already defined $g$ and $h$ in
\ganahare. In our notation, the orbifold limit of $K3$ will be a
$T^4$ modded out by the group $(1,gh)$. In the above chain
\dualchain, we have type I on $K3$ with some gauge bundle of
${{\rm Spin}(32)\over Z_2}$ that can belong to one of the three
topological classes characterized by an element ${\tilde w}_2$ of
$H^2(K3,Z)$. For our case we need: ${\tilde w}_2 \cdot {\tilde
w}_2 = 0$ mod 4. This is dual to the existence of discrete $B$
flux $\Lambda/2$ in the second chain of \dualchain. The precise
relation is ${\tilde w}_2 = \Lambda$. In the third chain of
\dualchain, there is no flux but now $K3$ is modded out by
$(1,g)$ and $g$ involves the Nikulin's involution $(2,2,0)$
\sensethi. Since we are more concerned about the  first and the
third chain of \dualchain, the effect of discrete flux in our
analysis will be irrelevant. To complete the chain in \dualchain\
we can go from the one heterotic to the other, i.e $E_8 \times
E_8$ theory, with instanton numbers $(12 - n, 12 + n)$. This is
dual to F-theory on $F_n$ with $n = 0, 1 ,4$.

Coming back to the issue of gauge bundle, we observed earlier that
the background axio-dilaton do not vanish and therefore we cannot
be in the constant coupling point of the Gimon-Polchinski model.
The background flux actually fixes the $\tau$ to a value given in
\gptau. This would determine the resulting positions of the
branes and planes in this scenario, and therefore the gauge bundle.

In an alternative scenario, the F-theory compactification on a
base $P^1 \times P^1$ is also connected to another Calabi-Yau with
Hodge numbers $(51,3)$. This connection was pointed out by
\dabholkarP and \blumZ. The difference between the two
compactifications: one with Hodge $(3,243)$ and the other with
Hodge $(51,3)$, is related to the number of tensor multiplets.
For the case studied, we saw that the heterotic dual has only one
tensor multiplet. To get more tensors in six dimensions, we have
to redefine the orientifold operation in the dual type I side. A
way to achieve this was shown in \dabholkarP. We can define the
orientifold operation in such a way that not only it reverses the
world sheet coordinate $\sigma$ to $\pi - \sigma$, but also flips
the sign of twist fields at all fixed points. In fact in the
closed string sector the two theories tally, but in the twisted
sector, we get 17 tensor multiplets instead of hypermultiplets.
Therefore even though the orientifolding action look similar in
both cases, the massless multiplets are quite different.

\newsec{Fluxes versus Branes: A Different Perspective}

In the previous section we have described six-dimensional flux
compactifications. However to make this discussion complete we
still have to justify the choice $m=2$ in \metans. In order to do
this we have to change the perspective slightly. This new
perspective will also allow us to argue for the validity of the
supergravity approximation.

The discussion in the previous section
can be interpreted in terms of branes
instead of fluxes. To illustrate this, consider M-theory
compactified on a four-fold ${\cal M}_8$. This background becomes
unstable, unless fluxes satisfying
$$
\int_{{\cal M}_8}G \wedge G = {\chi \o 24},
$$
are included. Here $\chi$ is the Euler characteristic of ${\cal
M}_8$. The background is
\eqn\beckback{ \eqalign{& ds^2 =  \tilde\Delta^{-1} ds^2_{012} +
{\tilde\Delta}^{1/2} ds^2_{{\cal M}_8} \cr & G_{012m} = \del_m
\tilde\Delta^{-3/2},\qquad  \quabla \tilde\Delta^{3/2} = {\rm
sources},}}
Imagine that we replace the fluxes by $n=\chi/24$ M2 branes. The
background then becomes
\eqn\metformrwo{\eqalign{& ds^2 = H^{-2/3}ds^2_{012} + H^{1/3}
ds^2_{{\cal M}_8} \cr & G_{012m} = {1\o 2} {\del_m H \o H^2},
\qquad \quabla H = {\rm sources},}} where $G$ is the source. The
two backgrounds above, i.e. equations \metformrwo\ and \beckback,
become equivalent if we identify $H = \tilde\Delta^{3/ 2}$.

This was an example of a more general statement which says that
if we can replace branes by fluxes (or vice-versa if the system is
compact), the backgrounds for the two systems become identical.
This has also been observed for Klebanov-Strassler type
geometries where the three-branes are replaced by fluxes on a
deformed conifold background. This was discussed in \adopt.

In the following we would like to find the brane interpretation
of the conformal $K3$ flux background of the heterotic string. On
the heterotic side we have ${\cal H}$-fluxes (RR three-form
fluxes for type I). We would like to interpret these fluxes as
heterotic $NS5$ branes (D5 branes for type I). The metric for a
$NS5$ brane at a point of a non-compact $K3$ is
\eqn\nsfivemet{\eqalign{&ds^2 = ds^2_{012345} + H_5~ds^2_{K3} \cr
& {\cal H}_{\mu\nu\rho} = \sqrt{g}~
\epsilon^\sigma_{~~\mu\nu\rho} ~\del_\sigma \phi, ~~~~~ e^\phi =
\sqrt{H_5}.}}
Under an S-duality \nsfivemet\ is mapped to the $D5$ brane metric
of type I. By redefining the harmonic function of the $NS5$ brane
as $H_5 = \Delta^2$ we will reproduce \metans\ (with $m=2$). In
order to understand how the sources are mapped we express
\nsfivemet\ in complex coordinates
\eqn\derivofh{{\cal H}_{ a \bar a b} = \left(g_{a \bar a} g_{b
\bar b} - g_{a \bar b} g_{\bar b a}\right)~g^{b \bar b}
~\del_b~\phi  = \del_b ~e^{2 \phi} = \del_b~ \Delta^2}
which agrees precisely with the torsional equation! After going
to the type I side, we obtain the type IIB background by
performing two T-dualities along the $K3$ directions. This
converts the $D5$ brane into a $D7$ brane. Therefore the axion
will be non-constant and together with the non-trivial dilaton
this will determine a non-trivial $\tau$ for the Gimon-Polchinski
model. This is of course consistent with the results of the
previous section.

Since the warp factor is assumed to be small everywhere, the
theory will be weakly coupled. The absence of cross terms in the
heterotic metric \metans\ implies that there is no $H_{NS}$ flux.
The above arguments give us a simple way to verify that
supersymmetry is preserved on the type IIB side. Indeed, by
reinterpreting the flux background in terms of branes we have
just obtained a $D7$ brane solution for type IIB. This, of
course, preserves supersymmetry. At this point it is instructive
to compare with the higher dimensional examples discussed in
\sav,\beckerD. Assuming we start on the heterotic side with the
metric described in \beckerD\ and perform the string dualities to
get to the IIB side, the metric we obtain is
\eqn\becmet{ds^2 = \Delta^{-1} ds^2_{0123} + \Delta~ds^2_{K3} +
\Delta~ds^2_{P^1}.}
Observe that all the internal directions are proportional to the
same power of the warp factor. By identifying $\Delta$ with the
harmonic function of a $D3$ brane, i.e $\Delta = \sqrt{H_3}$, the
metric \becmet\ becomes the $D3$ brane metric at a point on the
six manifold $K3 \times T^2/{\cal I}_2$. This implies $\tau = i$.
This is consistent because switching on fluxes creates a
background that simulates a $D3$ metric with a vanishing
axion-dilaton, giving us F-theory at {\it constant} coupling. On
the dual heterotic side we will have $NS5$ branes wrapping the
fiber of the non-K\"ahler manifold. This is similar to the
discussion in the previous paragraph in which we had a
non-K\"ahler space with ${\cal H}$ torsion being replaced by
$NS5$ brane metric. The $O3$ charge in the type IIB picture is
cancelled either by fluxes or by $D3$ branes. On the heterotic
side the $O5$ charge (due to the generator $\Omega {\cal I}_4$)
is cancelled either by the torsion or wrapped $NS5$ branes. The
above discussion supports the picture developed in \gauntlett.

There are two important questions we should answer before
proceeding. The first question is why for \metans, we are not at
constant coupling point of the Gimon-Polchinski model when the
$D3$ brane picture is consistent at constant coupling? This is
simply because of the construction. For the non-K\"ahler example
of \sav,\beckerD, the directions along which we were performing
dualities did not scale with the warp factor. Whereas in the
present case, because of the overall conformal factor, the duality
directions scale with the warp factor, giving us the metric
\metiintwob.

The second question is a little more subtle. In the brane picture on the
heterotic side we have $NS5$ branes wrapping the fiber of the
non-K\"ahler manifold. The fibration is non-trivial because on
the type IIB side we had $H_{NS}$ fields that T-dualise to the
metric. When we replace the three-forms $H_{RR}$ and $H_{NS}$ by
$D3$ branes, the naive expectation is that we will get $NS5$
brane wrapping a $T^2$ with trivial fibration. Why aren't we getting a
non-trivial fibration here? It turns out that to get a
non-trivial fibration we have to consider the abelian
instantons on the type IIB side. Indeed, the existence of abelian
instantons on the type IIB side can be verified as follows. The
$D3$ branes used to cancel the $O3$ charges, can be thought of as
coming from the $D7$ brane coupling $\int D^+ \wedge F \wedge F$,
where $F$ are the $D7$ brane gauge fields and $D^+$ is the type
IIB four form. Since we also require the warp factor equations
for both cases (one with fluxes and the other with branes) to be
identical, we need the gauge fields $A$ on the $D7$ branes to be
related to the $B = B_{NS}$ field of type IIB as
$$
B_{a \mu} = A_\mu,$$ with $x^a$ the direction of the fiber. These
$B$ fields dissolve on the heterotic side to appear as abelian
instantons on the base $K3$.

The above consideration of the three-form background in the
heterotic/type I case can also be verified using the analysis done
in \gauntlett. Here it was shown that the heterotic three-form is
determined in terms of generalized calibrations of \papado,
related to the G-structures of \gauntlett. According to these
references, there exist a generalized calibration-form $\Sigma$,
which determines the possible three-from background ${\cal H}$ in
the heterotic theory via the relation \gauntlett:
\eqn\relofgaun{\ast {\cal H} = d \Sigma +  \Sigma \wedge d\phi}
where $\phi$ is the dilaton. For the six dimensional non-K\"ahler
manifold studied in \sav,\beckerD,\bbdg\ this would reproduce the
relation that we derived in \bbdp\ via the superpotential. For the
present case of conformal $K3$, the first term of \relofgaun\
would vanish and we will have exactly the relation \torflisg.
Further confirmation of the choice of the metric that is taken
here (i.e the conformal $K3$) comes from analyzing the torsion
classes \carluest,\louisL,\gauntlett. For the conformal $K3$ case,
we get \eqn\otrclass{ {\cal W}_1 = {\cal W}_2 =  {\cal W}_3  = 0,
~~~~ {\cal W}_4 =  {\cal W}_5  =  2~d \phi} where ${\cal W}_i, ~ i
= 1, ...., 5$ are the five torsion classes. Since all these
details have already appeared in \carluest,\louisL,\gauntlett we
will not repeat them here. An additional insight that one can get
from \gauntlett\ is that the background \metans\ is actually a
heterotic five brane ansatz transverse to the $K3$, as observed
earlier.
 In fact taking a five brane $-$ wrapped on a
calibrated cycle of a given manifold $-$ we can reproduce the
torsional constraints in any dimensions. This is the basic idea
followed in the series of papers \gauntlett. Thus the derivation
given in sec. 3 of this paper will serve as a third way of
getting the background. Happily all these way of deriving the
result are mutually consistent.

This completes our discussion of six-dimensional compactification
with fluxes. We now turn to the discussion of new four dimensional
compactifications with fluxes and some related phenomenological
issues. We begin with an alternative derivation of the
non-K\"ahler manifold with zero Euler characteristics that are of
the form of $T^2$ bundles on $K3$.

\newsec{$T^2$ Bundles on $K3$: An Alternative Description}

In this section, we shall discuss the special case of non-K\"ahler
manifolds given by nontrivial $T^2$ bundles on $K3$ surfaces, of
the form discussed in \sav,\beckerD,\GP,\bbdg. These are examples
of complex threefolds with nowhere-zero holomorphic three-form
that, unless the $T^2$ bundle is trivial, are necessarily
non-K\"ahler.  It was argued, for example in \GP, that these
complex manifolds satisfy all the necessary conditions for a
consistent non-K\"ahler compactification.

These particular non-K\"ahler compactifications have another
interesting property: there is no warp factor on the
uncompactified directions \beckerD, \GP, making these models
especially amenable to methods of analysis developed for use with
Calabi-Yau manifolds. For other non-K\"ahler compactifications
there could be a warp factor on uncompactified directions.

The examples in this section deal with internal manifolds with a
vanishing Euler characteristic. This is, of course, not an
essential feature of consistent non-K\"ahler compactifications.
In later sections we shall discuss numerous examples with nonzero
Euler characteristic. Moreover, the number of generations is
computed in terms of the Chern classes of the gauge bundle and
not the Euler characteristic.

\subsec{Consistency Check via Duality Chasing}

In this section we will argue that heterotic compactifications on
$T^2$ bundles on $K3$'s are consistent using duality chasing, an
alternative approach to these backgrounds than previously
discussed in \sav,\beckerD,\bbdg. The previous analysis dealt with
constant three-form fields only. The argument in the present case
has the advantage that it will allow us to check consistency for
non-constant three-form fields on the heterotic (or even type
IIB) side.  The generic three-form we would like to consider
satisfies
\eqn\thrforgen{ {\cal H}_{ijk} = -{1\over 2} N_{[ijk]} -{1\over 12}
J^m_i~J^n_j~ J^r_k~J_{[mn,r]},}
where $N_{[ijk]}$ is the Nijenhuis tensor. If the manifold is
complex the Nijenhuis tensor vanishes; otherwise it is another
anti-symmetric tensor. Notice that, the derivation of the
superpotential, does not take into account whether the Nijenhuis
tensor vanishes or not. The generic three-form \thrforgen\ can be
non-constant, but should satisfy the torsional equation.

\vskip0.2cm

\centerline{$\underline{\rm{\bf Summary~ of ~Known~ Results}}$}

\vskip 0.2cm

Let us begin with a brief review of previous work on heterotic
compactifications on 3-folds of the form of $T^2$-bundles on
$K3$'s. Our work on this subject started with the construction of
a particular heterotic background for compactifications on such
manifolds using duality transformations of a consistent
background in M-theory. The particular M-theory background used
in \sav\ and \beckerD, was the four-fold $K3 \times K3$ and the
duality transformations were performed in several steps as
follows.

\item{1.} We include $G$-fluxes on $K3 \times K3$.
These should satisfy
\eqn\gflux{d*G = -{1\over 2} G \wedge G + (2\pi)^4 ~ X_8,}
Since this manifold satisfies $c_1=0$, (where $c_1$ is the first
Chern class), we can integrate \gflux\ over the four-fold
obtaining
$$
\int G \wedge G = {\chi \o 24},
$$
where $\chi\neq 0$ is the Euler characteristic.

\item{2.} This solution is lifted to F-theory
on $K3\times  {(T^2 \times T^2)/ {\cal I}_4}$, where ${\cal I}_4$
is an orbifold operation that reverses all the directions of
$T^4$.

\item{3.} The $T^2$ is shrunk to
zero size. This gives us type IIB on a compact six manifold of the
form $K3 \times T^2/ Z_2$, where $Z_2 = \Omega \cdot (-1)^{F_L}
\cdot \sigma$ and $\sigma$ reverses the two directions of the
torus $T^2$.

\item{4.} The M-theory fluxes become NS and
RR three-form fluxes on the type IIB side. These three-form fluxes
should have two legs on the $K3$ and one leg along the $T^2/Z_2$
direction. This is possible because $K3$ has 2-cycles but no
3-cycles. Furthermore, such a choice of three-form fluxes survives
the orientifold projection.

\item{5.} This type IIB background is transformed to the
heterotic side by two $T$-dualities and one $S$-duality. The
heterotic metric is given by \sav, \beckerD, \bem
\eqn\hemelo{ds^2 = \Delta^2 ds^2_{K3} + \vert dz + {\tilde f}
\vert^2,}
where $z$ is the complex coordinate of the fiber and ${\tilde f}$
depends on the coordinates of $K3$. For the choice of constant
fluxes ${\tilde f}$ takes the form ${\tilde f} = 2i~{\bar
z}^2dz_1 - (4 + 2i) {\bar z}^1dz_2$, where $z^{1,2}$ are the
complex coordinates of the $T^4/{\cal I}_4$ limit of the base
$K3$.

\vskip0.2cm

\centerline{$\underline{\rm{\bf Steps ~for ~Generating ~the
~Background}}$}

\vskip 0.2cm

\noindent Our duality argument for the consistency of
compactifications on non-K\"ahler manifolds of the form discussed
above begins with an orbifold point of the six manifold $K3 \times
T^2$, and works locally about a point close to one of the orbifold
singularities. This is equivalent to taking a non-compact six
manifold of the form $R^4 / {\cal I}_4 \times T^2$ which is
nothing but a torus fibered over a Taub-NUT base. This fibration
is trivial. To get the complete supergravity background we shall
consider the following steps:

\item{1.} Consider a type IIA background with a NS5 brane oriented
along the $x^{0, 1, 2, 3, 8, 9}$ directions and two D4 branes
oriented along $x^{0, 1, 2, 3, 8}$ and $x^{0, 1, 2, 3, 9}$
respectively. The directions $x^{6, 8, 9}$ lie on a $T^3$. Since
this system consists only of branes, the supergravity solution
can be easily written down.

\item{2.} We consider two of the cycles of the $T^3$ along
the $x^{6, 8}$ directions and make a twist on the coordinates
described by the matrix \origa\
\eqn\twistroj{\pmatrix{{\rm cos}~\alpha & -{\rm sin}~\alpha \cr 0
& {\rm sec}~\alpha}, }
where we have taken unit radii for all the three-cycles of the
$T^3$ (we consider equal radii for the cycles for simplicity).

\item{3.} We now T-dualize $R \to 1/R$ along the $x^6$
direction. This will lead us to our required non-compact six
manifold $R^4/{\cal I}_4 \times T^2$ with the TN oriented along
$x^{4, 5, 6, 7}$ and the torus oriented along $x^{8, 9}$. The
twist \twistroj\ will induce a threeform NS field strength with
components $H_{68r}$ where $r$ lies in the TN direction. The two
D4 branes will become two D5 branes along the $x^{0, 1, 2, 3, 6,
8}$ and $x^{0, 1, 2, 3, 6, 9}$ directions, respectively.

\item{4.} We now apply S-duality. This will convert the $H_{NS}$
field to a $H_{RR}$ field, along with the D5 branes to NS5 branes.
The NS5 branes will now be sources of $H_{NS}$ fields with
components $H_{8 r_1 r_2}$ and $H_{9 r_3 r_4}$, where $r_i$ are
the spatial directions along the TN space. At large distance from
the NS5 branes we see a configuration $R^4/{\cal I}_4 \times T^2$
with one leg of the NS-NS and the RR threeform fields along the
toroidal directions $x^{8,9}$. This configuration is analogous to
the type IIB configuration presented in \sav,\beckerD.

\item{5.} T-dualizing the $x^{8,9}$ directions of the
torus will convert the two NS5 branes into geometry. The fact
that the $x^{8,9}$ toroidal directions are non-trivially fibered
over the base can be easily seen from the action of the twisting
incorporated by the NS5 branes when they convert to geometry.

\vskip.1in

\noindent \centerline{$\underline{\rm {\bf Precise ~supergravity
~analysis}}$}

\vskip.1in

\noindent The steps mentioned above are not new and have appeared
in a different context in \papers. The brane configuration in
type IIA is a grid-like configuration and is somewhat similar to
the brane-box configuration developed in \hanany. The two D4
branes form a grid along the $x^{8,9}$ directions with NS5 branes
filled in it. See the figure below\foot{Since the above configuration is a
part of a bigger structure (not considered in details here), we expect susy
to be preserved in the fuller picture. And since susy is not
relevant to this analysis, we will ignore it.}.

\bigskip

\centerline{\epsfbox{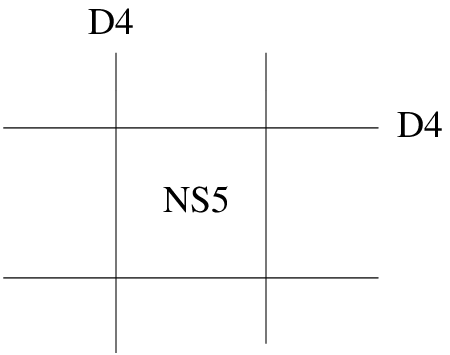}}\nobreak \centerline{{\bf Fig.
3}: {\it Grid diagram showing the D4 grids and NS5 branes
inside.}}

\bigskip

\noindent The supergravity solution for the system can only be
written if we delocalise some of the directions. We will assume
that all the harmonic functions $f_i$ depend on the $y^i =
x^{4,5,6,7}$ coordinates only. If we denote the harmonic function
of the NS5 branes as $h$, the metric of the system will be given
by
\eqn\metofsys{ds^2 = f_1 ds^2_{0123} + f_2 ds^2_8 + f_3 ds^2_9 +
h f_1^{-1} ds^2_{4567}.}
The above metric of the grid is untwisted. Now we will twist the
directions $x^{8,6}$ using the matrix \twistroj. The harmonic
function of the NS5 branes can be determined in terms of the
twist angle $\alpha$ as
\eqn\harm{ h = f_1 f_4 ~{\rm sec}^2~\alpha - f_1 f_2~ {\rm
tan}^2~\alpha,}
where $f_4 (y^i)$ measures the distance along the $x^6$ direction
in the twisted metric. The metric along the $x^{6,8}$ direction
can be written as
\eqn\metsixe{{\pmatrix{g_{66}&g_{68}\cr g_{86}&g_{88}}} =
{\pmatrix{f_4 & f_2{\rm tan}~\alpha\cr f_2{\rm tan}~\alpha &
f_2{\rm sec}^2~\alpha}},}
where we see that a cross-term has developed. The other directions
of the system do not change and therefore the metric remains the
same. The metric of the deformed grid is the first step of the
chain of dualities that we use to arrive at a heterotic
compactification on the non-K\"ahler manifold. Before we go into
more details it would be interesting to ask whether the above grid
model can have a partial decomposition in terms of different
numbers of NS5 branes placed in adjacent grids. This is where our
configuration differs from the usual brane box configuration
developed in \hanany. For the NS5 branes to break (or
alternatively if we want to keep $n_i$ number of NS5 branes in the
$ith$ grids) we would require a coupling on the NS5 branes of the
form $\int C_5 \wedge A$ where $A$ is a one form on the NS5. This
coupling is easily ruled out because the multiplets propagating on
the NS5 branes are (2,0) anti self-dual tensor multiplets that
have $B^-_{ab}$ as its propagating degrees of freedom. In terms of
M-theory this configuration is actually a single M5 configuration
since all the D4 branes become M5 branes when the type IIA
coupling become very strong. Now the grid structure in type IIA
can separate into two regions: one without twist and the other
with a twist. As we will see below, the regions inside the grid,
i.e the area $x^{8,9}$, will have the usual flat metric of a two
dimensional (compact) space. The regions outside will be twisted
by the twist generated from U-duality.

To see how this comes about, we take our deformed grid structure
in type IIA and perform a T-duality along the $x^6$ direction and
then a S-duality. Under a single T-duality, the type IIA
configuration will go to chiral type IIB theory with the NS5
branes inside the grid converting to a Taub-NUT manifold. The two
D4 grids will convert to two D5 branes sharing now four common
directions. Under a further S-duality the theory will go to a
brane box configuration with two NS5 branes forming the sides of
the box. See figure below.

\centerline{\epsfbox{ 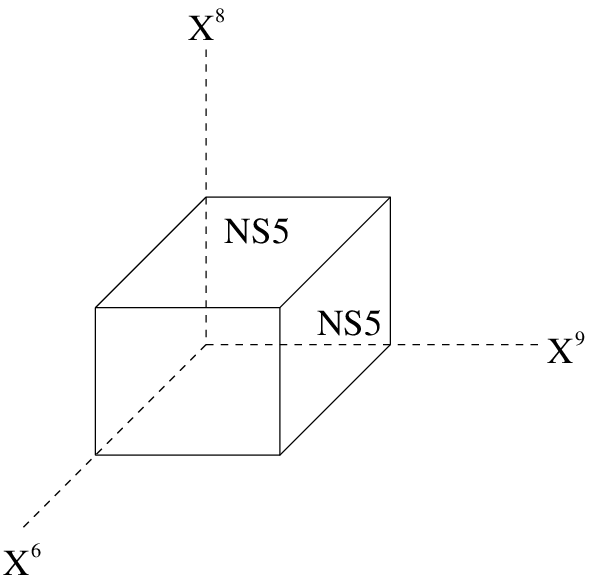}} \nobreak \centerline{{\bf
Fig. 4}: {\it Two NS5 branes forming a box embedded in a Taub-NUT
space.}}

\bigskip

\noindent This box-like configuration will be embedded in a
Taub-NUT space with a RR three-form field. The origin of this RR
three-from field is the deformation of the type IIA grid
structure by an angle $\alpha$. The two NS5 branes that form the
two sides of the box give rise to $H_{NS}$ fields whose lines of
force extend out of the box. These two NS three-forms twist the
$x^{8,9}$ torus. We can therefore view the configuration in the
following way: We take a two-dimensional slice (along $x^{8,9}$)
of the brane box. This separates the box into two regions. One of
the regions will be thread by the $H_{NS}$ sources and a $H_{RR}$
field with component
\eqn\hrr{H_{RR} = \del_r(f_2 f_4^{-1})~{\rm tan}~\alpha~dr \wedge
dx^6 \wedge dx^8.}
These sources will twist the geometry after applying two
additional T-dualities.

What remains to show is that the NS5 branes produce the necessary
twist in the geometry when we T-dualize twice. This has already
been discussed in some earlier work (see for example \dasmuk) so
we shall be brief. To see this let us first define the following
set of matrices \beckerD:
\eqn\Bmatrices{b \equiv b_{(mn)} =\pmatrix{B_{8m}&B_{8n}\cr
B_{9m}& B_{9n}}, ~~~ g = \pmatrix{g_{88}& g_{89}\cr g_{89}&
g_{99}}.}
Here $B$ are the two sources of the NS5 branes and $g_{\mu\nu}$
is the metric of the brane box configuration (we choose $\mu, \nu
= 8,9$ and $m,n$ to be the rest of the coordinates). The
corresponding T-dual metric will be denoted by $G_{\mu\nu}$. The
metric after two T-dualities is now related to the metric of the
IIB brane box configuration by the following simple relations:
\eqn\frst{\eqalign{& {\rm tr}~G = {\rm tr}~g^{-1},~~ \qquad ~~
{\rm det}~G = {\rm det}~g^{-1}\cr & G_{8m} = {\det (b\sigma_1 + g
\sigma_3)\over \det g}, ~~ \qquad ~~ G_{9m} = { \det (b\sigma_2 +
g \sigma_1)\over \det g}\cr & G_{mn} = g_{mn} + {{\rm tr}~(b
\sigma_4) ~ {\rm det}~(b\sigma_1 + g \sigma_3) + {\rm tr}~ (b
\sigma_3)~ {\rm det}~ (b \sigma_2 + g \sigma_1)\over {\rm
det}~g},}}
where we have only shown the possible components of the metric. A
similar analysis can be performed for the other fields in the
theory. This have been worked out in detail in \beckerD. The
matrices $\sigma_i$ are the Chan-Paton matrices and are given by
\eqn\chanpaton{ \sigma_1 = \pmatrix{1&0\cr 0&0}, \qquad  \sigma_2
= \pmatrix{ 0&1\cr 0&0}, \qquad  \sigma_3 = \pmatrix{0&0\cr 0&1},
\qquad
 \sigma_4 = \pmatrix{
0&0\cr 1&0}. }
{}From the above set of relations it is easy to see how the
$x^{8,9}$ torus gets twisted by the NS5 sources $B_{8m}$ and
$B_{9n}$. The $H_{RR}$ field in the type IIB framework is mapped
to $H_{NS}$ after two T-dualities and a S-duality (which we will
describe soon). But there is a possibility of the following cross
term
\eqn\cross{{\tilde B}_{mn} = 6 B_{[8m}B'_{n9]},}
where we have denoted the RR field as $B'$ and the $B$ field after
T-dualities as ${\tilde B}$. However for our configuration $-$
wherein we consider the region inside the box and compare with the
region far away outside the box $-$ the threeform sources could be
taken to be effectively constant and therefore \cross\ would
vanish by the same reasoning as we had in \beckerD. Using now the
set of relations in \frst\ we can easily see that the two sides of
the brane box $x^8$ and $x^9$ twist in the following way
\eqn\twising{\eqalign{& dx^8 ~\to ~ {1 \over \sqrt{{\rm tr}~(g
\sigma_1)}}
 \left( dx^8 + {\rm tr}~ (b \sigma_1) \cdot dx \right) \cr
& dx^9 ~\to ~ {1 \over \sqrt{{\rm tr}~(g \sigma_3)}} \left( dx^9
+ {\rm tr}~ (b \sigma_3) \cdot dx \right).}}
This is what we expect for the non-K\"ahler manifold. However, the
physical meaning of the twist can now be described in terms of the
NS5 brane action without going through the technicalities of
\frst. In fact, each of the NS5 branes converts to a Taub-NUT
geometry after one T-duality. For example, a T-duality along the
$x^8$ direction converts one side of the brane box to geometry
without affecting the other side. The second T-duality converts
the other side of the brane box. The twist mentioned above in
\twising\ comes from the source $\omega$ of the NS5 brane which
is related to the harmonic function $h$ of the NS5 brane in the
following way:
\eqn\twhar{ {\rm curl}~\omega = - {\rm grad} ~h,}
where the curl and the grad operations are along the $x^{4,5,6,7}$
directions with $x^6$ compact. Therefore until now we can
conclude: inside the box (which we take as a 2d slice) the torus
is a simple square torus (to be more precise it is $R^2$) fibered
over $R^4/{\cal I}_4$ base. Outside the box the $x^{8,9}$
directions get twisted by the TN twist given in \twising. This
twisted $R^2$ is fibered over a conformally scaled Taub-NUT base.

Part of the analysis above may seem a little confusing to an
attentive reader.  First, the discussion above only makes sense
for type IIB strings whereas we are more interested in type I (or
heterotic) strings.  To dualize into the desired string theories,
we must introduce orientifolds. Second, our analysis did not take
into account the possibility of non-abelian gauge multiplets. The
first issue can be easily tackled.  In our analysis, spacetime
was divided into a space inside the box and a space outside the
box.  The geometry inside the box was of the form $R^4/{\cal I}_4
\times R^2$.  However we always have the freedom to impose an
orientation reversal on the strings moving along the $R^2$
directions, {\it i.e.}
$$
R^2 ~ \to ~ {R^2 \over (-1)^{F_L} \cdot \Omega \cdot {\cal
I}_{89}}.
$$
Imposition of such an orientation reversal would have the effect
that the pair of global T-dualities would map a spacetime of the
form described inside the box into a type I background. A similar
argument can be given for the region outside the brane box. The
two D5 branes (forming the box) survive under the orientation
reversal and the duality chain lands in type I theory. Applying
an overall S-duality lands us in heterotic strings. Therefore we
identify the twist mentioned in \bbdp\ as the T-dual of two
intersecting NS5 branes, as outlined here. Surprisingly the above
analysis is not affected by the underlying theory and appears to
be background independent.  Our only concern here would be to
keep those fields that would survive under the orientation
reversal $\Omega$ as well as under the space orbifolding.
Happily, our choice of background is consistent with this
requirement.

The second issue is more serious.  Our analysis above is not
sufficiently general to take into account non-abelian gauge
mutiplets in the corresponding heterotic (or type I) setup.  This
is because the type IIA grid, or the brane box in M-theory (in
which the sides of the box are M5 branes), is a little too
naive.  In order to describe non-abelain gauge multiplets, one
must have additional singularities along the $R^2$ spanning the
$x^{8,9}$ directions. These singularities are similar to the
stringy cosmic strings of Greene et. al. \greene\ and therefore we
have to allow some {\it point-like} singularities on the 2d space
$R^2$.

Such a setup is realised in F-theory and therefore it should come
as no surprise that similar ideas must be applied here. The point
like singularities define monodromies of the form
\eqn\monod{M_{p,q} = M_{-p,-q} = \pmatrix{1-pq&p^2\cr -q^2&1+pq}}
where $p,q$ are the labels that is used to write any $p,q$
strings in this background. In fact these point like
singularities correspond to the ($p,q$) seven branes that
physically realize the gauge symmetry enhancement. As has been
discussed in \zwigab, the monodromies $M_{p,q}$ introduce branch
cuts in the two-dimensional space,
 and any string
crossing such a branch cut will be converted to its $SL(2,Z)$
cousin via the matrix $M_{p,q}$. For our case $-$ where we
requite a $D_4$ singularity at a point $-$ this has been worked
out in \zwigab.  We require three monodromy matrices of the form
\eqn\monomat{M_{1,0}=\pmatrix{1&1\cr 0&1}, ~~~M_{3,-1}=
\pmatrix{4&9\cr -1&-2}, ~~~ M_{1,-1} = \pmatrix{2&1\cr -1&0}.} We
know from \vafasen,\bbdg\ that we need four local D7 branes and
two ($p,q$) seven branes. This implies \eqn\monmatsat{M_{1,0}^4
\cdot M_{3,-1} \cdot M_{1,-1} = - I} where $I$ is the identity
matrix. The above relation is indeed realized by the choice
\monomat.  To understand this further we can choose to isolate the
singularities so that the monodromy action can act independently.
In terms of our earlier analysis in \bbdg\ this implies we would
be in a situation where the type IIB dynamics can not be described
perturbatively. In the figure below

\bigskip

\centerline{\epsfbox{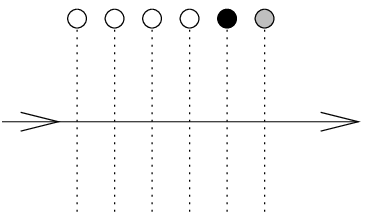}}

\bigskip

\noindent the four branch cuts associated with $M_{1,0}$ are
shown with white circles, whereas the other two branch cuts are
shown with dark circles. A string crossing this set of branch cuts
will undergo a total monodromy computed in \monmatsat. This is of
course only a local picture.  The global picture is not captured
here\foot{Globally there is an underlying Weierstrass equation $y
= \sqrt{(x-e_1)(x-e_2)(x-e_3)}$ with discriminant $\Delta =
(e_1-e_2)^2(e_2-e_3)^2(e_3-e_1)^2$. The polynomials $e_i$ are
functions of $z$, the coordinate of $R^2$. This story is very
well known, and so we will not discuss it further.}.  The
perturbative subgroup realised here is $SU(4)\times U(1)$, and
therefore if we allow a monodromy of $-M_{1,0}^n$ at infinity
instead of $-1$, i.e \eqn\monatinf{ M_{1,0}^{n+4} \cdot M_{3,-1}
\cdot M_{1,-1} = - M^n_{1,0},} where $n$ is an integer, a
perturbative subgroup of $SU(n+4) \times U(1)$ will be realised
in our framework. The full non-perturbative symmetry is of course
$D_{n+4}$ which is the type $I_n^\ast$ singularity of \morvaf.
The adjoint therefore decomposes as \zwigab \eqn\adjdecom{{\bf
(n+4)(2n+7)} = {\bf (n+4)}^2 - {\bf 1} + {1\over 2} {\bf
(n+4)(n+3)}_{-2} + {1\over 2} {\overline{{\bf (n+4)(n+3)}}}_{+2}
+ {\bf 1}} where $\pm 2$ are the possible $U(1)$ charges. For the
case discussed earlier the decomposition \adjdecom\ will tell us
how ${\bf 28}$ (the adjoint of $D_4$) is decomposed under
$SU(4)$.  The above analysis can be extended to the case when we
can have exceptional singularities. The existence of these points
in the moduli space was shown earlier (in the F-theory context) in
\dasF.  The monodromy matrices for these cases have all been
worked out in \zwigab. The result can be presented in the
following table:

\ \smallskip
\begintable
{\rm Group}|{\rm Monodromy}|{\rm Perturbative~Group}| {\rm
Degeneration}\eltt $D_4$|$M^4_{1,0}\cdot M_{3,-1}\cdot M_{1,-1} =
-1$ |$A_3 \times U(1)$|$Z_2$\elt $E_6$|$M^5_{1,0}\cdot
M_{3,-1}\cdot M^2_{1,-1} = (ST)^2$ |$A_4 \times A_1$ |$Z_3$\elt
$E_7$|$M^6_{1,0}\cdot M_{3,-1}\cdot M^2_{1,-1} = S^2$|$A_5 \times
A_1 \times U(1)$|$Z_4$\elt $E_8$|$M^7_{1,0}\cdot M_{3,-1}\cdot
M^2_{1,-1} = ST$|$A_6 \times A_1 \times U(1)$|$Z_6$
\endtable
\medskip
\noindent Note that we have also pointed out possible monodromies
on $R^2-\{0\}$ corresponding to the singularities.
The matrices $S$ and $T$ are given by \eqn\sandtmat{S =
\pmatrix{0&-1\cr 1&0},~~~~~T = M_{1,0} = \pmatrix{1&1\cr 0&1}.}
The reader might now ask what in our original type IIA grid
structure corresponds to the monodromies listed above. The grid
structure can be lifted to M-theory where it is a simple
configuration of M5 branes forming a box. The singularities
discussed above should lift to Atiyah-Hitchin (AH) space with our
M5 box embedded in it. The non-abelian monodromies from F-theory
tell us that we require a combination of AH and Taub-NUT spaces
on top of each other instead of just the AH spaces. For a single
TN space over a AH space, this is just the double cover of the AH
space \sewitten\ and as such supports an anti self-dual harmonic
two form $\Omega_o$.  A naive expectation would be that the
heterotic gauge fields are coming from the usual decomposition of
the $G$ fluxes over $\Omega_o$.  This is not quite the case,
partly because the M5 brane box structure will back-react on the
system and will change the analysis.  This back-reaction is not
too difficult to work out for some cases.  Observe that even
though in type IIB the analysis resembles the one given in
\sav,\beckerD, the T-duality to type IIA and its subsequent lift
to M-theory is completely different.  In both cases the seven
branes become six branes in the IIA picture, but are oriented
along different directions.  In \sav,\beckerD, the six branes
would wrap the $K3$ manifold and would stretch along the
remaining 2+1 dimensional spacetime. T-dualizing such a case
yields a four-fold in M-theory, whereas the T-dualities here give
rise to a grid structure in M-theory.  The other $K3$ in the
four-fold (constructed {\it \`a la} Borcea) comes from the
specific distributions of the seven branes that would curve the
space to make it compact. On the other hand, here we have an M5
brane inside a five-brane brane box.  In the figure below the
sides of the box are all M5 branes and the shaded cross-section
is another M5 brane inside the box.  The M-theory direction is
given by $x^{11}$.

\bigskip

\centerline{\epsfbox{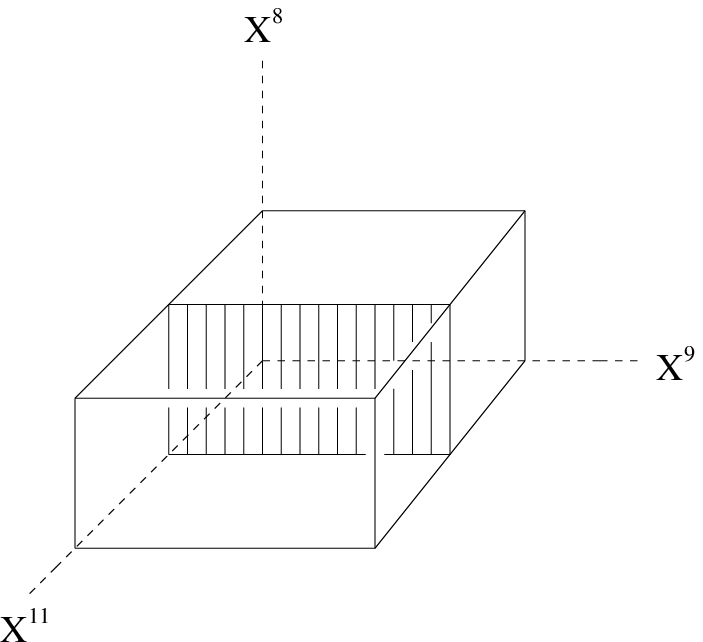}}\nobreak \centerline{{\bf Fig. 6}:
{\it An M5 brane inside a five-brane brane-box embedded in a
curved geometry.}}

\bigskip

\noindent The whole system would have to be embedded in a curved
geometry that is constructed out of the double cover of the AH
spaces and multi TN geometries. Whether it is possible to get a
fully compact version of this is an interesting question. Our
analysis only gives us a non-compact scenario (the branes being
all non-compact).  The gauge fluxes in this case would come from
the specific distribution of the harmonic forms $\Omega_o$ taking
into account the possible back-reactions from the brane-box.  For
the four-fold case, wherein we consider the $R^4/{\cal I}_4$
limit of one of the $K3$, the distribution of the localised gauge
fluxes also get affected by the presence of non-localised fluxes.
For the  generic case, when we are away from the orientifold
limit, the localised $G$-fluxes will be slightly different from
the one presented in \bbdp\ because of the back-reactions, and is
given by \eqn\locfluxl{{G \over 2\pi} = F \wedge dg \wedge d\psi
+ ..} where the dotted terms involve the other coordinates of the
TN space and $\psi$ is the compact circle with asymptotic radius
$R$.  We have denoted the abelian gauge field at one of the
orbifold points by $F$. The complete non-abelian generalisation
can be done following the procedure illustrated in \bbdp. The
value of $g(r)$, for some specific cases, has been  calculated in
\robbins. We can use a similar arguments here to get the
contributions from the fixed points.

\subsec{An Example with a $U(1)$ Bundle and a Nonzero Number of
Generations}

Having given an alternative way to get the non-K\"ahler space, it
is now time to do some explicit calculations of vector bundles on
this space. In this section we will only consider $U(1)$ bundles.

Heterotic compactifications on $U(1)$ bundles, in addition to
being phenomenologically nonviable, are also often plagued by
other difficulties, such as anomalous $U(1)$'s (typically Higgsed
via Dine-Seiberg-type mechanisms). For the specific purpose of
further illustrating aspects of $T^2$ bundles on $K3$'s, we shall
here count the number of generations in an oversimplistic example
with a $U(1)$ bundle, using general ideas described earlier in
section 2.4.

\noindent In particular, we will evaluate the number of generations,
$N_{\rm gen}$ when

\item{1.} The spin connection is the torsional connection.

\item{2.} The spin-connection is not embedded in gauge
connection.

\item{3.} The Euler characteristic may (or may not) be zero.

\item{4.} Non-trivial warp factor is taken into consideration.

\noindent Of course (2) and (3) are related.  When the spin
connection is embedded in the gauge connection, the number of
generations is given by half the Euler characteristic of the
manifold \candle, essentially a very special case of the
calculations in section 2.4. However, more generally the number
of generations is computed in terms of the gauge bundle, and not
the underlying space. In particular, when the torsional
connection is different from the gauge connection, vanishing of
Euler characteristics does not imply the vanishing of $N_{\rm
gen}$.

The next important issue is the existence of a non-trivial warp
factor.  Fortunately this is not a problem for the backgrounds
that were studied in \sav,\beckerD, because in those examples,
there is no warp factor for the the noncompact directions
corresponding to low-energy spacetime. In general, the metric for
a non-K\"ahler space constructed as the total space of a
$T^2$-bundle on $K3$ can be written in the form:
\eqn\genmitwa{ds^2 = \Delta^m ~ds^2_{0123} + \Delta^n~ ds^2_{K3}
+ \Delta^p~ (dz + f)(d{\bar z} + {\bar g}),} where $f,g$ are some
functions of the coordinates of $K3$.  For the cases studied in
\sav,\beckerD, $m = p = 0, n = 2$. This implies that the warp
factor will not effect the calculations of $N_{\rm gen}$.  In the
type I theory, $m = -1, n = 1, p = -1$, and therefore evaluation
of $N_{\rm gen}$ might be a little subtle.

Computing the number of generations, in both Calabi-Yau and
non-K\"ahler heterotic compactifications, was outlined in section
2.4. As mentioned previously, let us apply those methods to very
simple models with gauge group $U(1)^n$, for some integer $n$.
(Such examples are prone to certain technical difficulties, which
we will ignore as our interest here is in giving an easy overview
of some technology.) Recall the Chern classes of the gauge bundle
are given by\foot{The general formula for the Chern-class $c_n$
for a gauge bundle $F = T^a F^a$ (where $T^a$ is the generator of
gauge group $G$) is given by $$c_n =
(-1)^m~\sum_{k_1,..,k_n}~{\delta(n - k_1 - ...-n k_n)\o
k_1!..k_n!} \left[ \left({{\rm tr}~F\o 2\pi}\right)^{k_1}
\left({{\rm tr}~F^2 \o 8 \pi^2}\right)^{k_2} .... \left({{\rm
tr}~F^n \o n 2^n \pi^n}\right)^{k_n} \right]$$ where the traces
are taken in the fundamental representations and $m = n +
\sum_i^n k_i$.} \eqn\cherncl{\eqalign{& c_1 = {1\o 2\pi}  ~{\rm
tr} ~F  \cr & c_2 = -{1\o 8 \pi^2} \left[{\rm tr}~(F \wedge F) -
{\rm tr}~F \wedge {\rm tr}~F \right] \cr & c_3 = {1\o 48 \pi^3}
\left[ 2 {\rm tr}~(F \wedge F \wedge F) - 3 {\rm tr}~(F \wedge F)
\wedge {\rm tr}~F + {\rm tr}~F  \wedge {\rm tr}~F \wedge {\rm
tr}~F \right],}} where all the traces are in the fundamental
representations of the gauge group.  For the case of principal
$U(1)$ bundles these traces are very simple. For non-abelian
gauge fields these traces are computed using the structure
constant $f_{abc}$ or $d_{abc}$ of the gauge group, with
generators $T^a$.
As usual in heterotic compactifications, the gauge field must
satisfy \vecbund.  This would imply that a generic $U(1)$ bundle
has the following form: \eqn\vecumin{F = F_{i \bar j} ~dz^i
\wedge d{\bar z}^j,} with $i,j$ running over all the complex
three coordinates $z^{1,2,3}$.  Furthermore $F$ constructed above
should be real. The above form of the gauge bundle has also
appeared in \carluest\ with the components of $F$ restricted to
the base only.  We, on the other hand, will be working with
general $F$.  One can show that the components of $F$ are in
general of the form: \eqn\fone{ F_{a \bar b}= \pmatrix{ F_{1 \bar
1} & F_{ 1 \bar 2} & F_{1 \bar 3} \cr \noalign{\vskip -0.20 cm}
\cr F_{2 \bar 1} & F_{ 2 \bar 2} & F_{ 2 \bar 3}\cr
\noalign{\vskip -.20 cm}\cr F_{3 \bar 1} & F_{ 3 \bar 2} & F_{3
\bar 3} \cr } = \pmatrix{ i~ f_1 & f_{1\bar 2} &  f_{1\bar 3} \cr
\noalign{\vskip -0.20 cm}  \cr -{\bar f}_{1\bar 2} & i~f_2 & f_{2
\bar 3} \cr \noalign{\vskip -0.20 cm}  \cr
 -{\bar f}_{1\bar 3} & - {\bar f}_{2\bar 3} & i~ f_3 \cr}}
Here we have taken $f_1, f_2$ and $f_3$ to be real. All the other
$f_{i \bar j}$ are complex quantities. To evaluate the number of
generations for the above choice of the gauge bundle we have to
make sure that ${\rm tr} ~F \wedge F$ satisfy the necessary
Bianchi identity as we are not embedding spin-connection in gauge
connection.  Furthermore we will not have vanishing ${\rm tr}~R
\wedge R$ because of the existence of non-trivial warp factor. We
will tentatively take ${\rm tr}~R \wedge R = \zeta_4(\Delta)$,
where $\zeta_4$ is an appropriate four-form written in terms of
the warp factor $\Delta$. Our case will be different from the
analysis of \carluest\ in this respect.  For the non-K\"ahler
manifolds of the form of $T^2$-bundles on $K3$'s, with metric of
the form \genmitwa\ with $g = f$, ${\rm tr}~F \wedge F$ is given
by \eqn\trfwef{\eqalign{ {\rm tr}~F \wedge F & \equiv {\cal F}~
dz^1 \wedge d{\bar z}^1 \wedge dz^2 \wedge d{\bar z}^2 \cr & =
\zeta_4(\Delta) - {\bar\del}\del~f \wedge {\bar f} -
{\bar\del}\del \Delta^2 \wedge (dz^1 \wedge d{\bar z}^1 + dz^2
\wedge d{\bar z}^2).}} With the choice of $f$ given in \fiber,
the value of ${\cal F}$ in the above relation can be worked out
in complete detail.  What remains now is to evaluate the value of
the components $F_{i\bar j}$.The first requirement, for an
abelian bundle, is the equation $dF = 0$. Putting this condition
on \fone\ give rise to nine relations between the nine variables.
They can be written as: \eqn\frel{\matrix{ & i~\del_2 f_1 +
\del_1 {\bar f}_{1\bar 2} = 0, \hfill  &  i~\del_3 f_1 + \del_1
{\bar f}_{1\bar 3} = 0,  \hfill \cr \noalign{\vskip -0.20 cm}  \cr
&  i~\del_1 f_2 - \del_2 {\bar f}_{1\bar 2} = 0,  \hfill   &
i~\del_3 f_2 + \del_2 {\bar f}_{2\bar 3} = 0, \hfill \cr
\noalign{\vskip -0.20 cm}  \cr &  i~\del_1 f_3 - \del_3 {\bar
f}_{1\bar 3} = 0  , \hfill &   i~\del_2 f_3 - \del_3 {\bar
f}_{2\bar 3} = 0,  \hfill \cr \noalign{\vskip -0.20 cm}  \cr &
\del_3 f_{1 \bar 2} + \del_1 {\bar f}_{2\bar 3} = 0 , \hfill
&     \del_3 {\bar f}_{1\bar 2} - \del_2 {\bar f}_{1\bar 3} =
0,\cr}} \vskip -0.45 cm
$$ \del_2 f_{1 \bar 3} - \del_1 {\bar f}_{2 \bar 3} = 0$$
\noindent These relations are in addition to the constraint on
${\rm tr}~F \wedge F$ imposed by the Bianchi identity \trfwef.
But this is not enough.  There is yet another condition on the
components of fluxes from the Donaldson-Uhlenbeck-Yau equation
$F_{a \bar b} J^{a \bar b} = 0$. To evaluate this we require
$J^{a\bar b}$, which is given by: \eqn\jone{ J^{a \bar b}=
\pmatrix{ J^{1 \bar 1} & J^{ 1 \bar 2} & J^{1 \bar 3} \cr
\noalign{\vskip -0.20 cm}  \cr J^{2 \bar 1} & J^{ 2 \bar 2} & J^{
2 \bar 3}\cr \noalign{\vskip -.20 cm}\cr J^{3 \bar 1} & J^{ 3
\bar 2} & J^{3 \bar 3} \cr } =
 -{i\o \Delta^2} \pmatrix{
{1\o 2} & 0  & -i{\bar z}^2 \cr \noalign{\vskip -0.20 cm}  \cr 0
& {1\o 2} & (2+i){\bar z}^1\cr \noalign{\vskip -.20 cm}\cr -iz^2
& (2-i)z^1 & 2(\Delta^2 + 5 \vert z^1\vert^2 + \vert z^2 \vert^2)
\cr}} Multiplying \fone\ with \jone\ will give us another
condition on the gauge bundle. At this point it is instructive to
compare the analysis done in \carluest. The gauge field and the
inverse of the two form $J$ for the Iwasawa manifold is given by:
\eqn\luesexam{F_{a \bar b} = \pmatrix{ if_1 & f_{1\bar 2} &  0 \cr
\noalign{\vskip -0.20 cm}  \cr -{\bar f}_{1\bar 2} & if_2 & 0 \cr
\noalign{\vskip -0.20 cm}  \cr
 0 & 0 & 0\cr}, ~~~~~~~J^{a\bar b} = -i \pmatrix{
{1\o 2} & 0  & 0 \cr \noalign{\vskip -0.20 cm}  \cr 0 & {1\o 2} &
{z}^1\cr \noalign{\vskip -.20 cm}\cr 0 & {\bar z}^1 & 2(1 + \vert
z^1 \vert^2) \cr}} where the Donaldson-Uhlenbeck-Yau equation
yields $f_1 + f_2 = 0$ \carluest. Observe that \carluest\ have
not taken the warp factor into account in the metric and
therefore their results do not involve the third component of the
gauge field. Furthermore, in our case, even though we have
restricted our metric to the form \metans\ with the warp factor on
the $K3$ direction, we are no longer considering the case in
which the warp factor depends only on the coordinates $z^1, z^2$
and is completely independent of the coordinate $z^3$. This would
imply that the $F_{a 3}$ and $F_{a \bar 3}$ components are
non-zero. The Donaldson-Uhlenbeck-Yau equation restricts the
values of these components further. The additional constraint
equation is \eqn\addconst{ {f_1} + {f_2} + 4(\Delta^2 + 5\vert
z^1 \vert^2 + \vert z^2 \vert^2)f_3 + 4i {\rm {\bf Im}}~(z^2
{\bar f}_{1\bar 3}) + 4 {\rm {\bf Re}}~\left[(2i+1) z^1{\bar
f}_{2\bar 3}\right] = 0,} which, in the absence of the third
components of the gauge field, will become the condition
elaborated in \carluest. From the above analysis one can see that
the $U(1)$ bundle \fone\ on non-K\"ahler space of \sav,\beckerD\
have to satisfy a set of equations resulting from (a)
Donaldson-Uhlenbeck-Yau equation \addconst, (b) equation of
motion \frel, and (c)the  Bianchi identity \trfwef. The case when
\eqn\zetafour{ \Delta = 1, ~~~~~\zeta_4(1) = 0, ~~~~~ f_{a
3}=f_{a \bar 3} = 0, ~~~~~f_3 = 0,} has been worked out in
\carluest. For our case the situation is involved which we will
discuss in some details soon. Let us go back to the issue of the
number of generations for the case when the group is simple. Ten
dimensional heterotic theory has a gravity multiplet and a vector
multiplet with components
$$ (g_{mn}, \Psi_m^-, B_{mn}, \Psi^+, \phi)~~\oplus~~(A_m, \Lambda^-)$$
\noindent where $\pm$ denore the chiralities of the fermions which
are Majorana-Weyl. The vector multiplet\foot{We are suppressing
the gauge indices.} transforms in the adjoint of the gauge group
$SO(32)$ (or $E_8 \times E_8$). The spinor in the vector
multiplet will decompose in the usual way as $\Lambda^-(w) =
\psi(x) \otimes \eta(y)$, where $\psi(x)$ is the spinor on the
four-dimensional Minkowski space and $\eta(y)$ is the spinor on
the non-K\"ahler manifold with $x^\mu$ and $y^m$ as the
coordinates of these spaces respectively. Also let the ten
dimensional gauge group $G$ decompose as $G_1$ and $G_2$ with the
four dimensional spinors transforming under $G_1$ and the six
dimensional spinor transforming under $G_2$. The ${\bf 248}$
therefore decomposes under $G \to G_1 \times G_2$ as:
\eqn\numge{{\bf 248} = \sum_{I,J}~n_{IJ}~ ({\bf I,J})} where
$I,J$ are the representations of $G_1,G_2$ respectively and
$n_{IJ}$ is the multiplicity of a given generations (we are
following the notations of \barsV, which the readers can refer to
for greater details). The number of generations (in the
representation $I$) can now be evaluated from the six dimensional
Dirac index. The result is: \eqn\diracgen{N^I_{\rm gen} = {1 \o 8
\pi^3} \sum_J n_{IJ} \left( {1\o 6} \int {\rm tr}_J~ (F \wedge F
\wedge F) - {1 \o 1440} \int {\rm tr}_J ~F \cdot {\rm Tr}~(F
\wedge F) \right)} where the subscript $J$ and ${\rm Tr}$ stand
for $J$ representation and the trace in the adjoint
representation respectively. For any simple gauge group other
than $U(1)$, ${\rm tr}_J~F = 0$, and therefore the number of
generations is given by a simple formula. For these gauge groups
it really does not matter whether $c_2(F)$ vanishes or not. The
final result for any simple gauge group is given by the third
chern class $c_3(F)$, just as outlined in section 2.4.

Let us now return to the specific $U(1)$ example that we were
discussing earlier.  Take $f_1, f_2$ and $f_3$ to be constants.
{}From \addconst\ it is clear that we cannot take $f_{a3}$ or $f_{a
\bar 3}$ to be constants. From \frel\ it is also clear that
$f_{1\bar 3}$ should be independent of ${\bar z}^1$ and ${\bar
z}^3$.  Similarly $f_{2 \bar 3}$ should be independent of ${\bar
z}^2$ and ${\bar z}^3$. The other component $f_{1\bar 2}$ should
be independent of ${\bar z}^1$ and ${\bar z}^2$.  Therefore, one
specific solution to the set of equations \addconst\ and \frel\
would be to take the functions $f_{a\bar b}$ as:
\eqn\funabarb{f_{1\bar 2} \equiv f_{1\bar 2}(z^1, z^2),
~~~~~f_{1\bar 3} \equiv f_{1\bar 3}(z^1, z^3), ~~~~~f_{2\bar 3}
\equiv f_{2\bar 3}(z^1, z^2, z^3),} where \addconst\ is satisfied
because the warp factor is a function of all the coordinates.  Of
course this is only a specific solution and the not the most
generic one.  Furthermore, the choice of $f_{a\bar b}$ in
\funabarb\ also has to satisfy \trfwef. To evaluate this we need
the Riemann curvature in the presence of torsion. Recall that our
torsional connection is $\tilde\omega = \omega - {1\o 2} {\cal H}$.
Using this we get \eqn\riema{R_{mnpq}(\tilde\omega) =
R_{mnpq}(\omega) + {1\o 2} \nabla_{[p}{\cal H}_{q]mn} + {1\o 4}
{\cal H}_{sm[p}{\cal H}^s_{~~q]n}} where $\nabla =
\nabla(\omega)$. This also has the usual property of
antisymmetrization. The Ricci tensor is defined as
$R(\tilde\omega)^s_{~~msn}$, which does not vanish when we have
non-trivial dilaton and gauge fields. Now using \riema\ one can
evaluate $\zeta_4(\Delta)$ in \trfwef. Plugging in the values of
$f_{a\bar b}$ in \trfwef, one can therefore determine {\it a}
solution for the $U(1)$ gauge bundle.

Knowing the explicit form of the gauge field, the next step is to
evaluate the number of generations, as reviewed in section 2.4.
All the traces are determined in terms of $U(1)$ charge in the
$J$ representation, which we will call, following \barsV, as
$e_J$. Therefore ${\rm tr}_J F^3 = e_J^3 F^3, {\rm Tr}~F^2 =
\left( \sum_{I,J}n_{IJ}~{\rm dim}~I~e_J^2 \right) F^2 = ({\rm
Tr}~e^2)~F^2$.  The number of generations is now given by \barsV:
\eqn\nbargen{N_{\rm gen}^I = {1\o 48 \pi^3} \sum_J \left[
n_{IJ}~e_J^3 - {1\over 240} e_J~{\rm Tr} ~e^2\right] \int_{{\cal
M}_6} F \wedge F \wedge F,} with the integration being done over
the non-K\"ahler six manifold. Observe also that ${\rm tr}_J F$
is non-zero only for the $U(1)$ case. However if we also demand
that the first Chern class of the bundle to be always zero then,
from the analysis above $e_j = 0$, and therefore the number of
generations will have to vanish. For the non-abelian case, this
is not a problem because having $c_1(F) = 0$ does not imply
anyway $N_{\rm gen}$ to vanish.

This completes the unphysical $U(1)$ example on non-K\"ahler
manifolds of the form of $T^2$-bundles on $K3$'s. In the next
section we will review some additional smooth examples of
non-K\"ahler manifolds that could have non-zero Euler
characteristic.

\newsec{Smooth Examples with Nonzero Euler Characteristic}

In much of this paper we use string duality arguments to construct
orbifold compactifications with fluxes. However, in principle
smooth examples also exist.

Some easy examples of manifolds that can be used in flux
compactifications are the group manifolds underlying ungauged
(2,2) WZW models \spindel, such as for example $S^3 \times S^1$,
which are, by construction, flux compactifications, albeit in
strongly-coupled sigma models.

However, such examples all have vanishing Euler characteristic,
and although the number of generations is computed from properties
of the bundle, it is still nice to have examples with nonzero
Euler characteristic.  We shall review a few such examples next.

\subsec{Connected Sums of $S^3 \times S^3$}

The manifold $S^3 \times S^3$ is almost, but not quite, a
candidate for a non-K\"ahler compactification.  It has complex
structures \foot{ These complex structures are easy to
describe.  View each $S^3$ as a $U(1)$ bundle over $S^2$.  The
$S^2\times S^2$ base has a unique complex structure, and the $S^1
\times S^1 = T^2$ fiber has a one-parameter family of complex
structures. See \ce\ for more details.}, and cannot be K\"ahler,
but unfortunately does not\foot{There is an
interesting mathematical subtlety here. The tangent bundle of
$S^3\times S^3$ is topologically trivial, and hence has vanishing
$c_1$.  Now, on a simply-connected {\it K\"ahler} manifold with
vanishing $c_1$, one automatically has a nowhere-zero holomorphic
top form, but we see in this example that the K\"ahler condition
is required for that statement to be true.} have a nowhere-zero
holomorphic top form in any known complex structure.

However, although $S^3 \times S^3$ is not suitable, connected
sums of $S^3 \times S^3$'s {\it are} suitable. Connected sums of
$S^3 \times S^3$'s are complex non-K\"ahler manifolds, just like
$S^3 \times S^3$, but unlike $S^3 \times S^3$, connected sums do
have nowhere-zero holomorphic top-form \lutian.  Existence of
suitable pseudo-covariantly-constant spinors on manifolds of this
form has been discussed in the third reference of \papado.

The Betti numbers of a connected sum of $N$ copies of $S^3 \times
S^3$ are given by $b_0=1$, $b_1=b_2=0=b_4=b_5$, $b_3=2N$, and
$b_6=1$. The fact that $b_2=0$ tells us immediately that these
manifolds cannot be K\"ahler.

Curiously, it can be shown (see \reid, \wall) that {\it all}
complex non-K\"ahler six-manifolds with $b_1=0$ and $b_2=0$ are
diffeomorphic to a connected sum of $N$ copies of $S^3 \times
S^3$, for some $N$.

It is straightforward to construct extremal transitions from
ordinary Calabi-Yau's to connected sums of $S^3 \times S^3$'s: if
the Calabi-Yau has $b_2=2n$, then shrink $n+1$ rational curves to
points, and deform the resulting conifold singularities.
Unfortunately, such transitions necessarily involve taking the
Calabi-Yau to small size, and so target-space supergravity is not
relevant -- these mathematical extremal transitions cannot be
understood physically in terms of wrapped branes, unlike cases
studied in {\it e.g.} \gms, where the extremal transitions all
took place at large-radius, where target-space supergravity is a
good approximation.

\subsec{Flops of Calabi-Yau's}

Another way to generate examples of complex non-K\"ahler manifolds
with nowhere-zero holomorphic top forms is via flops. In cases
studied previously in the physics literature, flops generate
Calabi-Yau's; however, it is also true that in general, flops can
break the K\"ahler condition, and so (modulo questions of the
existence of spinors) could be used to give smooth examples.

Now, since we are interested in heterotic compactifications, we
are concerned with more than just the underlying manifold: we
also have a bundle on that manifold. Thus, one needs to determine
whether that bundle can follow the manifold through the flop. We
would also require minimal susy in four dimension. This can be
achieved if we allow the holomorphic ($3,0$) form to exist in the
final picture. Flops and a necessary condition for bundles to
follow a manifold through a flop allowing a nowhere vanising
(3,0) form will be discussed below.

A special class of ordinary Calabi-Yau's, important for F-theory,
are the elliptically-fibered Calabi-Yau's. A great deal of
technology has been developed to study bundles on
elliptically-fibered Calabi-Yau's, starting with \fmw.

In particular, starting with an elliptically-fibered Calabi-Yau,
it is sometimes possible to perform flops in the fibers, so that
the resulting non-K\"ahler manifold is again
elliptically-fibered. This might help us to connect to some of
the later examples that we will provide via dualities of orbifold
backgrounds. These examples are all elliptically fibered as they
come from consistent F-theory backgrounds. Of course the examples
that we will discuss in this section are all smooth, but the fact
that the elliptically fibered examples could be related to some
duality chased string backgrounds (by blowing up orbifold fixed
points) may provide a way to show that they could be solutions to
string equation of motion. We will discuss a way to construct
these manifolds via flops in the later part of this section. The
issue of duality chased orbifold examples will be discussed in
the next section. In this paper however we will {\it not} be able
to provide a precise connection between the elliptically fibered
manifolds in this section and the F-theory backgrounds of the
next one. More details on this will appear elsewhere.

The techniques described in \fmw\ and later work do not, in
principle, require that the manifold be K\"ahler, or even
Calabi-Yau, so one can now apply \fmw\ on the non-K\"ahler
manifolds generated by these flops to create holomorphic vector
bundles on non-K\"ahler spaces.  This program has been pursued in
\brin.

Let us now begin with the issue of flops and bundles on a
non-K\"ahler manifold that would also preserve some susy in four
dimensions.

\vskip.1in

\centerline{$\underline{\rm {\bf
A~few~necessary~conditions~to~pull~bundles~though~flops}}$}

\vskip.1in

\noindent Given that flops of Calabi-Yau manifolds can sometimes
be used to generate non-K\"ahler manifolds, the reader might then
ask under what circumstances bundles can be pulled through the
flops. Experience with $(0,2)$ linear sigma models teaches that
many bundles can be pulled through flops, but gives little
geometric insight into what constraints exist. In this section,
we shall work out some easy sufficient conditions, directly in
geometry, for bundles to pull through flops.

Begin by recalling that every holomorphic line bundle over
$CP^1(=S^2)$ is uniquely expressible, as the direct
sum of holomorphic line bundles, hereafter referred to as its
components, and can be coded as an unordered $n$-tuple of
integers representing the Chern numbers of the components. In
particular the tangent bundle to $CP^1$ has Chern number $2$, the
trivial line bundle has Chern number $0$, and a holomorphic
bundle all of whose components have the same Chern number is the
tensor product of a trivial bundle with a line bundle, and the
Chern number of the determinant bundle of any holomorphic vector
bundle over $CP^1$ is the sum of the Chern numbers of the
components.

We now consider a copy $C$ of $CP^1$ (otherwise known as a
rational curve) holomorphically embedded in a three dimensional
Calabi-Yau manifold $X$. We have the exact sequence of
holomorphic bundles \eqn\exactseq{ 0\rightarrow T_C\rightarrow
T_X \vert C\rightarrow N_C\rightarrow 0.} Since $X$ is
Calabi-Yau, the determinant bundle of $T_X$ is trivial. It
follows that the determinant bundle of $N_C$ has Chern number
$-2$. We will be interested in the case where both components of
the two dimensional bundle $N_C$ have Chern number $-1$.  Let us
write $h$ for the line bundle over $C$ with  Chern number $1$, so
that $h^k$ is the line bundle with Chern number $k$. Then the
exact  sequence above becomes \eqn\exbecom{ 0\rightarrow
h^2\rightarrow T_X \vert C\rightarrow h^{-1}\oplus h^{-1}.} It is
not true in general that an exact sequence of holomorphic line
bundles splits. However this one does, and it is worthwhile
reviewing the calculation that establishes this. To see what the
issues are, let us begin by backing off to a higher level of
generality. Let $Y$ be any complex manifold and let \eqn\whsec{
0\rightarrow A\rightarrow E\rightarrow B\rightarrow 0} be an
exact sequence of holomorphic vector bundles over $Y$. The
sequence splits, which means that $E$ can  be identified with
$A\oplus B$, precisely  if the fiberwise identity map from $B$ to
$B$ factors through $E$. To analyze the obstruction to this, we
write consider the holomorphic cohomology sequence associated to
the exact sequence \eqn\exsescno{ 0\rightarrow {\rm
Hom}(B,A)\rightarrow {\rm Hom}(B,E)\rightarrow {\rm
Hom}(B,B)\rightarrow 0.} The section of the cohomology sequence
that is of interest to us is \eqn\cohoseq{ \cdots\rightarrow
H^0(Y,{\rm Hom}(B,E))\rightarrow H^0(Y,{\rm Hom}(B,B))\rightarrow
H^1(Y,{\rm Hom}(B,A))\rightarrow\cdots} The fiberwise identity
map of $B$ lies in the middle term, and the sequence splits if is
the image of some element on the left Thus the obstruction lies
in the cohomology group on the right, $H^1(Y,{\rm Hom}(B,A))$.
Now let us specialise to the case $$Y=C, ~~~~ A=h^2, ~~~~
B=h^{-1}\oplus h^{-1}.$$ \noindent Then one can easily show,
\eqn\homolo{{\rm Hom}(B,A)=B^*\otimes A=h^3\oplus h^3.} Now we
have $H^0(C,h^i)$ has rank $i+1$ for $i$ non-negative and rank
$0$ otherwise.  Serre duality says that $H^1(C,h^i)$ is dual to
$H^0(C,h^{-i-2})$, from which it follows that $H^1(C,h^i)=0$ for
$i>-2$, so that the obstruction to a splitting must vanish in the
case we are considering.

The importance of the vanishing of this splitting obstruction is
that it is also the first obstruction to identifying a
neighborhood of the zero-section of the normal bundle with a
neighborhood of $C$ in $X$, which is essential to what follows.
Without going into further details, it is the case that all
obstructions to this identification vanish for similar reasons.

Now we proceed to define the flop. By the above remarks, we will
can restrict our attention to a suitable neighborhood of the zero
section of the normal bundle. We next observe that, since both
components of the normal bundle $N_C$ have Chern number $-1$,
$N_C$ is the tensor product of  $h^{-1}$ with a trivial two
dimensional vector bundle. It follows that the projective bundle
of $N_C$ is  trivial with fiber $CP^1$, so that its total space
has the form $C\times CP^1$.

Let us again back off to more generality, to better understand the
next move. Let $E$ be a trivial $n$-dimensional vector bundle
over $Y$. Then the projective bundle of $E$ is trivial with fiber
$CP^{n-1}$ Then $C^n-0$ is the total space of a line bundle over
$CP^{n-1}$, which we will call $h'$. It follows that the
complement of the zero section is the total space of a line
bundle over $E\times CP^{n-1}$ with fiber $\pi_2^\oslash(h')$,
where $\pi_2^\oslash(h')$ denotes the pullback of $h'$ by the
projection on the second factor. If we choose a line bundle $L$
ove $Y$ and replace $E$ by $E\otimes L$, the projective bundle is
trivial as before, but now the complement of the zero section is
also the complement of the zero section in the total space of
the  line bundle
$$\pi_1^\oslash(L)\otimes \pi_2^\oslash(h').$$
\noindent Returning to the case of interest, we have $Y=C$, where
$C$ is a copy of $CP^1$.Moreover $n=2$ so that the fiber of the
projective normal bundle is also $CP^1$. Finally both $h'$ and
$L$ have Chern number $-1$ and so are identified with $h^{-1}$,
so that we can write the line bundle over $C\times CP^1$ whose
total space (again minus the zero section)  is the complement of
the zero section of $N_C$ as $h_1^{-1}\otimes h_2^{-1}$, where
the subscript is keeping track of the factors.

\noindent It follows that the following three spaces are
identical:

\item{1.} The complement of the zero section in the total
space of $N_C$, as a two-dimensional vector bundle over $C$.

\item{2.} The complement of the zero section in the total
space of the line bundle $h_1^{-1}\otimes h_2^{-1}$ over $C\times
C'$, where we write $C'$ for the fiber of the projective normal
bundle.

\item{3.} The complement of the zero section in the total
space of a two-dimensional bundle over $C'$, obtained by
exchanging the roles of $C$ and $C'$.

\noindent For first and third forms, the Serre homology sequence
collapses and the homology of the total space is the tensor
product of the homology of $CP^1$ with the homology of a
three-sphere (actually a punctured $C^2$). The inclusion of the
zero-section, in either case, kills the three-dimensional
generator and leaves the homology of the base $CP^1$.

The flop consists of removing the zero-section of the first form
($C$) and including instead the zero-section of the second form
$C'$. {}From the second  form, we can see that the commn
complement of the zero section is, up to homotopy type, a circle
bundle over $CP^1\times CP^1$ and, in the Serre homology
sequence, the circle generator cancels the sum of the two
fundamental classes. It follows from this that, although the flop
does not change the homology of the total space, the fundamental
classes of $C$ and $C'$ represent opposite homology classes.

{}From this, it follows that if the same two-dimensional homology
class that is represented by the original rational curve is also
represented by some disjoint curve (rational or not), then the
manifold created by the flop is not K\"ahler.  This is because,
in a K\"ahler manifold, the integral of the K\"ahler class over
the fundamental cycle of any holomorphic curve must be positive.

The next question to be addressed is the extension of the
holomorphic three-form over the flop. The existence and uniqueness
of the extension is guaranteed by  Hartog's theorem, which states
that a holomorphic function can always be extended uniquely
across a submanifold of complex co-dimension at least two.
Replacing the function by its inverse, it also follows that the
extension of a nowhere vanishing function also does not vanish on
the flop.

Hartog's theorem also guarantees that any holomorphic line bundle
can be extended across the flop. This is because the transition
functions take their values in the non-zero complex numbers. We
can use this
 fact to address the question of whether a gauge bundle extends across
 the flop, but this turns out to be a more subtle question.

We know that the restriction of the gauge bundle to $C$ is, like
every holomorphic bundle, a direct sum of line bundles. If we
could extend this decomposition to a neigborhood of $C$, we could
extend each summand across the flop by Hartog's theorem and we
would be finished. What we must study, therefore, is the
obstruction to extending the direct sum decomposition to a
neighborhood of $C$.

Let us write $E$ for the gauge bundle and $E'=\sum L_i$ for the
 decomposition as a sum of line bundles. Because we can identify
  a neighborhood of $C$ with a neighborhood of the zero section of
  $N_C$, we can use the fiberwise projection to  extend $E'$ to
   a neighborhood of $C$. We have an isomorphism between $E$ and
   $E'$ along $C$. If we can extend this isomorphism, even as
   homomorphism, the extension will be an isomorphism sufficiently
   near $C$ and we will be finished.

Thus we must study the obstructions to extending a section of
${\rm Hom}(E,E')$ from $C$ to a neighborhood of $C$. There is a
sequence of obstructions which live in \eqn\obstr{H^1\left(C,{\rm
Hom}(S^i N), {\rm Hom}(E,E')\right)=H^1(C), \quad i\geq 1} where
$S^i(N)$ denotes the $i$th exterior power of $N$. Along $C$,
$S^iN^*\otimes E^*\otimes E'$ is a sum of many line bundles, each
of which has the form $h^i\otimes L^*\otimes L'$, where $L$ and
$L'$ are summands of $E$ and $E'$, respectively. Along $C$, we
may set $L=h^j$ and $L'=h^k$, so that the summand in question
becomes $h^{i-j+k}$. We recall next that
\eqn\rechval{H^1(C,h^{i-j+k})=0,  ~~~~~ \forall ~~i-j+k> -2,} so
that in order for some obstruction not to vanish, there must be
summands $L$ and $L'$ of $E$ and $E'$, and a positive integer
$i$, for which $k-j\leq -(2+i)\leq-3$. We recall now that $E$ and
$E'$ are isomorphic along $C$, and we reach the conclusion that
all obstructions must vanish provided the Chern numbers of
distinct summands of $E \vert C$ do not differ by more than 2. We
emphasize that this is a  sufficient condition, but not a
necessary one; the obstructions might vanish even if the
cohomology groups in which they must live do not.

\vskip.1in

\centerline{$\underline{\rm {\bf Flops ~on ~elliptically-fibred
~Calabi-Yau ~manifolds}}$}

\vskip.1in

\noindent
 As mentioned previously, a great deal of technology has
been developed to handle bundles on elliptic Calabi-Yau's, and so
flops of elliptic Calabi-Yau's, in which the K\"ahler property is
broken by extra fiber components, are of interest, as they
represent non-K\"ahler manifolds on which some well-developed
bundle technology could be quickly applied.  In this
sub-subsection, we shall comment further on this class of
non-K\"ahler manifolds.

To achieve an elliptically fibered non-K\"ahler manifold, our
task here will be to identify floppable rational curves on
elliptically fibred Calabi-Yau manifolds. We shall be interested
in those rational curves that are contained in singular fibres.
Let $B$ be the base manifold. We will assume that $K^*_B$, the
anticanonical bundle of $B$ is semi-ample, which means that it
has sufficiently many holomorphic sections for our purposes.
Next, we recall that the anticanonical bundle of $CP^2$ is $h^3$,
where we carry over our notation for line bundles on $CP^1$ to
$CP^2$. We will consider elliptically fibred varieties that can
be presented as the zero-locus in  $B\times CP^2$ of a
holomorphic  section of the bundle $K^*_B\otimes h^3$.  Such a
section gives a holomorphic map $f$ from $B$ to the
nine-dimensional projective space of homogeneous cubics on
$CP^2$. The space of singular cubics is an eight  dimensional
variety which intersects itself along the seven dimensional space
of decomposable  cubics. From these dimensions, it follows that
the image  $f(B)$ in the space of cubics should intersect the
space of singular cubics in a curve, which will cross itself at
finitely many points, in which $f(B)$ meets the space of
reducible cubics. We will call $b$ an exceptional point if
$\pi^{-1}(b)$ is singular, and a decomposable point if it is a
double point on the space of exceptional points.

Let $X$ be our Calabi-Yau manifold and $\pi:X\rightarrow B$ the
restriction to $X$ of the projection of $B\times CP^2$ on the
first factor. Then for $b\in B$ $\pi^{-1}b$ is the zero-locus in
$CP^2$ of the cubic $f(b)$. For $f(b)$ non-singular this is an
elliptic curve, reproducing the elliptic fibration. For $f(b)$
decomposable, $\pi^{-1}(b)$ is the union of a line and a quadric
in $CP^2$, both of which are rational curves.

We now need to investigate the normal bundle of such a curve. We
begin by observing that if $x\in \pi^{-1}(b)$ is any non-singular
point of any fiber, $\pi$ induces a linear isomorphism from the
normal bundle to the fiber at $x$ to $T_B(b)$, the tangent bundle
of $B$ at $b$. In particular, this trivializes the normal bundles
to all non-singular fibers.

At a singular point of fiber, the fiber crosses itself.It follows
that the image of the normal bundle to either branch of the fiber
in $T_B(b)$ is one dimensional, and coincides with the tangent
line to the curve of singular points at $b$.

Now let us consider a decomposable fiber $\pi^{-1}(b)$ of the form
$C\cup C'$, where $C$ and $C'$ are rational curves, meeting at
two points. At those two points, the map from the normal bundle
to either $C$ or  $C'$ to $T_B(b)$, the  tangent space to $B$ at
$b$ has a kernel, generated by the tangent space to the other
component. It follows that, at these points, the image of the
normal bundle to either component in $T_B(b)$ is one dimensional.
The images of the normal bundle to either component of the fiber
at these two points are the tangent lines to the two branches of
the space of exceptional points through $b$. These observations
allow us to conclude that the normal bundle to either component
of a decomposable fiber is $h^{-1}\oplus h^{-1}$. The argument is
as follows:

We know that the normal bundle has the form $h^{i}\oplus h^{j}$
with $i+j=-2$. The projection of the normal bundle to $T_B(b)$
induces a fiberwise map from $h^{i}\oplus h^{j}$ to a trivial
two-dimensional bundle, which is surjective on all but two
fibers. It follows that both $i$ and $j$ are non-positive. This
leaves only $i=j=-1$ and $i=0, j=-2$ (or vice versa) as
possibilities. In the latter case, however, the image of the
normal bundle in the trivial bundle would be the same at both
singular points. Since this is not the case, each component has a
$(-1,-1)$ normal bundle and is floppable.

This ends our discussion of elliptically fibered non-K\"ahler
manifolds. Since all these manifolds come from flops of Calabi-Yau
manifolds and are also elliptically fibered, they could in
principle be solutions to low energy equations of motion. In this
paper we will not check these details, but instead we will provide
explicit examples of non-K\"ahler manifolds that would satisfy
string equations of motion in the following section.

\newsec{Some Orbifold Examples}

In this section we will present more examples of non-K\"ahler
compactifications with nonzero Euler characteristic. All of the
examples presented in this section will be orbifolds, and will be
justified by string duality chasing from consistent orientifold
backgrounds in F-theory.

\subsec{The Duality Chains}

Let us first go to F-theory (or M-theory) where we will consider
an orbifold of $T^8$.  We will consider the orbifold action
denoted ${\cal I}_4 \times {\cal I}_4 \times {\cal I}_4$ where
the three generators of ${\cal I}_4$ are $(1, g_1) \times (1,
g_2) \times (1, g_3)$, where the $g_i$ are defined by
 \eqn\gigiven{\eqalign{&
g_1:~~~(z^1, z^2) ~~\to ~~(-z^1, -z^2)\cr & g_2:~~~(z^1, z^3)
~~\to ~~(-z^1, -z^3)\cr & g_3:~~~(z^2, z^4) ~~\to ~~(-z^2, -z^4)}}
and where $z^{1,2,3,4}$ are the coordinates of $T^8$. The above
fourfold is one of the six distinct orbifold points of ${K3
\times K3 \over Z_2}$ \gopmuk. The other five choices are given
in terms of orientation reversal and possible half-shifts of
coordinates \gopmuk. (The $Z_2$ action above will be elaborated
soon). In terms of F-theory description the four manifold will be
a Calabi-Yau fourfold and therefore in some appropriate limits
should reduce down to a Calabi-Yau threefold in the type IIB
theory \senF. This consideration, though correct, is rather naive
at this point. The issue is a little subtle due to the $Z_2$
ambiguities of the orientifold actions on the twisted sector
states. We will elaborate this as we go along.

{}From the above choice of the fourfold we see that we could go
to a point in the moduli space 
where the $Z_2$ action acts along the F-theory torus direction as
orientation reversal. This $Z_2$ action also reverses another
$T^2$ direction of the remaining orbifold of $T^6$. The four-fold
therefore looks like \eqn\ffold{{\cal M}_8 = {T^6/ \Gamma \times
T^2 \over Z_2}} where the $\Gamma$ action is such that the
orbifold $T^6/\Gamma$ is a blow-down limit of a (51, 3)
Calabi-Yau threefold.  In fact, the above fourfold is an example
of a Borcea fourfold \borcea.  The $Z_2$ action on each $K3$ is
given in terms of three integers $(r_i, a_i, \delta_i)$ where $i=
1,2$ label the two $K3$. These $Z_2$ actions reverse the $(2,0)$
forms of each $K3$ but preserve the $(4,0)$ form.  The Euler
characteristics of these manifolds are given by\foot{A very short
summary of Nikulin three-folds \nikulin\ and Borcea four-folds
\borcea\  can be given as follows:  Define two quantities $g_i$
and $f_i$ as $g_i = {1\o 2}(22 - r_i - a_i), ~f_i \equiv 1 + k_i
= 1 + {1\o 2}(r_i - a_i)$. In terms of these, Nikulin manifolds
have Hodge numbers $h^{1,1} = 5 + 3 r_1 - 2a_1,~ h^{2,1} = 22 -
3r_1 - 2a_1$ with Euler characteristics $\chi = 12(r_1 - 10)$. On
the other hand, Borcea four-folds have Hodge numbers
 $h^{1,0} = h^{2,0} = h^{3,0} = 0$ and $h^{1,1} = r_1 + r_2 + f_1 f_2,
~h^{2,1} = f_1g_2+f_2g_1,~ h^{3,1} = 40 - r_1-r_2+g_1g_2,~
h^{2,2} = 2[102+(r_1-10)(r_2-10)+f_1f_2+g_1g_2]$. The mirror
symmetry in these manifolds is given by the operation: $r_i
\leftrightarrow 20-r_i$.} \eqn\eulborc{\chi =  6(r_1 - 10)(r_2 -
10) + 288,} so in particular their Euler characteristics can be
nonzero.  We will exploit some of the properties of these
fourfolds to construct, via duality, examples of non-K\"ahler
six-manifolds that have non-zero Euler characteristic.  As
discussed above, we go to a point in moduli space where the
Borcea fourfolds are realized as ${\cal I}_4^3$ orbifolds of
$T^8$. This will help us to connect smoothly to the other
examples studied in this paper and earlier in \sav,\beckerD. Let
us elaborate on this connection briefly. The first duality chain
used was of the form \eqn\chainone{F~~{\rm on}~~{T^4\o {\cal
I}_4}~~{}^{~\cal R}_{\longrightarrow}~~IIB~~{\rm on}~~{T^2\o
(1,g)}~~ {}^{~\cal T}_{\longrightarrow}~~IIB~~{\rm on}~~{T^2\o
(1,\Omega)}} where $g$ was defined in \ganahare, ${\cal R}$
denotes the reduction from F-theory to type IIB and ${\cal T}$
denotes the two T-dualities to go from type IIB to type I. In the
presence of fluxes this argument can be used to give us a
six-dimensional non-K\"ahler manifold.  However, since $T^2$ has
zero Euler characteristics, the resulting non-K\"ahler space also
has zero Euler characteristic\foot{An alternative approach would
be to consider type IIB on ${T^6\o (1, {\cal I}_4) \times (1,g)}$
and then follow the T-dualities to get type IIB on ${T^6\o
(1,\Omega)\times (1, {\cal I}_4)}$.  This would give the same
result, with anomaly cancellation related to the $O3$ charge
cancellation from the generator $g{\cal I}_4$.}.

The next chain of dualities was used in the section 3, in
the context of conformal $K3$'s. Specifically, consider the
duality chain \eqn\chaintwo{F~~{\rm on}~~{T^6\o {\cal I}_4 \times
{\cal I}_4}~~{}^{~\cal R}_{\longrightarrow}~~IIB~~{\rm on}~~
{T^4\o (1,g) \times (1,h)}~~ {}^{~\cal
T}_{\longrightarrow}~~IIB~~{\rm on}~~{T^4\o (1,\Omega) \times
(1,gh)},} where the last sequence can be S-dualised to describe
heterotic strings on an orbifold limit of $K3$.  In the presence
of fluxes, as we saw in section 3, the $K3$ manifold
picks up a conformal factor $\Delta^2$. However one has to be
slightly careful while defining the $Z_2$ operation. It turns out
that there is a {\it two-fold} ambiguity in the $Z_2$ action. The
F-theory orbifold that we are using here is a blow-down limit of
the $(3,243)$ Calabi-Yau.  The final picture in type IIB is a set
of intersecting seven branes and orientifold seven planes, which
in type I become $D5$ branes and $D9$ branes (with corresponding
orientifold planes). As discussed in \polci,\gopmuk,\dabholkarP,
the action of $\Omega$ on the twisted sector states is ambiguous.
There is another choice of action that flips the signs of twist
fields at all the fixed points, and yields multiple tensor
multiplets in six dimensions. An alternative way to see this
ambiguity in the action of $\Omega$ on twist fields would be to
observe how the ${\cal N} = 2$ six dimensional multiplets
decompose under ${\cal N} = 1$. The sixteen tensor multiplets of
${\cal N}= 2$ decompose as 16 tensor multiplets {\it plus} 16
hypermultiplets of ${\cal N} = 1$.  Therefore we can define
$\Omega$ in two ways: either preserving the hypermultiplets (and
projecting out the tensor multiplets) or preserving the tensor
multiplets (and projecting out the hypermultiplets).  The $GP$
model discussed earlier, realizes the former.

Motivated by the two examples above, the duality chain that we
will use to get a six-dimensional manifold with non-zero Euler
characteristic is: \eqn\chainthree{\eqalign{F~~{\rm on}~~{T^8\o
{\cal I}_4 \times {\cal I}_4
 \times {\cal I}_4}~~
&{}^{~\cal R}_{\longrightarrow}~~IIB~~{\rm on}~~{T^6\o (1,g)
\times (1,h) \times (1,k)}~~\cr &{}^{~\cal
T}_{\longrightarrow}~~IIB~~{\rm on}~~{T^6\o (1,\Omega) \times
(1,gk) \times (1,hk)},}} where the action of $k$ is defined
parallel to the actions of $g$ and $h$, i.e $k$ would be like
$\Omega\cdot (-1)^{F_L} \cdot \sigma$ where $\sigma$ reverses the
other $T^2$. Now there is a {\it four-fold} ambiguity \gopmuk,
which is related to the ${\cal N} = 4$ vector multiplets
decomposing as a ${\cal N} = 1$ vector multiplet and three chiral
multiplets.

We can now relate the operation $\Gamma$ in \ffold\ to the third
step of \chainthree.  Specifically, $\Gamma$ is the same as $(1,
gk) \times (1, hk)$.  Performing an S-duality should map the
compactification to heterotic theory on ${T^6\o (1,gk) \times
(1,hk)}$, which by construction has non-zero Euler characteristic.
This naive expectation is however complicated by orientifold
ambiguities. In particular, there are two distinct choices of
$T^6$ orbifold in this case. One choice of $T^6$ orbifold in the
type IIB theory is the blow-down limit of a $T^2$ fibration over
a $P^1 \times P^1 \times P^1$ base, and the other choice is more
complicated. This 2-fold choice is where the ambiguity discussed
above manifests itself. The blown up version where we have a
F-theory picture of the manifold is dual to the type I orbifold
constructed in \berkoozL\ (as first proposed in \gopmuk).  So
even though the brane configurations look similar, the open
string spectra of these models will be different. The full
F-theory picture has to be worked out case by case, {\it a-la}
\senF.  In short, these fourfolds behave analogously to the
$(3,243)$ and $(51,3)$ Calabi-Yau 3-folds.

Despite all the differences, there are many similarities between
the two
 models.  For example, the orientifold planes in both the models are
identical. Fixed points under the generators $g,h$ and $k$ give
rise to intersecting $O7$ planes that are arranged according to
their respective $\sigma_i$ actions on the tori. Fixed points
under the generator $ghk$
 give rise to $O3$ planes. These $O3$ planes are space filling and are
therefore oriented along the directions $x^{0,1,2,3}$. Fixed
points under $gh$, $hk$ and $gk$ give rise to orbifold fixed
points. The twisted and untwisted closed string states are also
identical.  The spectrum is given by \eqn\neqonesug{(g_{\mu\nu},
\psi_\mu)~~\oplus~~ 55(\phi, \lambda)~~\oplus~~{\rm
open~string~states}} (where $\phi$ is a complex scalar in the
chiral multiplet) out of which 7 chiral multiplets come from the
untwisted sector and 48
 chiral multiplets
come from the fixed points of the $T^6$ orbifolds. The difference
between the two models come from the open string states. These
have been elaborated in \berkoozL\ and in \gopmuk, so we refer
the reader to them for more detail.  It is however interesting to
observe that the type I framework will now have three
intersecting five branes, the third set coming from the $D3$
branes in type IIB, put in to cancel the $O3$ charges.

The F-theory fourfold compactification discussed in connection
with the two models also defines an M-theory compactification, by
doing a circle reduction along an orthogonal direction.  We can
now use the full techniques of M-theory analysis $-$ along the
lines of \rBB,\sav\ $-$ to analyse the system when we switch on
$G$-fluxes.  There are various consistency conditions that depend
on the reality condition on the fluxes and also the precise
orbifold group.  Switching on fluxes will also freeze many of the
moduli, in particular all the complex structure moduli. Previously
in \beckerD, we saw that this problem can be tackled easily if we
go to type IIB theory instead of working solely in M-theory.  We
will therefore do the analysis from the IIB point of view
and try
to derive a consistent picture in M-theory using the results of
our IIB analysis.

\subsec{Analysis of Type IIB Background}

The manifold
that we consider in type IIB is ${T^6\o (1,gk)
\times (1,hk) \times (1,k)}$ with a choice of $H_{NS}$ and
$H_{RR}$ fluxes that fix the toroidal complex structure moduli to
$\tau_{ij} = {\rm diag}~(i,i,i)$.\foot{Alternatively, we can also consider
the orientifold ${\cal M}_6 = {{\cal N}_6 \o (1,k)}$ where ${\cal N}_6$
is the blow-up of the orbifold ${T^6\o \Gamma}$. This is somewhat similar
to the case studied in the earlier sections where we had a four-dimensional
manifold which is an orientifold of $K3$ with the orientifold action
given by ${K3 \o (1,k)}$.}
This way we can combine the two
real coordinates of the torus and define $g, h$ and $k$
generically as $\Omega \cdot (-1)^{F_L} \cdot \sigma_i$ with
$\sigma_i: z^i \to -z^i$. Since the orientifolding action is
ambiguous we will keep the Euler characteristics $\chi$ and the
background axion dilaton $\varphi$ arbitrary.

Before going into the subtleties of the orbifold and the
orientifold projections, let us first tackle easier issues. The
first thing to do is to see the possible choices of background
fluxes in this model. Remembering that this should keep $\tau$
and $\varphi$ fixed at some values, the result could be
interpreted in terms of type IIB superpotential.  Using the
notation of \kst, the background three-forms can be written in
the form: \eqn\bcakthform{\eqalign{&H_{RR} = a_0 \alpha_0 + {\rm
tr}~(a^\top \alpha) + {\rm tr}~(b^\top \beta) + b_0 \beta_0,\cr
&H_{NS} = c_0 \alpha_0 + {\rm tr}~(c^\top \alpha) + {\rm
tr}~(d^\top \beta) + d_0 \beta_0,}} where the matrices $\alpha$
and $\beta$ are defined in \kst\ (see for example eq. 2.17 in the
second reference)
and the matrices
$a,b,c$ and $d$ are defined in \beckerD\ (see eq. 4.13). For a
generic choice of $\tau \equiv {\rm diag}~(\tau_1, \tau_2,
\tau_3)$ and background axio-dilaton $\varphi$, the
superpotential is given by: \eqn\iibsuper{\eqalign{W & = (a_0 -
\varphi c_0)~ {\rm det}~ \tau - {\rm det}~\tau~{\rm tr}[(a -
\varphi c)^{\top}{\tau}^{-1}] - {\rm tr}[(b - \varphi d)^{\top}
\tau] - (b_0 - \varphi d_0)\cr & = (a_0 - \varphi c_0)\tau_1
\tau_2 \tau_3 - (a_1 - \varphi c_1) \tau_2 \tau_3 - (a_2 -
\varphi c_2) \tau_1 \tau_3
 - (a_3 - \varphi c_3) \tau_1 \tau_2 \cr
& ~~~ - (b_1 - \varphi d_1) \tau_1 - (b_2 - \varphi d_2) \tau_2
 - (b_3 - \varphi d_3) \tau_3 - (b_0 - \varphi d_0).}}
At this point let us survey the situation. We require all the
complex structure moduli to be fixed to $i$ and therefore the
$T^6$ torus factorizes as $T^2 \times T^2 \times T^2$, with equal
complex structures for each of the tori.  On the other hand we
will keep the axion-dilaton to be $\varphi$ which may or may not
be fixed at $i$. This situation will remind readers of the GP
model. For the conformal $K3$ example the background axio-dilaton
was not fixed to $\varphi = i$.  For the present case, this would
imply that the local charges of the seven branes and planes do
not cancel. But the manifold admits complex coordinates: $z^1 =
x^4 + i x^5, z^2 = x^6 + i x^7$ and $z^3 = x^8 + i x^9$ with
their complex conjugates ${\bar z}^k$ ($k = 1,2,3$), where we
have each of the tori as $x^{4,5}, x^{6,7}$ and $x^{8,9}$. In
terms of the notations of \kst, the three complex coordinates
would be $z^1 = x^1 + i y^1, z^2 = x^2 + i y^2$ and $z^3 = x^3 +
iy^3$.  We will interchangably use either of the coordinate
systems (this should hopefully be clear from the context).

The equations of motion that result from the superpotential
\iibsuper\ have already been discussed in \kst\ (see eq. 3.22 in the second
reference).
These equations are written in terms of a generic $\tau$ and a
generic $\varphi$.  Let us consider, for simplicity, a case in
which we fix both $\varphi$ and $\tau$ in the form discussed
earlier, {\it i.e.}, $\varphi = i$, along with $\tau = {\rm
diag}~(i,i,i)$. Earlier we stated that one can choose three-forms
that fix $\varphi$ and $\tau$ to these values; precisely which
three-forms do this? It is not too difficult to work out the
three-forms using the expressions \bcakthform. The result can be
written in terms of $z^i$ above as \foot{This arrangement is
partially motivated by similar results in \beckerD. The fact that
this is consistent will become obvious soon.}:
\eqn\btfnow{\eqalign{&H_{NS} = {1\o 4}{\Big [}
\left(d_1-a_3+i(b_2+c_3)\right)~d{\bar z}^1\wedge dz^2 \wedge
d{\bar z}^3 + \left( d_2-b_0 +i(a_0-c_1)\right)~ d{z}^1\wedge
d{\bar z}^2 \wedge d{\bar z}^3 \cr &~~~~~~~~ + \left(a_1 -c_0
+i(b_1-d_0)\right)~d{ z}^1\wedge d{\bar z}^2 \wedge d{z}^3 +
\left(d_3-a_2 + i(b_3 - c_2)\right)~ d{\bar z}^1\wedge dz^2
\wedge d{z}^3{\Big ]} \cr &H_{RR} ={1\o 4}{\Big [}\left(
-d_0-b_2+i(a_2-c_0)\right)~d{\bar z}^1 \wedge dz^2 \wedge d{\bar
z}^3 -\left( c_3+b_1+i(a_3+d_1)\right)~ d{z}^1\wedge d{\bar z}^2
\wedge d{\bar z}^3  \cr &~~~~~~~~ + \left(
b_3-c_1+i(a_1-d_3)\right)~ d{ z}^1\wedge d{\bar z}^2 \wedge
d{z}^3 + \left(-a_0-c_2+i(b_0-d_2) \right)~ d{z}^1\wedge d{\bar
z}^2 \wedge d{\bar z}^3 {\Big ]}.}} Note that the (constant)
three-forms above are written in terms of harmonic $(2,1)$ and
$(1,2)$ forms.  The reader might well ask, why are these not
projected out? The answer is simple:  the combination of orbifold
and the orientifold operations preserve these forms. Recall that
the group action on $T^6$ is $(1,gk) \times (1,hk) \times (1,k)$.
Both the $gk$
 and the $hk$ elements are orbifold operations, whereas
the generator $k$ involves orientifolding. Therefore all
components that have a leg along the direction of the generator
$k$ will survive the full action. This is similar to the case
encountered in \sav,\beckerD. In particular, this means that the
corresponding $B_{NS}$ and $B_{RR}$ fields should also have a leg
along the $z^3$ direction. (In the blown-up picture of the
${T^6/\Gamma}$ orbifold, this is also obviously true.)
 The three-forms written above can be
expressed more concisely in the form \eqn\thfoconc{H_k =
\sum_{l=1}^4 \alpha_{kl}~ \Omega_{kl},} where $\alpha_{kl}$ are
constants and $H_1 \equiv H_{NS}, H_2 \equiv H_{RR}$. The
summation is over the four harmonic forms that could be written as
\eqn\harmoni{\eqalign{& \Omega_{k1} = (-i)^{k+1}~d{\bar
z}^1\wedge dz^2 \wedge d{\bar z}^3,~~~ \Omega_{k2} =
\Omega^\ast_{k1} = i^{k+1}~ d{z}^1\wedge d{\bar z}^2 \wedge
d{z}^3 \cr & \Omega_{k3} = (-i)^{k+1}~ d{\bar z}^1\wedge dz^2
\wedge d{z}^3, ~~~\Omega_{k4} = \Omega^\ast_{k3} = i^{k+1}~
d{z}^1\wedge d{\bar z}^2 \wedge d{\bar z}^3.}} At this point one
could try to find the connection between the $\alpha_{kl}$ and
the constants $a,b,c,d$ appearing in \btfnow. The constants
$a,b,c,d$ are real whereas $\alpha_{kl}$ could be complex.
Defining two vectors $V_1$ and $V_2$ as \eqn\vectrosv{V_1 =
\pmatrix{\alpha_{k1} &\alpha_{k2} &
 \alpha_{k3} & \alpha_{k4}}, ~~~~~~ V_2 = \pmatrix{d_3 & ic_3 & ic_1 &
d_1},} we can relate them via a $4 \times 4$ matrix ${\cal P}$ as
$V_1^\top = {\cal P} ~V^\top_2$. The existence of such a matrix
${\cal P}$ signifies many things. First, the constants appearing
in \btfnow\ are not all independent; they are related to each
other through the matrix ${\cal P}$. Second, we can write all the
three-forms using fewer variables that would solve the equations
of motion deduced from the superpotential \iibsuper.  A
straightforward, but tedious analysis, reveals that the matrix
${\cal P}$ is in fact unique and is given by: \eqn\matmis{{\cal
P} = {1\o 4} \pmatrix{1&~~1&~~1&~~1 \cr 1&-1&-1&~~1 \cr
1&-1&~~1&-1 \cr 1&~~1&-1&-1}.}  One can easily check that ${\cal
P}$ is invertible and $4 {\cal P}^2 = I$.  Knowing ${\cal P}$
tells us that a background with vanishing axio-dilaton and $\tau
= i$ can be realised.  However, that constraint does not fully
specify the background, because our analysis no way fixes the
values of $\alpha_{kl}$ or equivalently $a,b,c,d$.  A more
detailed analysis is required, which we shall describe next.

Since the background involves toroidal orbifolds, a more detailed
analysis is indeed possible for this case. Instead of doing the
most general possible analysis (which can nevertheless be done),
for purposes of simplifying the exposition we will make the
assumption that $\tau = i$.  This is required because
we want to interpret all the complex structures (in type IIB) as
$i$. We will, however, keep the axion-dilaton arbitrary. With
this assumption, the background equations of motion can be
written down explicitly. They are given by a set of thirteen relations 
between sixteen variables:
\eqn\baceomcanbe{\eqalign{& (1)~~~ a_0 + b_1 - \tilde\phi (c_0 +
d_1) + e^{-\phi} (c_2 + c_3) = 0,\cr & (2) ~~~ a_2 + a_3
-\tilde\phi (c_2+c_3) - e^{-\phi}(c_0+d_1)=0, \cr & (3) ~~~ a_0 +
b_2 -\tilde\phi (c_0+d_2) + e^{-\phi} (c_1 + c_3)= 0, \cr & (4)
~~~ a_1 + a_3 - \tilde\phi (c_1 + c_3) - e^{-\phi} (c_0 + d_2) =
0, \cr & (5) ~~~ a_0 + b_3 - \tilde\phi (c_0 + d_3) + e^{-\phi}
(c_1+c_2) = 0,\cr & (6)~~~ a_1 + a_2 - \tilde\phi (c_1+c_2) -
e^{-\phi} (c_0 + d_3) = 0, \cr &  (7) ~~~ c_1+c_2+c_3 = d_0,
~~~~~~~~ (8)~~~ d_1+d_2+d_3= -c_0, \cr &  (9) ~~~
b_1+b_2+b_3=-a_0, ~~~~ (10)~~~ a_1+a_2+a_3 = b_0, \cr & (11)~~~
d_1-a_3 \pm d_2 \mp b_0 = 16 m,~~ (12) ~~~ b_2+c_3 \pm a_0 \mp
c_1 = 16 n, \cr & (13)~~~ (d_0+b_2)^2 + (c_0+a_2)^2 + (b_1+
c_3)^2 + (a_3+d_1)^2 = {\chi \o 3} - 8p,}} where $m,n$ and $p$
are integers and $\chi$ is the Euler charcteristics of the
four-fold in F-theory (or M-theory) which is an orbifold of $T^8$
torus with the orbifold group ${\cal I}_4 \times {\cal I}_4
\times {\cal I}_4$ (see \chainthree). The solutions of the above
set of equations will determine a possible consistent background
in type IIB theory. In the most general case this is rather
involved, so we impose the following simplifying condition:
$\tilde\phi = \phi = 0$. This is basically the condition $\varphi
= i$. There is now a very simple solution to the system:
\eqn\simplesol{\pmatrix{a_0&a_1&a_2&a_3\cr b_0&b_1&b_2&b_3\cr
c_0&c_1&c_2&c_3\cr d_0&d_1&d_2&d_3\cr } = \pmatrix{~~0&0&0&-8 \cr
-8&0&0&~~0 \cr -8 &0&0&~~0  \cr ~~0&0&0&~~8 \cr}} where we have
inherently chosen a Borcea four-fold whose Euler
characteristic\foot{$(r_i, a_i, \delta_i) =(18, 4, 0)$.} is $\chi
= 672$ for illustrative purpose.  One can verify that the above
result matches the conditions specified in \sav. We have also
learned something more here. The choices of $a,b,c,d$ given above
in \simplesol\ freeze all the complex structure moduli to $i$,
while also freezing the axion to zero and the coupling constant
to one. Therefore this background is a strong coupling point in
type IIB theory. The radius of the six-manifold is however not
fixed, as expected.

Further analyzing the set of equations in \baceomcanbe, along with
the matrix ${\cal P}$, we observe that most of the constants
$a,b,c$ and $d$ are in fact related. This is similar to the case
observed in \beckerD, and looks like a generic property. The
integers $m,n$ and $p$ are also not arbitrary. And so are the
coefficients $\alpha_{kl}$ in \thfoconc. They are given by:
\eqn\abcdalpha{m = 1,0, ~~~~~ n = 0, ~~~~~ p = 12, ~~~~~
\alpha_{kl} = 2.} Comparing with \sav, we see that this is what
we expected.  The background is anomaly free with 12 $D3$ branes
as well as $H_{NS}$ and $H_{RR}$ fluxes. The value of $p$
measures the number of freely propagating $D3$ branes in our
framework. In terms of M-theory, this will correspond to freely
propagating $M2$ branes. The $M$-theory background can now be
written in terms of the constants evaluated above, as:
\eqn\mthbackre{\eqalign{& {G\o 4\pi} = \sum_{l=1}^4
\left(\Omega_{1l} \wedge d{\rm x} + \Omega_{2l}\wedge d{\rm
y}\right),\cr
 & ds^2 = \Delta^{-1}~ ds^2_{012} + \Delta^{1/2}~ ds^2_{{\cal M}_8}}}
where the M-theory torus is denoted by $dz \equiv d{\rm x} +
\varphi ~ d{\rm y} = d{\rm x} +
 i ~ d{\rm y}$ and ${\cal M}_8$ is the corresponding
toroidal orbifold in \chainthree. Observe that we can explicitly
write down the metric of ${\cal M}_8$ because of its orbifold
form. Therefore \mthbackre\ gives the complete background in
M-theory for the case that we are studying here.

\subsec{Analysis of the Heterotic Background}

Knowing the M-theory background \mthbackre, we can immediately
infer the metric for the type IIB case. The behavior is
essentially given as in \sav,\beckerD, so we will not repeat the
details here. The three-form backgrounds have already been given
in the previous section.  Therefore it is now time to duality
chase the type IIB background and infer the corresponding
heterotic scenario.  As usual we will set the axio-dilaton to the
non-zero value $\varphi = \tilde\phi + i~e^{-\phi}$.  The other
fields with non-zero vevs are the two forms $B$ and $B'$ that are
derived\foot{These $B$-fields are not globally defined on the
six-manifold.} from the three-forms $H_{NS}$ and $H_{RR}$
respectively, by keeping track of the fact that they should have
a leg along the duality directions (which we denote as
$x^{8,9}$)\foot{The analysis in the previous section was for
$\chi = 672$. However we will continue to have arbitrary $\chi$
to avoid the subtleties of orientifold ambiguities.  Notice
however that for both choices of orientifolding there are
corresponding heterotic duals: the dual of $\chi = 672$ model is
given in \font\ and the dual of the other foufold is given in
\urangafont.}.

The first important point to note here is that the torsion in the
heterotic theory is now no longer determined just by $H_{RR}$.
Because of the non-trivial axion-dilaton, the torsion gets
contributions from all the background fields that include the
axion, meaning not just $B^{RR}$ but also $B^{NS}$.  If we divide
the six-dimensional space as $x^{8,9} \oplus x^{m,n}$, then the
non-zero ${\cal B}$ field in the heterotic theory is given by:
\eqn\torbfield{{\cal B} = \pmatrix{~~{\cal B}_{8m}& {\cal
B}_{89}\cr \noalign{\vskip 0.02 cm}  \cr -{\cal B}_{89}& {\cal
B}_{9m}} = \pmatrix{-B'_{9m}+\tilde\phi~B_{9m}& -\tilde\phi\cr
\noalign{\vskip 0.02 cm}  \cr \tilde\phi & B'_{8m} -
\tilde\phi~B_{8m}}.} There is also a non-zero component ${\cal
B}_{mn}$.  This component was not present in the case studied in
\beckerD.  However now we have to take this into account also,
because of the background axion and possible cross term in the
type IIB metric $g_{89}$ which we denote as $r$.  In fact the
generic form of the component is {\it quadratic} in the type IIB
$B$-fields, and is given by \eqn\bmnvalue{{\cal B}_{mn} =
\alpha_1~B_{9[m}B_{n]8} +
 \alpha_2~B_{8[m}B'_{n]8} + \alpha_3~B_{[9m}B'_{n8]},}
where for the choice of constant three-form fields in type IIB
theory the last term in \bmnvalue\ would vanish. This has been
explicitly checked in \beckerD.  For our case, we will keep this
and the other terms for completeness. The explicit values of
$\alpha_i$ are given by \eqn\alpvalye{\alpha_1 = -2 \tilde\phi,
~~~~~\alpha_2 = r, ~~~~~ \alpha_3 = 6.} Therefore the combination
of ${\cal B}$ in \torbfield\ and ${\cal B}_{mn}$ above, along
with the gauge fields plus the curvature terms would determine
the total torsion ${\cal H}$ for our case here.

Next, we shall determine the background metric. In the absence of
fluxes the heterotic dual is compactified on a `manifold' of
nonzero Euler characteristic that can be described as an orbifold
of $T^6$ by the group $(1,gk) \times (1,hk)$. In the presence of
fluxes, the non-K\"ahler nature is clear: $dJ$ is non zero
precisely because of the warp factor. Using the matrices $g$ and
$b$, defined in \Bmatrices,  the heterotic metric along the
duality directions $x^{8,9}$ has the form\foot{A derivation of
this is given in \beckerD. Interested readers may want to look
there for details.}: \eqn\hetmetdualdir{\eqalign{ds^2 = ds^2_{89}
& + {{\rm det}~(b \sigma_1 + g \sigma_3)\cdot dx \o {\rm det}~g}~
\left[ dx^8 + {\rm tr}~(b\sigma_4)\cdot dx \right]\cr & + {{\rm
det}~(b \sigma_2 + g \sigma_1)\cdot dx \o {\rm det}~g} ~ \left[
dx^9 + {\rm tr}~(b\sigma_3)\cdot dx \right],}} where $\sigma_i$
are the Chan-Paton matrices given in \chanpaton.  An important
thing to observe here is the absence of any warp factor. The
directions $dx^{4,5,6,7}$ are denoted by $dx$. Also, in the
absence of fluxes, the additional terms in \hetmetdualdir\ will
vanish and we will be left with the dual heterotic metric
$ds^2_{89}$.

There are couple of points that we should mention regarding the
heterotic metric derived above. This non-K\"ahler heterotic
compactification was determined by string duality chasing
involved the simplest of the four actions of $\Omega$. Physically
this non-K\"ahler manifold will look like a $Z_2$ action of a
torus fibration over a compact base.  This fibration is
non-trivial and is given above in \hetmetdualdir.  If we choose
another action of $\Omega$ that gives a $P^1$ fibration over $P^1
\times P^1$ base in the type IIB side (in the absense of fluxes),
then the heterotic manifold will be a non-trivial six manifold
that is a $Z_2$ action of a torus fibration over a $P^1 \times
P^1$ base. This fibration is non-trivial and will look similar to
the fibration that we have given above \hetmetdualdir, though the
details may differ. More on this will be discussed elsewhere.
Thus, with the dilaton proportional to the warp factor, the
metric \hetmetdualdir\ along with the ${\cal B}$-fields
\torbfield\ and \bmnvalue, will therefore determine a new
four-dimensional background for heterotic theory.

\newsec{Discussion and Future Directions}

In this paper we gave new examples of heterotic compactifications
with fluxes that have non-zero Euler characteristics. For the six
dimensional case, we showed how the conformal $K3$ example is a
consistent compactification by (a) duality chasing to a F-theory
model, and (b) by mapping to a $NS5$ brane configuration. These
mappings appear to be generically useful in this kind of set-ups,
as they allow us to check, if a particular non-K\"ahler complex
manifold is a consistent solution to string theory. This could not
have been done in the early literature, as string dualities were
not developed at that time. Such dualities have been helpful to
construct many heterotic examples and show their consistency.

Whenever a flux compactification can be {\it replaced} by an
appropriate brane configuration, the supergravity backgrounds can
be made to coincide, by relating the harmonic functions of the
brane set-up to the warp factors in the flux configuration. We
gave an alternative way to study the models discussed earlier in
\sav, \beckerD, \GP, \bbdp. Some of the new four dimensional
models studied here can be realized as flops of Calabi-Yau
manifolds. We have presented a detailed mathematical discussion of
this. The manifolds considered herein are phenomenologically
useful because they are heterotic models with fluxes in which many
moduli can be stabilized. We studied the DUY equation for the
vector bundles, solved it for the case of a local $U(1)$ bundle and
computed the number of generations. Let us now discuss some open
questions and future directions.

\subsec{New Issues on Non-K\"ahler Manifolds and Open Questions}

We have addressed many aspects regarding compactifications of the
heterotic string on non-K\"ahler complex manifolds with torsion.
Nevertheless, we clearly feel that these issues need a much more
detailed analysis, than the one that is provided in here. There
are also many new issues that remain to be elaborated upon. We have
shown, how a brane box like configuration, when properly
U-dualized, can result in a configuration that simulates some of
the interesting properties of the non-K\"ahler manifolds
constructed in \sav\ and \beckerD. As we mentioned in section 5,
the brane configuration can actually give us a torsional
background that is no longer a constant. 
This configuration divides the scenario into two regions. The first region
is {\it inside} the box. Here we have a six manifold with following 
topological numbers:
\eqn\topnumb{b_i~ = ~1,~ 2,~ 23,~ 44,~ 23,~ 2,~ 1; ~~~\chi ~= ~0.}
The second region is far {\it outside} the box. Now we have a six manifold 
with topological numbers:
\eqn\topnutwo{b_i ~=~ 1,~ 0,~ 20,~ 42,~ 20, ~0, ~1; ~~~\chi ~= ~0.} 
These two regions are connected by an interpolating metric whose 
details\foot{A slightly more precise analysis can be performed by 
noticing that the {\it delocalised} harmonic functions of the two $D4$ branes in the
grid can be given in terms of $f_i$ in \metofsys\ as $F_1 \equiv f_1^{-1}f_2^{-1}$ and $F_2 \equiv f_1^{-1}f_3^{-1}$
respectively with the condition that $f_2f_3 \sim 1$. After a series of $U$-dualities, we 
have the following metric: $$ds^2 = ds^2_{~0123} + F_2^{-1} (ds_8 + \alpha)^2 + F_1^{-1} (ds_9 + \beta)^2 + 
hF_1 F_2 ~d{s}^2_{~457} + h^{-1} (ds_6 + \gamma)^2,$$ \noindent where  $\alpha, \beta$ determine the non-trivial $T^2$ fibration over 
a Taub-NUT base given by the metric $$ds^2_{\rm Taub-NUT}= h ~d{\tilde s}^2_{~457} + h^{-1} (ds_6 + \gamma)^2,$$ 
\noindent with $d{\tilde s}$ defined as the scaled metric with a
scaling factor, in the delocalised limit, given by
 $F_1F_2$, and $\gamma$ determines the usual fibration of $S^1$ over a $R^3$ base for the Taub-NUT space. Observe that, the metric in the 
above form is precisely the non-K\"ahler metric that we have (in the non-compact case), as the one-forms $\alpha, \beta$ are determined 
by the $H_{NS}$ fields in the dual picture. 
At large distances, $F_i \to 1$ and therefore
reproduces the fibration correctly. One might wonder about the warp factor in front of the Taub-NUT metric. At large distances, in 
our frame work, the warp factor will be of order 1 and therefore we will not see the warping. In the fully localised picture we should
be able to see the exact metric from this analysis. Notice also that the existence of non-constant three-form fluxes have not 
changed the form of the metric that we had in \sav, \beckerD. This is again consistent with the fact that the metric comes from 
duality chasing a consistent F-theory background.} 
we have provided in sec. 5.1. At the two regions supersymmetries are
of course preserved, but we haven't checked the susy of the 
interpolating configuration though we believe that susy will be there once 
the complete background (with all the fluxes) is considered.  

Though this configuration reproduces the geometry of non-K\"ahler manifolds,
it is a subtle issue to study the gauge bundles. As we saw in sec. 5.1, it
requires a more detailed analysis of the F-theory
monodromies to fully obtain the complete picture. Furthermore, it
is a very interesting question to ask, how the superpotential
\superinhet\ term appears in this context. Presumably one has to
check, how the five-brane charges jump, as we cross the brane-box
configuration. The calibration picture developed in \gauntlett\
suggests, that this should be the case but does require a
more careful analysis.

Let us mention next that, the new six-dimensional example studied
in section 3 is incomplete in one crucial sense: we haven't
derived this background from first principles. Recall, that the
four dimensional examples studied so far (with zero Euler
characteristics) were {\it derived} from a fully consistent
dual warped M-theory background \rBB, so that it was easy to prove
their consistency. On the other hand, for the conformal $K3$
example we have reversed the duality chasing in order to map this
heterotic model to a solution of type IIB theory (and also to M-theory).
As mentioned
earlier in section 3, there are some inherent open questions in
this picture. First, the duality chasing cannot use the usual
rules of Buscher, as some of the fields depend on the directions
along which we are doing dualities. We have taken a safer path by
choosing backgrounds, that are completely independent of the
duality directions.
 But clearly a first principle proof of the existence of a
consistent type IIB background that preserves supersymmetry is
required at this point and it would be important, to do this in a
near future. Another open question appears, if we take into
account that, we haven't exposed fully the {\it two-fold}
ambiguities, that are inherent in this framework. Of course,
demanding, that the model that we are studying is dual to a
Gimon-Polchinski type model may seem to resolve the ambiguity, but
this will fix only one
 particular orientifold operation.

We have also seen, that the corresponding Gimon-Polchinski model
is no longer at constant coupling. Rather, there is a non-trivial
axion-dilaton background, along with possible non-zero $H_{RR}$.
The combination of the $H_{RR}$ and the axion T-dualizes to the
torsion on the heterotic side. Furthermore, it is known that the
constant coupling scenario in this framework is rather subtle
\senF\ because of the existence of half integer $B_{NS}$ fluxes
at the singularities \aspinwall. These fluxes are responsible to
give us $SU(4)$ gauge symmetries, instead of $SO(8)$ gauge
symmetries from the $O$-planes. Also, because of the $B_{NS}$
fields, there are no tensionless strings in this scenario \senF.
Clearly all these issues need to be worked out carefully. Our
analysis has therefore simply scratched the surface.

Before we go on, we should mention yet another thing that we
discussed in section 3. In \scali\ we briefly mentioned, that
there {\it could} be a possible supergravity background
describing our compactifications to six dimensions because we can
compactify on a four dimensional manifold that has a large volume
and a weak six dimensional string coupling constant {\it both} on
the heterotic, as well as on the type I side. We also observed an
interesting point. The size of one of the $P^1$ is of order
$\alpha'$, even though the total volume is large. This situation
parallels the one that we encountered in \bbdp, namely, the size
of the fiber is of order $\alpha'$. But, because we were
able to map this compactification to the $NS5$ brane set-up, we
showed that a supergravity solution for the system can be
expected. Therefore, this issue already appearing in \bbdp\ can
now be addressed with a more detailed investigation. This will be
discussed elsewhere.

We have constructed new four dimensional compactifications of the
heterotic theory on non-K\"ahler manifolds, that have non-zero
Euler characteristics. As we mentioned in section 7, the
orientifolding operation is again ambiguous in this set-up. We have
chosen the simplest action. For another choice of orientifold
operation, this manifold takes the form of a non-trivial $P^1$
fibration over a $P^1 \times P^1$ base. For this kind of
construction, the type IIB dual will be a product of three tori
$T^2 \times T^2 \times T^2$, with the singularities distributed
as in this figure

\vskip.2in

\centerline{\epsfbox{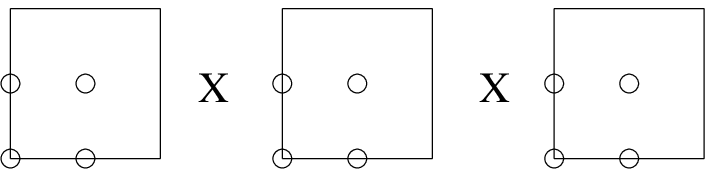}}

\vskip.2in

\noindent Here all the complex structures of the different tori
take the same value $\tau_i$. This configuration is similar to the
$(3,243)$ three-fold, that we studied in section 3 and therefore
should have a nice F-theory
 description.
Our analysis of section 3 is more useful for this example, than
for the manifold with Euler character $\chi = 672$, that we
briefly discussed in section 7. The F-theory description is given
in terms of $f(u,v,w)$ and $g(u,v,w)$, where $u,v$ and $w$
represent the three tori (see figure above). Some special choice
of $f,g$ (given in \senF), will give us a description in terms of
orientifold planes and $D$- branes. {}From there the full
non-perturbative description can be studied. Since this is related
to non-K\"ahler manifolds in the presence of fluxes, the knowledge
of the behavior of $O$-planes will be very crucial to study the
enhanced gauge symmetries in this set-up.

\subsec{De Sitter spacetimes}

The non-K\"ahler manifolds, that we studied till now all preserve
${\cal N} = 1$ supersymmetry. Therefore, the four dimensional
external space-time is Minkowski. To attain de Sitter spaces we
have to break supersymmetry\foot{Some aspects of the discussion
in this section were developed a couple of years ago by one of us
(K.D), in collaboration with C. Herdeiro and S. Hirano.}. This
can be achieved, for example,  by turning on non-primitive fluxes,
as described in \bbsb. First indications, that a cosmological
constant could be generated in the type IIB context appeared in
\BeckerNN, as the no scale structure of the type IIB theory gets
broken by quantum corrections to the K\"ahler potential, so that
after supersymmetry breaking a cosmological constant is
generated. This question has been addressed and solved recently
in an interesting set of papers \kklt, where a positive cosmological
constant in the IIB theory was found by incorporating
non-perturbative effects coming from gluino condensation. This is
rather exciting, as it has been known for some time, that our
universe has a positive cosmological constant. It is rather
important to see if such a scenario can be realized in the
phenomenologically more interesting heterotic theories considered
herein. One way to do this could be the following.

The required
supersymmetry breaking can be achieved by switching on fluxes
at the beginning of the duality chain, i.e in M-theory, which can
generate time dependence in the system \renata. In other words,
the warp factor will become time-dependent. A simple ansatz for
the metric (on the type IIB side), for example, will now be \eqn\desi{ds^2
= \Delta_1^{-1}(r, t)~(-dx_0^2 + ...+ dx_3^2) + \Delta_2(r,
t)~\left( ds^2_{K3}
 + \vert dz_3 \vert^2 \right),}
where $\Delta_{1,2}$ are the time dependent warp factors, $z_3,
{\bar z}_3$ are the coordinates of the fiber torus and $r$ is the
internal coordinate. One way to get a four dimensional
compactification with {\it positive} cosmological constant
$\Lambda$ in the heterotic theory, would be to require, that the
warp factors satisfy \eqn\warbecm{\Delta_1(r, t) = {\Delta(r) \o
\Lambda t^2}, ~~~~~~~~~~ \Delta_2(r, t) = \Delta(r),} where
$\Delta(r)$ is the warp factor for the supersymmetric case in
\becmet. The metric in the heterotic theory therefore will take
the {\it usual} form \eqn\hetmetdesitp{ds^2 = {1 \o \Lambda
t^2}(-dx_0^2 + ...+dx_3^2) + \Delta^2 ds^2_{K3} + \vert dz_3 +
{\tilde f}(z_1, z_2, {\bar z}_1, {\bar z}_2) \vert^2} where ${\tilde f}$
is defined earlier in \hemelo. This is a very simple ansatz in
which we have put the time dependence only in the $x^{0,1,2,3}$
directions. The metric in the internal directions is still time
independent. What about the other fields? As we discussed in
earlier sections, in the presence of {\it primitive} fluxes the
axion and dilaton in type IIB theory vanish, so that the complex
structure is described by  $\tau = {\tilde\phi} + i e^{-\phi} =
i$. We will assume, that this continues to hold even in the
presence of non-primitive fluxes. This is a strong assumption,
because incorporating time dependence implies, that the $D3$
branes, in the scenario of \renata,
are moving towards the $D7$ branes. Since the
axion-dilaton value is determined by the $D7$-$O7$ system, the
$\tau \approx i$ value would imply that the constant coupling
scenario is somewhat unaffected by the moving $D3$ branes. This
is not so unrealistic. If we have a single $D3$ brane in this
set-up, then this is possible, because the $D7$ branes would be
heavier than the $D3$ brane. Of course, in the exact situation we
need to consider time dependence of the whole system. Introducing
time dependence in the whole system is somewhat involved, because
as soon as the $D7$ branes start moving, non-perturbative effects
convert the orientifold seven planes into ($p,q$) seven branes.
The net effect will be to have the full F-theory picture with
additional time dependences. The system shifts from the simple
 orientifold models, that we
have been studying in the earlier sections to a more complicated
one, that has non-trivial $\tau$. Some aspects of this have
appeared in \renata. If we assume that $\tau \approx i$, it would
immediately imply that the corresponding four-fold in F-theory
will have an asymmetric time dependent warping of the form
\eqn\mbecmety{ds^2 = {\Delta^{-4/3} \o (\Lambda t^2)^{4/3}}
~ds^2_{012} + {\Delta^{2/3} \o (\Lambda t^2)^{1/3}}~ds^2_{K3
\times T^2} + (\Delta \Lambda t^2)^{2/3} \left\vert dz_4 + {\rm
f}(z_1, z_2, z_3) \right\vert^2,} where ${\rm f}(z_1, z_2, z_3)$
is a function of the three complex coordinates $z_{1,2,3}$ of the
base $K3 \times T^2$, that degenerates at 24 points on the
subspace spanned by $z_3$. In the presence of primitive fluxes,
this manifold is therefore simply $K3 \times K3$
\rBB, \sav. The $G$-flux develops time dependence {\it only}
along the directions $x^{0,1,2,a}$ as \eqn\gbeme{ G_{012a} =
{\del_a \Delta^{-2} \o (\Lambda t^2)^2},} where $a$ is a
coordinate of the internal manifold. Comparing the above results
with the two solutions \beckback\ and \metformrwo, that we had in
section 4 we see that, in terms of $M2$ branes, we have a system
of point charges on the fourfold that are in motion whose
geodesic is given by \eqn\geodbec{\quabla X^m + h_{\mu\nu}~
\del_\mu X^n ~\del_\nu X^p~ \Gamma^m_{np} - {1\o 3!}
\epsilon^{\mu\nu\rho} ~ \del_\mu X^q ~\del_\nu X^n ~\del_\rho
X^r~ G^m_{qnr} = 0,} where $X^k$ are the coordinates of the $M2$
branes. This is basically one possibility to realize a de-Sitter
 solution in the torsional compactification of the heterotic
theory for the inflationary type model of \renata.
What we haven't really verified, is whether all the set of
solutions are consistent with respect to each other. This
requires the complete solution of the system and is beyond the
scope of this paper. It will also be interesting to relate this
discussion to the recent developments on de-Sitter spacetime done
in \carlos,\renata,\kklt. In these papers the inflation present in
our universe is described in the context of string theory. In the
heterotic theory that we considered herein, this could be
developed along the lines of \renata, where there appears a model
that is specifically suited for the type of heterotic
compactifications that are studied here.

Finally let us comment about a phenomenologically promising
approach to derive de Sitter vacua from string/M-theory by going
to the strongly coupled heterotic string \howi,\howii. Here the
dominant non-perturbative open membrane instanton effects \mps\
lead to a stabilization of all model-independent moduli, the
orbifold length (dilaton), the average Calabi-Yau volume and, if
present the position of an M5 brane along the orbifold \cukri.
Indeed, one finds that the open membrane instantons break
supersymmetry spontaneously in the effective four dimensional
theory and one remains with de Sitter vacua \cukri. Moreover, the
computed value for Newton's constant is generically close to its
measured value \curioLu. In order to fully study the moduli
potential at orbifold lengths beyond that corresponding to the de
Sitter minima, one would have to use the proper non-linear
extensions of the backgrounds of \wittenstrong, that were derived
in \cukriii, \curioLu\ and include all field theory corrections
required by eleven dimensional supersymmetry \bckr. We hope to
return to these interesting questions in a near future.


\noindent \centerline{\bf Acknowledgements}

\noindent We would like to thank Edward Goldstein and Sergey
Prokushkin for initial collaboration and helpful comments. We
would also like to acknowledge useful discussions with
Andrei~Caldararu, Gottfried~Curio, Ron~Donagi, Jarah~Evslin, Simeon~Hellerman,
Stefan~Ivanov, Shamit~Kachru, Renata~Kallosh, Sheldon~Katz, Axel~Krause,
Dieter~Luest, Varghese~Mathai,
Sunil~Mukhi, Burt~Ovrut, Jogesh~Pati, Tony~Pantev, Joe~Polchinski, Ashoke~Sen,
Savdeep~Sethi, Mohammad~M.~Sheikh-Jabbari,
Andy~Strominger, Raman~Sundrum, Cumrun~Vafa, Shing-Tung~Yau and
Edward~Witten. M.B, K.D and E.S would like to thank the
organizers of QTS3 Cincinnati, where some of the results were
presented, for a stimulating workshop. The work of K.B is
supported by NSF grant PHY-0244722, an Alfred Sloan Fellowship
and the University of Utah. The work of M.B is supported by NSF
grant PHY-01-5-23911 and an Alfred Sloan Fellowship. The work of
K.D is supported by a David and Lucile Packard Foundation
Fellowship 2000-13856. The work of E.S is partially supported by
NSF grant DMS 02-96154.

\appendix{A} {Massless Spectra in Non-K\"ahler Compactifications}

In this appendix we shall describe how one can formally compute
massless spectra of large-radius non-K\"ahler compactifications
on the world-sheet. The reader should note that such calculations
assume that one can consistently work at large-radius, and in
particular necessarily assume that $\alpha'=0$, which as shown in
\PapaDI\ can not actually happen.  Thus, the resulting massless
spectrum calculation should itself probably not be taken too
seriously, although the result for the number of generations is
surely more reliable.

One other factor that should be taken into account in massless
spectrum computations is the existence of a warp factor on
uncompactified directions.  However, these warp factors are of
the form $1 + {\cal O}(\alpha')$, and as the massless spectrum
calculation we shall outline here only makes sense when
$\alpha'=0$, we shall ignore such warp factors.

We begin by reviewing the construction of sigma models with
torsion and (0,2) supersymmetry, then discussion the calculation
of massless spectra, following \disgreone\ and \rohmW.

\subsec{Heterotic Sigma Model with $(0,1)$ Supersymmetry and
Torsion}

A heterotic lagrangian with $(0,1)$ world-sheet supersymmetry is
\eqn\helaoi{\eqalign{S ~ = ~& G_{\mu \nu} \partial X^{\mu}
\overline{\partial} X^{\nu} ~ + ~ 2 {\cal B}_{\mu \nu} \partial
X^{\mu} \overline{\partial} X^{\nu} ~ ~ + ~ i G_{\mu \nu}
\psi_+^{\mu} D_{\overline{z}} \psi_+^{\mu} \cr & ~ + ~ i g_{AB}
\lambda_-^A D_z \lambda_-^B ~ + ~ \frac{1}{2} (F_{\mu \nu})_{AB}
\psi_+^{\mu} \psi_-^{\nu} \lambda_-^A \lambda_-^B}} where the
various quantities appearing above are defined as \eqn\defofevr{
\eqalign{& D_{\overline{z}} \psi_+^{\nu}  = ~ \overline{\partial}
\psi_+^{\nu} ~ + ~ \left( \overline{\partial} X^{\rho}
\right)\left( \Gamma^{\nu}_{\rho \kappa} ~ - ~ {\cal H}^{\nu}_{~
~ \rho \kappa} \right) \psi_+^{\kappa} \cr & D_z \lambda_-^B  = ~
\partial \lambda_-^B ~ + ~ \left( \partial X^{\mu} \right) \left(
A_{\mu C}^B \right) \lambda_-^A \cr & {\cal H}_{\mu \nu \lambda}
= ~ - \left( {\cal B}_{\mu \nu, \lambda} ~ + ~ {\cal B}_{\nu
\lambda, \mu} ~ - ~ {\cal B}_{\lambda \mu, \nu} \right) }} and
$\mu, \nu$ are target space coordinate indices, $A, B$ are vector
bundle indices, $G_{\mu \nu}$ is the target space metric, and
$g_{AB}$ is the hermitian fiber metric that determines the gauge
connection $A_{\mu B}^A$. The $(0,1)$ supersymmetry
transformations are given by \eqn\sustra{ \eqalign{& \delta
X^{\mu}  = ~ i \epsilon_- \psi_+^{\mu} \cr & \delta \psi_+^{\mu}
= ~ - \epsilon_- \partial X^{\mu} }}

This lagrangian is taken from \ct.

Note that classically in the sigma model we only see $d{\cal H} =
0$. The anomaly-matching condition $d{\cal H} \propto
c_2(T)-c_2(V)$ emerges in the sigma model at one-loop \ct, which
is why the right-hand-side contains a factor of $\alpha'$. Also,
for $(0,1)$ supersymmetry, we do not have a constraint relating
${\cal H}$ to the metric -- that will emerge as part of demanding
that the action possess $(0,2)$ supersymmetry.

\subsec{Heterotic Sigma Model with $(0,2)$ Supersymmetry and
Torsion}

A heterotic lagrangian with $(0,2)$ supersymmetry is \eqn\hetotwo{
\eqalign{S~  = ~ G_{\mu \nu} \partial X^{\mu} \overline{\partial}
X^{\nu} & + ~ 2 {\cal B}_{i \overline{\jmath}} \partial X^i
\overline{\partial} X^{ \overline{\jmath}} ~ - ~ 2 {\cal B}_{i
\overline{\jmath}} \partial X^{\overline{\jmath}}
\overline{\partial} X^i ~ + ~ 2 i G_{i \overline{\jmath}}
\psi_+^{ \overline{\jmath}} D_{ \overline{z}} \psi_+^i i  \cr & +
~ i g_{AB} \lambda_-^A D_z \lambda_-^B ~ + ~ \left(F_{i
\overline{\jmath}}\right)_{AB} \psi_+^i
\psi_+^{\overline{\jmath}} \lambda_-^A \lambda_-^B }} where
various quantities appearing above are defined now as:
\eqn\suttapo{ \eqalign{& D_{ \overline{z}} \psi_+^i  = ~
\overline{\partial} \psi_+^i ~ + ~ \left( \overline{\partial}
X^{k} \right) \left( \Gamma^i_{k l} - {\cal H}^i_{~ ~ k l}
\right) \psi_+^l ~ + ~ \left( \overline{\partial} X^k \right)
\left( \Gamma^i_{k \overline{l}} ~ - ~ {\cal H}^i_{~ ~ k
\overline{l}} \right) \psi_+^{\overline{l}} i \cr & ~~~~~~~~~~~~
+ ~ \left( \overline{\partial} X^{ \overline{k} } \right) \left(
\Gamma^i_{ \overline{k} l } ~ - ~ {\cal H}^i_{ ~ ~ \overline{k} l
} \right) \psi_+^l ~ + ~ \left( \overline{\partial} X^{
\overline{k} } \right) \left( \Gamma^i_{ \overline{k}
\overline{l} } ~ - ~ {\cal H}^i_{ ~ ~ \overline{k} \overline{l}
}\right) \psi_+^{\overline{l} } \cr & D_z \lambda_-^B  = ~
\partial \lambda_-^B ~ + ~ \left( \partial X^i \right) \left(
A_{i C}^B \right) \lambda_-^C ~ + ~ \left( \partial
X^{\overline{\imath}} \right) \left( A_{ \overline{\imath} C}^B
\right) \lambda_-^C \cr & {\cal B}_{i j}  = ~ 0 \cr & {\cal B}_{
\overline{\imath} \overline{\jmath} }  = ~ 0 }} and $i,
\overline{\imath}$ are holomorphic and antiholomorphic indices
(previously $\mu, \nu$ were mere $C^{\infty}$ indices). The
$(0,2)$ supersymmetry transformations are \eqn\otsuta{ \eqalign{&
\delta X^i  = ~ i \epsilon \psi_+^i \cr & \delta
X^{\overline{\imath}}  = ~ i \tilde{\epsilon} \psi_+^{
\overline{\imath} } \cr & \delta \psi_+^i  = ~ - \tilde{\epsilon}
\partial X^i \cr & \delta \psi_+^{\overline{\imath}}  = ~ -
\epsilon \partial X^{\overline{\imath}} }} {}From demanding
$(0,2)$ supersymmetry we find we must demand \eqn\demanwhat{
{\cal H}_{\mu \nu \lambda} ~ = ~ - \left( {\cal B}_{\mu \nu,
\lambda} ~ + ~ {\cal B}_{\nu \lambda, \mu} ~ + ~ {\cal
B}_{\lambda \mu, \nu} \right)} just as for $(0,1)$ supersymmetry,
and in addition, \eqn\denmnow{ {\cal H}_{i \overline{\jmath} k }
~ = ~ {1 \o 2} \left( G_{i \overline{\jmath}, k} ~ - ~ G_{k
\overline{\jmath}, i} \right)} which the reader will recognize
from \rstrom. Again, the full anomaly-cancellation condition only
emerges at one-loop. Define also \eqn\dehhit{ \eqalign{&
H^{(2,1)} = ~ {\cal H}_{i \overline{\jmath} k} dz^i \wedge d
\overline{z}^{ \overline{\jmath}} \wedge dz^k \cr & H^{(1,2)} = ~
{\cal H}_{\overline{\imath} j \overline{k} } d
\overline{z}^{\overline{\imath}} \wedge dz^j \wedge d
\overline{z}^{ \overline{k} } }} Equivalently, if $J  = i~G_{i
\overline{\jmath}} dz^i \wedge
d\overline{z}^{\overline{\jmath}}$, then \eqn\aftjdef{ \eqalign{&
H^{(2,1)} = ~ \partial J  \cr & H^{(1,2)}  = ~
\overline{\partial} J }} Then, as an additional condition for
$(0,2)$ supersymmetry, we find that $\overline{\partial}
H^{(2,1)} = 0$, {\it i.e.} the strong K\"ahler torsion condition
$\overline{\partial} \partial J = 0$.

In any event, note that so long as $\overline{\partial} H^{(2,1)}
= 0$, then we have a nilpotent operator $\overline{\partial} +
H^{(2,1)}$, that can be used to define a ${\bf Z}$-graded
Dolbeault-type cohomology theory, in close analogy with the $d+H$
cohomology that appeared in \rohmW. Dolbeault-type bundle-valued
cohomology, twisted by $H^{(2,1)}$ in this form, we shall refer
to as $H$-twisted sheaf cohomology, and will denote $H$-twisted
sheaf cohomology valued in ${\cal E}$ by $H^*_H(X, {\cal E})$.

Noether currents are straightforward to determine. The variation
of the lagrangian under the supersymmetry transformations above
is given by \eqn\noether{\eqalign{ \delta_{\epsilon,
\bar\epsilon} ~S & ~ =~ 2 i G_{i \overline{\jmath}} \left(
\overline{\partial} \epsilon\right) \left( \partial
X^{\overline{\jmath}} \right) \psi_+^i ~ + ~ 4 \left(
\overline{\partial} \epsilon \right) \psi_+^{\overline{\jmath}}
\psi_+^k \psi_+^i H_{k \overline{\jmath} i}\cr & ~~~~ + ~ 2 i
\left( \overline{\partial} \tilde{\epsilon} \right) G_{i
\overline{\jmath}} \left( \partial X^i \right)
\psi_+^{\overline{\jmath}} ~ - ~ 4 \left( \overline{\partial}
\tilde{\epsilon} \right) \psi_+^{\overline{k}}
\psi_+^{\overline{\jmath}} \psi_+^i H_{\overline{\jmath} i
\overline{k}}}} (Strictly speaking, we have omitted a three-fermi
term involving the heterotic gauge field, but it is not important
for what follows.)

In any event, we now have all we need to repeat the spectrum
analysis of \disgreone[section 3]. The analysis seems to work the
same, except that every occurrence of $\overline{\partial}$ is
replaced by $\overline{\partial} + H^{(2,1)}$.

The (large-radius) massless spectrum computation now follows
exactly the same form as \disgreone[section 3], with the
replacement above.  We shall simply refer the reader to
\disgreone rather than repeat the details of the analysis here.

\subsec{Interpretation}

How do we interpret the states counted by this $H$-twisted sheaf
cohomology?

The first point that should be made is that the usual counting of
complex \& K\"ahler moduli is no longer relevant. Low-energy
fields come from infinitesimal deformations of supergravity
fields, and the complex and K\"ahler moduli of Calabi-Yau
compactifications arise as {\it metric} moduli.

On a Calabi-Yau, since we usually cannot write the metric
explicitly, we instead use Yau's theorem, which says that a
choice of complex and K\"ahler modulus is equivalent to a
specific choice of Ricci-flat metric, and so metric moduli are
encoded in complex and K\"ahler moduli.

In the present non-K\"ahler circumstances, however, Yau's theorem
does not apply. Thus, for non-K\"ahler manifolds, complex and
K\"ahler moduli no longer need count metric moduli, and so no
longer need have anything to do with massless states at all.

An issue that might puzzle the reader somewhat more is how {\it
bundle} moduli appear in our spectrum calculations. For a given
holomorphic vector bundle ${\cal E}$, it is well-known that
moduli of the bundle are calculated by the group $H^1(X, {End }~
{\cal E})$, literally adjoint-valued one-forms corresponding to
deformations of the gauge field, preserving the holomorphic
structure, modulo deformations equivalent to gauge
transformations.

Since the K\"ahler condition is irrelevant mathematically for the
interpretation of $H^1(X, {End }~ {\cal E})$, the reader may well
wonder how it can be consistent to have twisted $H^1_H(X, {End }~
{\cal E})$ in the low-energy spectrum, but not $H^1(X, {End }~
{\cal E})$ itself.

To help understand this apparent problem, let us study the
$H$-twisted class $H^1_H(X, {End }~ {\cal E})$ more closely. A
representative of such a cohomology class has the form
\eqn\cohocla{ A^{(1)}_{ \overline{\imath} } d \overline{z}^{
\overline{\imath} } ~ + ~ A^{(3)}_{ \overline{\imath} j k } d
\overline{z}^{\overline{\imath}} \wedge dz^j \wedge dz^k ~ + ~
\cdots} where the $A^{(n)}$ are adjoint-valued forms.
BRST-closure of these representatives gives a set of conditions
on the $A^{(n)}$: \eqn\hclosure{\eqalign{& \overline{\partial}
A^{(1)}  = ~ 0 \cr & \overline{\partial} A^{(3)} = ~ - H^{(2,1)}
\wedge A^{(1)} \cr
 & \cdots
}} The first equation has an obvious interpretation:  $A^{(1)}$ is
an infinitesimal deformation of the gauge field, that preserves
the holomorphic structure.  Phrased another way, by selecting out
$A^{(1)}$ we have a map \eqn\cohmap{ H^1_H(X, {End } ~{\cal E}) ~
\longrightarrow ~ H^1(X, {End } ~{\cal E} )} which makes it clear
that $H^1_H(X, {End }~ {\cal E})$ does encode bundle deformations.

However, it should also be clear that $H^1_H(X, {End } ~{\cal E})$
encodes more than just bundle deformations. Intuitively, the
meaning of that extra data in $H^1_H(X, {End }~ {\cal E})$ should
be clear -- in backgrounds with $H$ flux, bundle deformations
cannot be considered in isolation, but rather consistency with
the supergravity equations of motion implies that bundle
deformations induce deformations of other fields. In Calabi-Yau
heterotic compactifications, infinitesimal bundle moduli decouple
from other moduli, but in compactifications with nonzero $H$
flux, such a decoupling no longer happens. In principle, the
extra information in $H^1_H(X, {End }~ {\cal E})$ should be
encoding deformations of other fields. We are currently working
to make this description of the other data more explicit.

In any event, as this massless spectrum calculation necessarily
assumes that $\alpha'=0$, {\it e.g.} that the metric satisfies
the strong K\"ahler torsion condition, which was shown in
\PapaDI\ to not be compatible with spacetime supersymmetry,
perhaps one should not read too much into the results of the
calculation.

\subsec{Bismut's Index Theorem}

How do we compute the number of generations in such models? As
discussed elsewhere in this paper, the usual trick is to first
count massless states with sheaf cohomology \disgreone, then
apply the Hirzebruch-Riemann-Roch theorem \hirze, and for
Calabi-Yau three-folds, the result is proportional to the third
Chern class.

Here, however, we no longer have honest sheaf cohomology, we have
$H$-twisted sheaf cohomology. Luckily for us, the relevant
analogue of Hirzebruch-Riemann-Roch is known, and was worked out
in \bismut.

Bismut expresses his analogue in terms of differential forms
realizing the various cohomology classes.  The Todd class
appearing in \hirze is replaced by an A-roof genus, multiplied by
a function of $c_1$ of the tangent bundle (for spaces with
trivial canonical bundle, the Todd class and A-roof genus are the
same). In particular, he uses an A-roof genus involving the
twisted Riemann curvature.

In more detail, the usual A-roof genus can be expressed in the
form \eqn\aroof{ \eqalign{ { R/2 \o \sinh( R/2 ) } ~ = & 1 ~ - ~
{1 \o 3!} \left( {1 \o 2} \right)^2 R\wedge R ~ + ~ \left( \left(
{1 \o 3!} \right)^2 - {1 \o 5!}\right) \left( {1 \o 2} \right)^4
R\wedge R \wedge R \wedge R  \cr & + ~ \left( {2 \o  3! 5! } - {1
\o 7!} - \left( {1 \o 3!} \right)^3 \right) \left( {1 \o 2}
\right)^6 R \wedge R \wedge R \wedge R \wedge R \wedge R ~ + ~
\cdots }} Here, by contrast, the usual A-roof genus is replaced
by the A-roof genus of a twisted curvature $R_H$, namely the
Riemann curvature form associated to the twisted connection
$\Gamma + H$ rather than the usual Levi-Civita connection
$\Gamma$. (The same substitution that's standard in supergravity
theories, in other words.) Thus, the new index theorem is phrased
in terms of \eqn\newind{ { R_H / 2 \o \sinh( R_H / 2 ) }} which
has the same expansion as before.

Now, calculating this A-roof genus is less useful in practice
 -- expanding out $R_H$ in terms of $R$ and $H$ gives $\Gamma \cdot H$
cross terms, for example.

However, in the case at hand, matters simplifies enormously.

Note that the A-roof genus expansion above contains only forms
whose degrees are a multiple of 4 -- this is true both for
ordinary and the $H$-twisted A-roof genus.

That fact means that when evaluating either the ordinary or
Bismut's version of Riemann-Roch on a complex {\it three}-fold,
with a gauge bundle with vanishing first Chern class, the only
contribution can come from the Chern character of the gauge
bundle -- there cannot be any contribution at all from the
$R$'s.  This is why the number of generations on a three-fold
involves only $c_3$ of the gauge bundle, and not any curvature of
the tangent bundle.

In particular, detailed knowledge of the $H$-twisted A-roof genus
is completely irrelevant for counting the number of generations
on a threefold with trivial canonical bundle.

Thus, the number of generations is counted by ${1 \o 2} c_3$,
both for the standard Calabi-Yau case, as well as for
non-K\"ahler cases, for gauge bundles with vanishing first Chern
classes, on complex three-folds.

Mathematically, this result should not be too surprising. As
pointed out to us by \mathaipriv, the operator
$\overline{\partial} + H^{(1,2)}$ is homotopic to
$\overline{\partial}$ (just multiply $H$ by $t$ and send $t$ to
zero), as a result of which, the two operators should have the
same index theory. So not only is the number of generations
counted by $c_3$ in the case $c_1=0$, but in general as well.

For the same reasons, the cohomology class of the $\hat{A}$-genus
of $\Gamma+H$ is the same as that of the Levi-Civita connection
$\Gamma$ -- so actually doing more general calculations with
Bismut's result is easy, as one does not have to worry about how
precisely $H$ enters equations.





\listrefs

\bye